\documentclass[12pt]{book}
\usepackage{setspace}
\usepackage[inner=1in, outer=1in, top=1in, bottom=1in]{geometry}
\usepackage[bottom]{footmisc}

\begin{document}

\centerline{\bf QUANTUM STRUCTURES: A View of the Quantum World}

\centerline{J. Jekni\' c-Dugi\' c$^1$, M. Arsenijevi\' c$^2$, M.
Dugi\' c$^2$}

\bigskip

$^1$Faculty of Science and Mathematics, Ni\v s, Serbia

$^2$Faculty of Science, Kragujevac, Serbia

\pagebreak

\pagenumbering{roman}

\noindent\textbf{\large Preface}

\bigskip

\noindent Quantum mechanics provides a striking and
counterintuitive observation in that it is easier to describe a
quantum whole than its constituent parts. In other words, there is
more uncertainty about quantum subsystems than about the total
system composed of the quantum subsystems. This observation
suggests that there is more to the idea of "structure"
(decomposition into parts, subsystems) in the quantum world than
there is in the classical world.

\smallskip

\noindent For example, we can smash a rock into little bits, and
each of those bits continues to follow classical physics, but
ultimately when we smash those bits to the quantum scale we no
longer see "material" so much as we see something that is better
described as "pure behavior".  And this behavior is so fast and
fleeting that we have only statistical methods with which to
continue to presume that there is "material stuff".  Odd as this
is, classically, Quantum Mechanics is so accurate that we have to
presume that the statistical methods have a lock on something very
real.  In experiments and in math it is possible to select and or
to describe different and equivalent possible views of the same
thing.  The maths of quantum physics are at a point where they
have captured the lessons of many different experiments and we can
most readily describe these views in terms of mathematical
decomposition "structures" that achieve the whole.

\smallskip

\noindent Ultimately questions arise such as:  Is there a unique
fundamental structure of a composite quantum system (is one view
any more "real" or 'better" than any other)? How do classical
structures (and intuition) appear from the quantum substrate? Can
the structural variations available in the quantum world be of any
practical use that is not known to classical physics?  These are
the main questions raised in this book. Some of the answers that
we provide are so far only partial. Nevertheless, the results
suggest that the topic of "quantum structures" will be of
significant interest to fundamental physics going forward.

\smallskip

\noindent The bulk of the results presented here have been
obtained by the authors over the last seven years. Emphasis is
placed on the appearance of "structure" in different contexts and
particular attention is given to the distinction between the
classical and quantum mechanical concept of structure. The
contents are arranged so as to provide a coherent and self
contained reading. We have endeavored to include enough
supplemental material to be "reader-friendly", while of course
providing references for more detailed investigation of the topics
covered.

\smallskip

\noindent We benefited much from discussions and encouragement
that came from Allen Francom, Nate Harshman, David Steglet,
Stephen P. King, Hitoshi Kitada and Dejan Rakovi\' c. Some lucid
observations provided by Allen Francom in a few last years have
constantly enriched our inspiration and significantly influenced
our thinking and the view of the quantum world.

\bigskip

\noindent  Ni\v s/Kragujevac, Summer 2013 \hfill J. Jekni\' c
Dugi\' c

\hfill M. Arsenijevi\' c

\hfill M. Dugi\' c

\newpage

\noindent\textbf{\large Abbreviations}

\bigskip

BP -- {\it Brownian particle}

ChM -- {\it Chemical model} [of molecule]

CM -- {\it Center of mass}

DISD -- {\it Decoherence-induced suppression of decoherence}

EM -- {\it Electromagnetic} [field]

ER -- {\it Entanglement relativity}

lhs --{\it Left hand side} [of an equation]

LOCC -- {\it Local operations and classical communication}

PFP -- {\it Protein folding problem}

POD -- {\it Parallel occurrence of decoherence}

QBM -- {\it Quantum Brownian motion}

QCM -- {\it Quantum chemistry model} [of molecule]

QCR -- {\it Quantum correlations relativity}

QDR -- {\it Quantum discord relativity}

QRF -- {\it Quantum reference frame}

RF -- {\it Reference frame}

rhs--{\it Right hand side} [of an equation]

SSM -- {\it Solid-state model } [of molecule]

\cleardoublepage

\noindent\textbf{\large Contents}

\vspace*{55mm}

\noindent {\bf 1 Introduction} \hfill 1

\bigskip

\noindent {\bf 2 The transformations of variables} \hfill 6

\smallskip

2.1 Classical physics prejudice on the transformations of
variables
 \dotfill 7

2.2. Classifications of the canonical transformations \dotfill 9

{\emph {2.2.1 Some classifications of the LCTs and structures}}
\dotfill 9

{\emph {2.2.2 Mutually irreducible global structures}} \dotfill 13

2.3 Quantum mechanical structures \dotfill 14

2.4 Quantum relativity of "system" and "locality" \dotfill 16

\bigskip

\noindent {\bf 3 Quantum Correlations Relativity} \hfill 18

\smallskip

3.1 Quantum entanglement relativity \dotfill 20

3.2 Quantum discord relativity \dotfill 22

3.3 Some mathematical remarks \dotfill 25

3.4 Some physical remarks \dotfill 26

\bigskip

\noindent {\bf 4 Quantum Molecule Structures} \hfill 28

\smallskip

4.1 Mutual relations of the molecule structures \dotfill 29

4.2 The protein folding problem \dotfill 34

{\emph {4.2.1 The statistical-thermodynamic approach}} \dotfill 35

{\emph {4.2.2 A quantum decoherence approach}} \dotfill 36

{\emph {4.2.3 Overview}} \dotfill 43

4.3 Quantum structures in context \dotfill 44

\bigskip

\noindent {\bf  5 Realistic Physical Structures} \hfill 47

\smallskip

5.1 Relativity of "local operations" \dotfill 47

5.2 Manipulating the center of mass \dotfill 50

5.3 Manipulating the relative positions \dotfill 52

5.4 Quantum correlations relativity in use\dotfill 54

5.5 Outlook \dotfill 57

\bigskip

\noindent {\bf 6 Parallel Occurrence of Decoherence} \hfill 58

\smallskip

6.1 The task \dotfill 60

6.2 The obstacles \dotfill 60

6.3 Quantum Brownian motion \dotfill 62

{\emph {6.3.1 The LCTs and the structures of interest}} \dotfill
62

{\emph {6.3.2 The Caldeira-Leggett model}} \dotfill 64

{\emph {6.3.3 $S'$ is a Brownian particle}} \dotfill 67

{\emph {6.3.4 A limitation of the Nakajima-Zwanzig projection
method }} \dotfill 72

6.4 The LCTs preserve linearity of a composite-system's model
\dotfill 80

6.5 More than one "classical world" \dotfill 81

6.6 A few general notions \dotfill 83

\bigskip

\noindent {\bf 7 Decoherence-Induced Preferred Structure} \hfill
84

\smallskip

7.1 Decoherence based  classicality \dotfill 84

7.2 Asymptotic dynamics of a two-mode system \dotfill 87

{\emph {7.2.1 Original degrees of freedom }} \dotfill 88

{\emph {7.2.2 Alternative degrees of freedom }} \dotfill 93

{\emph {7.2.3 Preferred structure}} \dotfill 96

7.3 Atom in electromagnetic field \dotfill 97

7.4 Outlook \dotfill 99

\bigskip

\noindent {\bf  8 Some Interpretation-Related Issues} \hfill 103

\smallskip

8.1 Global irreducible structures with decoherence \dotfill 103

 8.2 Local structures and classicality \dotfill 105

8.3 A unifying physical picture \dotfill 107

{\emph {8.3.1 Microscopic vs macroscopic domain }} \dotfill 108

{\emph {8.3.2 The quantum reference frame issue }} \dotfill 110

{\emph {8.3.3 The unifying picture }} \dotfill 113

8.4 Some interpretational issues \dotfill 113

{\emph {8.4.1 Non-branching of the Everett worlds }} \dotfill 114

{\emph {8.4.2 Emergent structures and decoherence }} \dotfill 116

8.5 There are no "particles" \dotfill 118

{\emph {8.5.1 Delayed choice experiments }} \dotfill 119

{\emph {8.5.2 Interaction-free quantum measurements }} \dotfill
120

{\emph {8.5.3 Relativistic quantum processes }} \dotfill 121

 8.6 The universally valid and complete quantum theory \dotfill
 121

{\emph {8.6.1 Why universally valid and complete quantum theory?
 }} \dotfill 121

{\emph {8.6.2 Completing the picture}} \dotfill 123

\bigskip

\noindent {\bf  9 Outlook and Prospects} \hfill 124

\bigskip

{\bf References} \hfill 127

\bigskip

{\bf Index} \hfill 137

\bigskip

{\bf Supplement} \hfill 139

\cleardoublepage

\pagenumbering{arabic}

\setcounter{page}{1}

 \noindent{\bf Chapter 1}

 \noindent\textbf{\large Introduction}

\vspace*{55mm}

\noindent In non-relativistic physics, "system" is defined by the
set of its degrees of freedom ($x_i$) and the related conjugate
momentums ($p_i$), as well as by the set of the parameters (such
as the mass, electric charge etc.). If the system is not
elementary, it's said to be "composite", i.e. to be decomposable
into parts (subsystems), which define the system's {\it
structure}. Realistic systems, ranging from  mesons to the
Universe as a whole, are--composite.

\smallskip

\noindent In classical physics, "structure" is pre-defined and
assumed to be basic. Subsystems of a composite system can be
further decomposed ("fine-graining"), or grouped
("coarse-graining"), while the other variations of the system's
structure are often regarded as physically artificial--a
mathematical artifact.

\smallskip

\noindent However, in the quantum (non-relativistic) theory, the
things look different. Solution to the quantum hydrogen atom (HA)
Schr\" odinger equation is a cornerstone of the quantum theory
that provides an outstanding observation. Actually, the
classically paradoxical discrete energy-spectrums, which are
experimentally observed, refer precisely to the atomic {\it
internal} degrees of freedom (denoted $R$).

\smallskip

\noindent Only recently some elaborate attempts of describing HA
as "electron+proton" ($e+p$) system have been made (Tomassini et
al 1998, Dugi\' c and Jekni\' c 2006, Dugi\' c and Jekni\'
c-Dugi\' c 2008, Jekni\' c-Dugi\' c et al 2012). The conclusions
are classically non-describable. Actually, as distinct from the
atomic "center-of-mass+relative position" ($CM+R$)  structure, the
$e+p$ structure is endowed by quantum entanglement and {\it
cannot} provide a proper theoretical explanation of the
experimental evidence of the discrete atomic
spectra\footnote{\small See eq.(21) below. An up to date
presentation can be found in (Jekni\' c-Dugi\' c et al 2012).}.

\smallskip

\noindent On the other hand, the atomic $CM$ and $R$ systems
appear {\it directly} accessible in the {\it realistic} physical
situations such as e.g.  atomic cooling, Bose condensation,
interference, atomic lithography etc. The atomic $CM$ and $R$
formal subsystems are exactly decoupled from each other and (for
the atom considered as a "closed" quantum system)  can have the
quantum states of their own--independent of each other. This is an
important observation, which suggests that we can "directly"
observe the (sub)systems only if there is the (sub)systems' {\it
individuality}. However, the choice of such "preferred" degrees of
freedom (i.e. of the preferred structure) of a composite system is
not established by any general rule or condition. Is there such a
rule or condition? If there is such a rule or condition, what
might be its/their kinematic or dynamic quantum-mechanical basis?
In other words: what constitutes a "system" accessible for an
observer? These are the main questions of interest for us that are
occasionally posed in the contemporary physics-research papers,
see e.g. (Dugi\' c and Jekni\' c 2006, Dugi\' c and Jekni\'
c-Dugi\' c 2008, Harshman 2012a, Fel'dman and Zenchuk 2012,
2014a,b 2014a,b, Lychkovskiy 2013, Lim et al 2014).

\smallskip

\noindent These questions are of {\it universal} interest and
importance in quantum theory, independently of the model of the
composite system, which can be open or isolated, finite- or
infinite-dimensional. This is of interest not only for the
foundations of quantum theory but also for certain applications as
well as for some interpretational reasons. The fact that this
topic is as yet weakly appreciated may be a consequence of the
classical prejudice and intuition, as well as of the widespread
 scientific attitude, which is described by Zurek's (Zurek
2003):

\smallskip

\noindent "{\it Quantum mechanics has been to date, by and large,
presented in a manner that reflects its historical development.
That is, Bohr's planetary model of the atom is still often the
point of departure, Hamilton-Jacobi equations are used to "derive"
the Schr\" odinger equation, and an oversimplified version of the
quantum-classical relationship (attributed to Bohr, but generally
not doing justice to his much more sophisticated views) with the
correspondence principle, kinship of commutators and Poisson
brackets, the Ehrenfest theorem, some version of the Copenhagen
interpretation, and other evidence that quantum theory is really
not all that different from classical--especially when systems of
interest become macroscopic, and all one cares about are
averages--is presented}."

\smallskip

\noindent In  different applications of quantum theory,  numerous
"one-particle" (i.e. noninteracting particles) models--which
include the  "virtual particles"--have been developed in order to
avoid the difficulties in description of the many-particle
systems. However, there is a {\it caveat}\footnote{\small See also
the arguments in (Zeh 2005).}: there is not any guarantee that the
results  can be straightforwardly interpreted in terms of the
"original" constituent particles. It seems that this subtle point
remained virtually unnoticed until recently (Dugi\' c and Jekni\'
c 2006, Dugi\' c and Jekni\' c-Dugi\' c 2008, Stokes et al 2012,
Fel'dman and Zenchuk 2012, 2014a,b, Harshman 2012a,  Dugi\' c et
al 2013, Arsenijevi\' c et al 2013a,b, Lychkovskiy 2013, Jekni\'
c-Dugi\' c et al 2014). Presence of quantum correlations and
related lack of classical individuality of quantum subsystems
precludes a straightforward transfer of the results between the
different structures of the composite system. Having this in mind,
careful analysis of the composite system's structures and their
behaviors becomes  an emerging need of the modern quantum
(non-relativistic) theory. This way come both a fresh insight into
well known methods and their results as well as emergence of a new
methodological basis of quantum theory. As a matter of fact, we
are yet to start working in that direction. So, the main purpose
of this book is to try to overcome the classical prejudice, and to
promote a fresh view of the quantum world.

\smallskip

\noindent The contents of this book is based mainly (but not
exclusively) on the authors' contributions that are made  in
approximately the last seven years. It tackles all of the
above-posed questions and carefully emphasizes subtlety of the
topic of quantum structures. In Chapter 2 we provide a general
conceptual basis for the remainder of the book. In Chapter 3, we
introduce relativity of the concept of "quantum [non-classical]
correlations" in composite quantum (closed or open) systems. Both
the quantum entanglement and the quantum discord relativity are
presented in  detail. In Chapter 4 we give some details regarding
the different molecule structures in use in the different fields.
We emphasize a new qualitative proposal of interest for resolving
the long-standing problems known as the Hund's paradox (Hund 1927)
in chemistry, and the Levinthal paradox (or the "protein folding"
problem), (Levinthal 1968), in the foundations of the
macro-molecules dynamics  (such as e.g. molecular recognition). In
Chapter 5 we highlight some experimental evidence, which clearly
stresses physical importance of the "non-fundamental" structure of
the atomic and molecule species. Chapter 6 is one of the central
parts of this book. It tackles the above-posed questions regarding
the Universe as the isolated (closed) quantum system. There we
present a recently obtained result of the "parallel occurrence of
decoherence" for a specific model of the quantum Brownian motion.
Physically, the results are striking: the model-Universe hosts
some mutually irreducible and, physically and
information-theoretically, mutually independent (autonomous),
simultaneously evolving-in-time structures, which are endowed by
the decoherence-induced quasi-classical structures. Some
interpretational aspects of our findings, as well as the important
issues raised by the "quantum reference frame", are the subject of
Chapter 8. In Chapter 7, we provide a few models, which strongly
suggest the following answer to the above questions for  {\it
open} systems:  environment singles out the "preferred" (i.e.
directly accessible for an observer) structure of the composite
system. This finding can be described (see also Harshman 2012a) by
the condition of the minimum quantum correlations  in the
preferred structure. Chapter 9 collects the questions, the offered
answers and their subtleties in one place.  Regarding application
of our results, we hope for prospects anticipated in (Fel'dman and
Zenchuk 2012): "{\it Using different bases, we may choose the
preferable behavior of quantum correlations which allows a given
quantum system to be more flexible in applications}."

\noindent To ease the exposition, we provide  Supplement, which
completes and partially, technically, extends what is told in the
body text. A comprehensive list of papers can be found at http://physics.kg.ac.rs/fizika/prilozi/quantumStructures/QS%20Comprehensive%20List.html .

\newpage

\noindent {\bf Chapter 2}

\noindent\textbf{\large The transformations of variables}

\vspace*{55mm}

\noindent Transformations of variables are ubiquitous in
mathematical methods and manipulations. They are so common that
sometimes are not explicitly distinguished. For instance, the
equality: %%
\begin{equation}
\cos (\alpha + \beta) \sin (\alpha - \beta) = \sin \alpha \cos
\alpha - \sin \beta \cos \beta
\end{equation}

\noindent involves a linear transformation of the set $\{\alpha,
\beta\}$ into the set $\{u, v\}$, where $u=\alpha + \beta$ and $v
= \alpha - \beta$. In the spirit of our considerations, eq.(1)
reads as: %%
\begin{equation}
\cos u \sin v = f_1(\alpha) g_1(\beta) + f_2(\alpha) g_2(\beta);
\end{equation}

\noindent $f_1(\alpha) = \sin \alpha \cos \alpha$, $f_2(\alpha) =
- 1$, $g_1 (\beta) =  1$, $g_2 (\beta) =  \sin \beta \cos \beta$,
while $\alpha, \beta \in [0, 2\pi]$.

\smallskip

\noindent Product of two Gaussian functions, $F(x_1) = \exp \{-
x_1^2/2\}$ and $F(x_2) = \exp \{- x_2^2/2\}$: %%
\begin{equation}
F(x_1) F(x_2) =  F(x_-) F(x_+),
\end{equation}

\noindent for the new variables $x_{\pm} = (x_1 \pm x_2)/\sqrt{2}$
and
 $x_i \in (-\infty, \infty), i=1,2$. As it can be easily shown:
 for Gaussian functions, in general, there are also the sums
 like in eq.(2)--there is more than one term on at least one side of eq.(3).

 \smallskip

\noindent The linear {\it canonical} transformations are of
general use in
 physics. For a pair of one-dimensional systems described by the
 respective position and momentum variables, $(x_1, p_1)$ and $(x_2,
 p_2)$, one can define the center of mass ($CM$) and the "relative
 position"($R$) degrees of freedom:
 \begin{equation}
X_{CM} = {m_1x_1 + m_2 x_2 \over M}, \quad r_R = x_1 - x_2,
 \end{equation}

 \noindent
 where $m_i, i=1,2$, are the masses and $M=m_1+m_2$. The new
 degrees of freedom, $X_{CM}$ and $r_R$, define the respective
 conjugate momentums, $P_{CM} = p_1 + p_2$ and $p_R = (m_2p_1 - m_1p_2)/M$.
 Typical physical interactions are distance dependent,
 $V(\vert x_1 - x_2 \vert)$ for the pair of systems $1$ and $2$.
 Then the transformations of variables, $\{x_1, x_2\} \to \{X_{CM},
 r_R\}$, give rise to the variables separation. The total
 Hamiltonian, $H$, does not involve any coupling of the
new variables:
 \begin{equation}
{p_1^2 \over 2m_1} + {p_2^2 \over 2m_2} + V(\vert x_1 - x_2 \vert)
= H = {P_{CM}^2 \over 2M} + {p_R^2 \over 2\mu} + V(\vert
r_R\vert),
 \end{equation}

\noindent where $\mu = (m_1^{-1} + m_2^{-1})^{-1}$ is the "reduced
mass". Separation of the new variables (of the $CM$ and the $R$
formal systems) is at the root of exact solvability of the
classical two-body problem\footnote{\small Which is illustrated in
Chapter 1 by the hydrogen atom.}.

\smallskip

\noindent Mathematical spirit of the transformations of variables
can blur their physical contents. The transformations of variables
are often considered as a purely mathematical tool, a mathematical
artifact not having any physical meaning. This classical prejudice
is going to be challenged, and then removed, in the quantum
mechanical context, starting from the next section.

\bigskip

\noindent  {\bf 2.1 Classical physics prejudice on the
transformations of variables}

\bigskip

\noindent Let us start with the "obviously" correct observations.
Center of mass for the pair "the Earth and the Venus" is an empty
point in space, not a physical object. That is, for the classical,
macroscopic bodies, the transformations of variables are simply
mathematical artifacts\footnote{\small Sometimes it is said that
the "relative positions" in eq.(4) do not have the same physical
 meaning as the original degrees of freedom. To this end, it is
 important to stress: {\it all the variables linked mutually via some proper variables
 transformations}
 have  the same {\it mathematical meaning}. They are all vectors in the same vector space thus providing
 the {\it universal
 mathematical basis} of the physical considerations.}. For this
reason,  the results presented via the "artificial" (the new)
variables ($X_{CM}$ and $r_R$), are typically transformed and
presented via the "realistic" (the original) variables ($x_1$ and
$x_2$).

\smallskip

\noindent As we show in Section 8.2, this classical interpretation
of the transformations of variables is quantum mechanically sound.
However, its extrapolation may be misleading. As an illustration,
we borrow from (Dieks 1998):

\smallskip

\noindent "{\it This suggests the following necessary condition
for factorizations to correspond to physically real systems. The
factor Hilbert spaces should carry a representation of the
space-time group... in the same way as the factor spaces of the
original factorization, with the usual identification of
generators of the space-time group and dynamical variables.}"

\smallskip

\noindent It is suggested above that the [linear canonical]
transformations not preserving the space-time symmetry are not
physically "real". The new degrees of freedom are required to be
describable by the exactly the same physics (symmetries and the
particles interactions) as the original ones.  However, the later
is not fulfilled already for the $CM+R$ system, the rhs of eq.(5),
which, in turn, is the standard quantum mechanical model of the
hydrogen atom as well as of the whole of atomic physics and
quantum chemistry (see Chapter 5 for some phenomenological facts).
Requiring the same symmetry rules for the "new" variables is
mathematically misleading--the symmetry rules for the new
variables {\it cannot be chosen} but are (uniquely) defined
(induced) by the symmetry rules for the original ones (Anderson
1993, 1994, Harshman 2012b, Manzano et al 2013).

\smallskip

\noindent So we conclude: Every set of the physical degrees of
freedom is formally {\it equal} to any other--i.e. it's subject of
the same formalism. On the other hand, physical {\it reality} of
the degrees of freedom is a separate issue that, as we show
starting from Chapter 3, is a bit more subtle than in the
classical physics context\footnote{\small Of course, mathematical
consistency, as described above, is required.}. To this end, {\it
phenomenology} plays substantial role--the purely theoretical
analysis is not sufficient (see Section 8.2).

\bigskip

\noindent {\bf 2.2.  Classifications of the canonical
transformations}

\bigskip

\noindent In the remainder of this book, we stick to the {\it
linear canonical transformations} (LCTs)\footnote{\small For
non-canonical transformations see e.g. (Lychkovskiy 2013).}.

\smallskip

\noindent For a composite system $\mathcal{C}$ defined by the
conjugate variables, $x_i$, $p_j$, where the commutator $[x_{ i},
p_{j}] = \imath \hbar \delta_{ij}$, the LCTs are formally defined:
\begin{equation}
\xi_m = \sum_i c_{im} x_i + \sum_n d_{jm} p_j , \pi_n = \sum_i
c'_{in} x_i + \sum_n d'_{jn} p_j
\end{equation}

\noindent while $[\xi_m, \pi_n] = \imath \hbar \delta_{mn}$. The
constants appearing in eq.(6) are mutually constrained; for an
example, see eq.(127) below. In a slightly different form, one can
introduce analogous expressions
 for the finite-dimensional systems (e.g. the qubit systems) or
 regarding
 the Fock space (e.g. via the Bogoliubov transformations)--see Supplement.

 \smallskip

\noindent {\bf Def.2.1}: {\it By} structure {\it of a composite
system it is assumed the set of the composite system's degrees of
freedom,i.e. the relate tensor factorization of the system's
Hilbert state space.}

\smallskip

\noindent This descriptive terminology should ease the exposition
as well as to support physical intuition. e.g. In eq.(6) appear
two structures, $\mathcal{S} = \{x_i, p_j\}$ and $\mathcal{S}' =
\{\xi_m, \pi_n\}$, that refer to one and the same composite
system, $\mathcal{C}$, and are mutually related by some LCTs.

\bigskip

\noindent {\bf {\emph {2.2.1 Some classifications of the LCTs and
structures}}}

\bigskip

\noindent We introduce a few criteria for classification of the
LCTs (and of the related structures) of interest for our
considerations. All of them apply to the finite- as well as the
infinite-dimensional (the continuous variable),  open or closed,
classical or quantum systems.

While an LCT {\it induces} a structure, we will use the same
terminology  for both LCTs and structures.

\smallskip

\noindent (A) If no   "new" variable,  in eq.(6), can be expressed
by more than one "old" ("original") variable, we refer to such
transformations (and the related structures) as {\it trivial}.
Otherwise they are {\it nontrivial}. Regarding the trivial LCTs,
we distinguish the particles re-ordering or permutations, grouping
of the particles (the "coarse graining" of the composite system's
structure). In a simplified form, the later can be
illustrated\footnote{\small We simplify notation: instead of the
degrees of freedom, we simply use the particles labels .}: %%
\begin{equation}
1+2+3 \to 1+ (2+3) \equiv 1 + S,
\end{equation}

\noindent where the bipartite system $S=2+3$; if the $2$ system
represents the electron and the $3$ system represents the proton,
then $S$ may be the hydrogen atom. So for the total system,
$\mathcal{C}$, the following two structures are distinguished
above: $\mathcal{S} = \{1,2,3\}$, which is tripartite, and the
bipartite structure $\mathcal{S}' = \{1, S\}$. An example of
non-trivial LCTs is given by eq.(4).

\smallskip

\noindent (B) LCTs are {\it global} if they target {\it all}
degrees of freedom of a composite system.
 Otherwise, they are  {\it local} (non-global).
  As it can be easily
shown, global/local character for a pair of structures is not
transitive. If $\mathcal{S}_2$ is local relative to both,
$\mathcal{S}_1$ and $\mathcal{S}_3$, $\mathcal{S}_1$ and
$\mathcal{S}_3$ may still be global to each other. Similarly, if
$\mathcal{S}_2$ is global relative to both, $\mathcal{S}_1$ and
$\mathcal{S}_3$, $\mathcal{S}_1$ and $\mathcal{S}_3$ may still be
local to each other. This relation for a pair of structures is
symmetric: if $\mathcal{S}_1$ is global (or local) relative to
$\mathcal{S}_2$, then also $\mathcal{S}_2$ is global (local) to
$\mathcal{S}_1$.

\smallskip

\noindent (C) If the "fine graining" (i.e. splitting into smaller
parts) of a structure $\mathcal{S}$ can lead to a structure
$\mathcal{S}'$, then the $\mathcal{S}$ structure is said to be
{\it reducible} to the $\mathcal{S}'$ structure. Otherwise, the
$\mathcal{S}$ structure is said to be {\it irreducible} to the
$\mathcal{S}'$ structure. By definition, two structures having the
same set of the degrees of freedom are mutually reducible. In
eq.(7), the $\mathcal{S}'$ structure is reducible to the
$\mathcal{S}$ structure, but not the other way
around--reducibility is not symmetric. Reducibility is transitive
yet: if $\mathcal{S}_3$ is reducible to $\mathcal{S}_2$, and
$\mathcal{S}_2$ is reducible to $\mathcal{S}_1$, then
$\mathcal{S}_3$ is reducible to $\mathcal{S}_1$.

\smallskip

\noindent Below, we provide some illustrative examples.

\smallskip

\noindent {\it Example 1.} Let us  consider a composite system
$\mathcal{C}$ consisting of three particles, $1$, $2$ and $3$. We
are interested in the following structures of $\mathcal{C}$:
$\mathcal{S} = \{1,2,3\}$, $\mathcal{S}_1 = \{1,3,2\}$,
$\mathcal{S}_2 = \{1, S\}$ and $\mathcal{S}_3 = \{S', 3\}$, where
the bipartite systems, $S=2+3$ and $S' = 1+2$. These structures
are obtained from each other by the trivial operations of grouping
the systems, or decomposing, or reordering/permutations. In the
set of the structures, only $\mathcal{S}_2$ and $\mathcal{S}_3$
are mutually global structures, while the other are  mutually
reducible: $\mathcal{S}_i \to \mathcal{S}_1, i=2,3$. The global
transformation: %%
\begin{equation}
\mathcal{S}_2 = \{1, S\} \to \mathcal{S}_3 = \{S', 3\}
\end{equation}

\noindent is characteristic for quantum teleportation. The
composite system's state is endowed by entanglement--"entanglement
swapping"; for some details see Section 3.1.

\smallskip

\noindent {\it Example 2.} Consider a composite system consisting
of four subsystems. To be specific, let us consider two electrons,
formally presented as $1e, 2e$, and two protons, $1p, 2p$. The
structures of interest are: $\mathcal{S} = \{1e, 2e, 1p, 2p\}$,
$\mathcal{S}_1 = \{1e, 1p, 2e, 2p\}$, $\mathcal{S}_2 = \{1H, 2e,
2p\}$, $\mathcal{S}_3 = \{1H, 2H\}$, $\mathcal{S}_4 = \{1CM, 1R,$
$2CM, 2R\}$ and $\mathcal{S}_5 = \{CM, R\}$; $CM$ and $R$, cf.
eq.(4), represent the center-of-mass and the "relative particle",
while e.g. the $1H$ represents the hydrogen atom composed of the
$1e$ and $1p$, symbolically $1H = 1e+1p$. By  definition, eq.(4),
the $CM$ and $R$ "systems" do not consist of (cannot be separated
into) the "original" systems (the electrons and protons), and {\it
vice versa}.
 The $\mathcal{S}_3$
structure refers to a pair of the hydrogen atoms, each atom being
presented as a pair  "electron and proton" ($H = e+p$). The
$\mathcal{S}_4$ structure represents a pair of the hydrogen atoms,
each of which decomposed as $CM+R$. The $\mathcal{S}_5$ represents
a pair $CM+R$ for the total $\mathcal{C}$ system.

\smallskip

\noindent We strongly emphasize: $\mathcal{S}_3 = \{1H, 2H\} \neq
\mathcal{S}'_4 = \{1H, 2H\}$, where the later is obtained by
grouping in the $\mathcal{S}_4$ structure, $H = CM+R$. This
non-equality is a consequence of the fact that the hydrogen atoms
are differently {\it built} starting from the smaller
"pieces"--from the $e$ and $p$, or from the $CM$ and $R$ systems,
respectively. On the other hand, if the atom's structures are not
specified, then formally the structure $\{1H, 2H\}$ can be
differently decomposed also as (is reducible to) $\mathcal{S}_1$,
${\mathcal{S}_4}$ or $\mathcal{S}_5$.

\smallskip

\noindent In quantum chemistry, the $\mathcal{S}$ structure is
considered to be the most fundamental [non-relativistic]
definition of the hydrogen molecule. To this end, more precisely,
the bipartite structure $\mathcal{S}' = \{E, P\}$ is considered,
where $E=1e+2e$ and $P=1p+2p$. On the other hand, the
$\mathcal{S}_3$ structure is of interest in some condensed-matter
considerations. There, the hydrogen molecule is considered as a
pair of electrically neutral, oscillating particles (the atoms,
$1H$ and $2H$). The $\mathcal{S}_5$ is of interest  in
investigating the large-molecules interference/decoherence effects
as well as in the Stern-Gerlach-like experiments. All  these
structures are physically realistic in the respective physical
situations--see Chapter 5 for  details.

\smallskip

\noindent Regarding Example 2: the structures follow from each
other, e.g., as %%
\begin{equation}
\mathcal{S} \stackrel{re-order}{\longrightarrow} \mathcal{S}_1
\stackrel{grouping}{\longrightarrow} \mathcal{S}_2
\stackrel{grouping}{\longrightarrow} \mathcal{S}_3
\stackrel{non-trivial}{\longrightarrow} \mathcal{S}_4
\stackrel{non-trivial}{\longrightarrow} \mathcal{S}_5.
\end{equation}

\noindent The $\mathcal{S}_2$ structure is local relative to both,
$\mathcal{S}_{1,3}$, while $\mathcal{S}_1$ is global relative to
$\mathcal{S}_3$. There is a chain of reducibility: $\mathcal{S}_3
\to \mathcal{S}_2 \to \mathcal{S}_1$.   For certain definitions of
the relative positions, $R$ [e.g. $R=1R\bigcup 2R$]: the
$\mathcal{S}_3$ structure is global relative to both
$\mathcal{S}_{4,5}$, but $\mathcal{S}_4$ is local relative to
$\mathcal{S}_5$. The structures $\mathcal{S}_i, i=3,4,5$, are
mutually irreducible.

\smallskip

\noindent It is important to emphasize: every structure
$\mathcal{S}$ uniquely determines the sets of the structures of
the composite system that are global/local, reducible/irreducible
relative to $\mathcal{S}$.

\pagebreak

\noindent {\bf {\emph {2.2.2 Mutually irreducible global
structures}}}

\bigskip

\noindent Instead of delving into the mathematical subtleties of
the LCTs forming the symplectic group for a composite system, we
proceed with the less formal yet  physically more intuitive
presentation.

\smallskip

\noindent In this section, we consider the {\it mutually global
and irreducible} structures, which cannot be obtained from each
other via the trivial LCTs. By excluding the trivial
transformations, this kind of structures are mutually global if
they do not have even a single degree of freedom in common. In
Example 2: the $\mathcal{S}_{4,5}$ structures are of the kind
relative to all other structures, as well as to each other.

\smallskip

\noindent For a set of such structures, the following are  direct
implications of eq.(6):

\smallskip

\noindent (1) For every bipartite structure of a composite system,
$\mathcal{C} = A+B$, the  $A$ subsystem is defined if and only if
the $B$ subsystem is defined. The subsystems $A$ and/or $B$ may
have their own structures.

\smallskip

\noindent (2) Subsystems belonging to  different (not necessarily
bipartite) structures are mutually irreducible. To this end, the
transformations eq.(4) are paradigmatic: the $CM$ or $R$ systems
cannot be decomposed into the original systems $1$ and $2$, and
{\it vice versa}.

\smallskip

\noindent (3) Subsystems belonging to  different structures
mutually do not  interact. Therefore there is not any correlation
and hence there is no information flow between them, Lemma 2.1.

\smallskip

\noindent (4) Every structure is defined by its own "elementary"
particles and their interactions. The symmetry rules for one
structure are in a unique (the LCTs-defined) relation with the
symmetry rules for any other structure (Anderson 1993, 1994,
Harshman 2012b, Manzano et al 2013).

\smallskip

\noindent Now, for the mutually global and irreducible structures,
we provide:

\smallskip

\noindent {\bf Lemma 2.1} {\it Subsystems belonging to different
structures are mutually infor\-ma\-tion-theoretically separated.
There is not any information flow between them.}

\noindent {\it Proof}: Without loss of generality, let us consider
bipartite structures, $\mathcal{S}_1=A+B$ and $\mathcal{S}_2=D+E$;
each subsystem, $A,B,D$ or $E$ may be a composite system itself.
The probability distributions for e.g. $A$ and $D$,
denoted\footnote{\small  In order to simplify notation, we do not
use the rigorous form, $\rho(x,x')\equiv \langle x \vert \rho
\vert x' \rangle$.} respectively $\rho(x_A)$ and $\rho(x_D)$, are
not mutually related. The $\rho(x_D) = \rho(x_D(x_A, x_B))$ cannot
be used to derive $\rho(x_A)$: integrating $\rho(x_D)$ over $x_B$
does not provide $\rho(x_A)$: $\int \rho (x_D) dx_B \neq \rho
(x_A)$. The only way to obtain $\rho (x_A)$ via integrating over
$x_B$ is to use the total system's, $\mathcal{C}$'s, probability
density, $\rho_C \equiv \rho(x_A, x_B) = \rho (x_D, x_E)$:
$\rho(x_A) = \int \rho_C dx_B$. On the other hand, presenting
$\rho_C = \rho(x_A, x_D)$ cannot help as the linear dependence of
$x_A$ and $x_D$ makes the integration $\int \rho(x_A, x_D) dx_B$
ill defined. So,  knowledge about one subsystem (e.g. about the
$A$ system), {\it in principle}, does not provide any information
about a subsystem (e.g. the $D$ system) belonging to an alternate
structure. Finally, due to the above point (3), there is not
exchange of information or correlation between  subsystems, which
belong to  different structures--which completes the proof. \hfill
Q.E.D.

\smallskip

\noindent Of course,  subsystems belonging to the {\it same}
structure (e.g. $A$ and $B$) may in principle  provide description
of each other. Due to  correlation between the subsystems, a local
measurement performed on one subsystem may provide some
information regarding the other subsystem. Regarding quantitative
measures of correlations, see Sections 3.1 and 3.2.

\smallskip

\noindent Everything told above equally refers to the "closed" as
well as to the "open" quantum systems.

\bigskip

\noindent  {\bf 2.3 Quantum mechanical structures}

\bigskip

\noindent It is essential, yet probably trivial, to note: every
composite quantum system $\mathcal{C}$ is defined by {\it unique}
Hilbert state space, $\mathcal{H}_C$,  Hamiltonian, $H$, and
quantum state in every instant in time.

\smallskip

\noindent The "coarse-graining" and the "fine-graining" operations
formally do not preserve the number of the composite system's
degrees of freedom. Of course, the degrees of freedom as stil
there, but "buried" in the degeneracy of the coarse-grained state.
LCTs induce re-factorization of the composite system's Hilbert
state space, $\mathcal{H}_C$ %%
\begin{equation}
\Pi_{i=1}^{\otimes N} \mathcal{H}_i = \mathcal{H}_C =
\Pi_{\alpha=1}^{\otimes N'} \mathcal{H}_{\alpha},
\end{equation}

\noindent where $\mathcal{H}_i$ is a factor space pertaining to
the original, while $\mathcal{H}_{\alpha}$ pertains to the new
structure of the composite system. For  bipartite decompositions,
which is our main subject, $\mathcal{C} = A+B$, or
$\mathcal{C}=D+E$, eq.(10) reads as: %%
\begin{equation}
\mathcal{H}_A \otimes \mathcal{H}_B = \mathcal{H}_C =
\mathcal{H}_D \otimes \mathcal{H}_E.
\end{equation}

\noindent Needless to say, tensor product (as well as  scalar
product) of states belonging to the factor spaces of the different
factorizations, is not defined.

\smallskip

\noindent The composite system's  Hamiltonian, $H$, has different
forms for the different structures, e.g.: %%
\begin{equation}
\sum_{i=1}^N {p_i^2 \over 2m_i} + V(\{x_i\}) = \sum_{\alpha=1}^N
{\pi_{\alpha}^2 \over 2\mu_{\alpha}} + V(\{\xi_{\alpha}\}).
\end{equation}

\noindent For the bipartite structures introduced  above, the
Hamiltonian reads, in general, as: %%
\begin{equation}
T_A + V(x_A) + T_B + V(x_B) + V_{AB} = T_D + V(x_D) + T_E + V(x_E)
+ V_{DE},
\end{equation}

\noindent where "$V$" denotes the possible classical external
fields for the subsystems, while the double-subscript terms
represent the interactions.

\smallskip

\noindent It is worth repeating: for a composite system, the
Hilbert state space and the set of the observables, including the
system's Hamiltonian, are {\it unique}. According to the Schr\"
odinger law, [normalized] quantum state of the system is [up to
arbitrary phase] also {\it unique} in every instant in time. The
same applies to the open systems not describable by the Schr\"
odinger law. An open system's state is defined by the tracing out
operation, e.g. $\rho_A (t) = tr_B \rho_C(t)$, which implies
unique density matrix for every subsystem in every instant in time
[as illustrated by eq.(75)].

\bigskip

\noindent{\bf 2.4 Quantum relativity of "system" and "locality"}

\bigskip

\noindent Quantum structures point out, that the very basic
physical concept of "system" is {\it relative}. Lemma 2.1, Section
2.2.2, exhibits that  subsystems belonging to certain different
structures are mutually physically and information-theoretically
separated. In practice this means that the concept of "system" is
{\it structure dependent} (Dugi\'c and Jekni\' c 2006, Dugi\' c
and Jekni\' c-Dugi\' c 2008): a system belonging to one structure
can be unobservable for an observer that belongs to some
alternative structure.

\smallskip

\noindent An open composite system is determined by the set of the
observables accessible  to measurement and manipulation (the
"preferred" observables). In principle, every observable of a
composite system can be measured. In practice, it is usually the
case that only some observables can be easily measured (Zanardi
2001). In effect, experimenter acquires only small fraction of
information about the composite system. Most of the remaining
degrees of freedom (subsystems) remain undetermined or poorly
known.

\smallskip

\noindent Every structure is uniquely defined by a tensor product
factorization eq.(10). The orthonormalized bases and observables
{\it adapted} to a factorization are {\it structure specific}.
e.g. For the $\mathcal{H}_1 \otimes \mathcal{H}_2$ factorization,
an orthonormalized basis $\{\vert m\rangle_1 \otimes \vert n
\rangle_2\}$ as well as an observable $A_1\otimes I_2$ are
structure specific. It is important to stress: the observable
$A_1\otimes I_2$ and its measurement are {\it local} only for the
$\mathcal{H}_1 \otimes \mathcal{H}_2$ factorization. Relative to
some alternative structure, the observable $A_1 \otimes I_2$ is a
"collective" observable whose measurement is non-local (Dugi\' c
and Jekni\' c 2006, Dugi\' c and Jekni\' c-Dugi\' c 2008). e.g.
Measurement of the atomic $CM$ position is a one-particle
observable, $x_{CM} \otimes I_R$, for the $\mathcal{H}_{CM}
\otimes \mathcal{H}_{R}$ factorization. However, regarding the
alternate $e+p$ structure, its measurement is collective--eq.(4)
emphasizes that the $CM$ position is determined by the positions
of {\it both} $e$ and $p$.

\smallskip

\noindent Therefore, measurement of the $CM$'s position (or
momentum) affects both $e$ and $p$, but only {\it partially}--the
$R$ system remains unaffected by the measurement. On the other
hand, simultaneous measurement of both the $CM$ and $R$ systems is
equivalent to a simultaneous measurement of both the $e$ and $p$,
and {\it vice versa}. Such measurements are the measurements
performed on the atom as a whole. Thereby the concept of
"collective" ("composite") observables and measurements also
become relative (structure dependent). So we may say, that the
atomic $CM$ system is not  more "collective" than the atomic
electron. Needless to say: every observable local to a structure
is an observable of the total system $\mathcal{C}$.

\smallskip

\noindent Hence, in formal terms, "locality" (of a subsystem,
observable or of a measurement, or of any action exerted on the
composite system $\mathcal{C}$) is defined by the tensor product
structure of the Hilbert state space and by the corresponding
"one-particle" observables. Phenomenological aspects of these
findings are presented and discussed in Chapter 5.

\smallskip

\noindent Finally, the structure-induced notion of locality as
described above does not incorporate the relativistic notion of
locality, which is of interest for the Bell inequalities tests.
Relativistic locality can be  introduced once the tensor-product
structure (i.e. the structure-induced locality) is defined. For
approaches that unite the structure-induced locality and the
relativistic locality see e.g. (Zanardi et al 2004, Harshman and
Ranade 2011).

\newpage

\noindent {\bf Chapter 3}

\noindent\textbf{\large Quantum Correlations Relativity}

\vspace*{55mm}

\noindent In quantum teleportation (Bennett et al 1993), three
qubits are specifically prepared. A qubit $1$ is in unknown state
$\vert u \rangle_1$, while the pair $2+3$ is in a Bell state, e.g.
$\vert \Phi^- \rangle_{23} = {1 \over \sqrt{2}} (\vert 0 \rangle_2
\vert 1 \rangle_3 - \vert 1 \rangle_2 \vert 0 \rangle_3)$. The
total system's quantum state can be presented in the different
forms for the following structures\footnote{\small We omit the
tensor-product symbol.}: %%
\begin{eqnarray}
&\nonumber&  \mathcal{S} = 1 + (2 + 3) : \quad \vert
\Psi\rangle_{123} = \vert u \rangle_1 {1 \over \sqrt{2}} (\vert 0
\rangle_2 \vert 1 \rangle_3 - \vert 1 \rangle_2 \vert 0 \rangle_3)
\\&&
\mathcal{S}_1 = 1 + 2 + 3 : \quad  \vert \Psi\rangle_{123} = {1
\over \sqrt{2}} (\vert u \rangle_1  \vert 0\rangle_2  \vert 1
\rangle_3 -
 \vert u \rangle_1 \vert 1 \rangle_2 \vert 0 \rangle_3)
\nonumber \\&&
 \mathcal{S}_2 = 2 + 1 + 3 : \quad \vert
\Psi\rangle_{123} = {1 \over \sqrt{2}} (\vert 0 \rangle_2\vert u
\rangle_1  \vert 1 \rangle_3 - \vert 1 \rangle_2 \vert u \rangle_1
\vert 0 \rangle_3).
\end{eqnarray}

\noindent However, quantum teleportation is {\it not} possible due
to any of the structures presented in eq.(14), but rather to the
$\mathcal{S}_3 = (1 + 2) + 3$ structure for which: %%
\begin{equation}
\vert \Psi\rangle_{123} = {1 \over 2 } [\vert \Psi^-\rangle_{12}
\vert u_1\rangle_3 + \vert \Psi^+\rangle_{12} \vert u_2\rangle_3 +
\vert \Phi^+\rangle_{12} \vert u_3\rangle_3 + \vert
\Phi^-\rangle_{12} \vert u_4\rangle_3].
\end{equation}

\noindent The  states $\{\vert \Psi^{\pm}\rangle, \vert
\Phi^{\pm}\rangle\}$ form the so-called Bell basis, while the
$\vert u_i \rangle$ states are not mutually orthogonal (for
further details see (Benett et al 1993) or (Nielsen and Chuang
2000)).

\smallskip

\noindent The structures are related by the trivial
transformations of variables as follows: %%
\begin{equation}
\mathcal{S} \stackrel{decomposing}{\longrightarrow} \mathcal{S}_1
\stackrel{grouping}{\longrightarrow} \mathcal{S}_3.
\end{equation}

\noindent So, a simple redefinition of the composite system's
structure provides  entanglement swapping from the pair $2+3$ to
the pair $1+2$, for the quantum state $\vert \Psi\rangle_{123}$ in
an instant in time. A composite measurement on the subsystem $1+2$
and the classical communication between Alice and Bob allow
teleportation of the unknown state $\vert u \rangle$ from qubit
$1$ to  qubit $3$.

\smallskip

\noindent This dependence of the form of a quantum state on the
composite system's structure can be easily demonstrated for the
hydrogen atom, which is a {\it continuous variable} (CV) system.

\smallskip

\noindent Hydrogen Atom (HA) is defined as a pair
"electron+proton" ($e+p$) that mutually interact via Coulomb
interaction. Introducing the atomic $CM$ and $R$ variables (while
neglecting the spin) provides   separation of variables and the
exact solution to the quantum HA model. Relation of the two
structures is of the kind considered in Section 2.2, see also
eq.(4).

\smallskip

\noindent Formally, the Hamiltonian, $H$, of the atom reads as: %%
\begin{equation}
{{\vec p}_e^2 \over 2m_e} + {{\vec p}_p^2 \over 2m_p} - {e^2 \over
4\pi\epsilon_{\circ} \vert \vec r_e - \vec r_p \vert} = H = {{\vec
p}_{CM}^2 \over 2M} + {{\vec p}_{R}^2 \over 2\mu} - {e^2 \over
4\pi\epsilon_{\circ} \vert \vec r_R \vert}
\end{equation}

\noindent where we borrow notation from eq.(4). Due to
non-interaction between the $CM$ and $R$ systems,  the state for
the $CM+R$ structure of the atom can be assumed to be tensor
product: %%
\begin{equation}
CM+R : \vert \psi\rangle_{atom} = \quad \vert \chi\rangle_{CM}
\vert n l m_l m_s \rangle_R;
\end{equation}

\noindent in eq.(18), we employ the standard notation  of quantum
theory for the hydrogen atom.

\smallskip

\noindent However, due to the Coulomb interaction between the
electron and the proton, one should expect entanglement: %%
\begin{equation}
e+p : \quad \vert \psi\rangle_{atom} = \sum_i c_i \vert i\rangle_e
\vert i \rangle_p.
\end{equation}

\noindent This entanglement can be estimated to be weak as
follows. Let us apply the standard {\it adiabatic approximation}
to the pair $e+p$. In the zeroth order of approximation, the
proton is "frozen" in a spatial position $\vec r_{p}^{\circ}$.
Then the atomic Hamiltonian, the lhs of eq.(17), reduces to the
electron's effective Hamiltonian and the related Schr\" odinger
equation: %%
\begin{equation}
\left\{ {{\vec p}_e^2 \over 2m_e} - {e^2 \over
4\pi\epsilon_{\circ} \vert \vec r_e - \vec r_p^{\circ} \vert}
\right\} \vert \phi_n\rangle_e = E_{en} \vert \phi_n\rangle_e.
\end{equation}

\noindent where $\vec r_p^{\circ}$ is a $c$-number, not a
dynamical variable. It is obvious that eq.(20) is of the same form
as  for the $R$ system--cf. the rhs of eq.(17). So, the solutions
to the rhs of eq.(17) and  eq.(20) are formally the same. With
some simplification of notation, the state of the atom is of the
form (Gribov and Mushtakova 1999, Atkins and Friedman 2005): %%
\begin{equation}
\sqrt{1 - \kappa^2} \vert \phi_n\rangle_e \vert \vec
r^{\circ}\rangle_p + \vert O(\kappa)\rangle_{ep},
\end{equation}

\noindent where $\kappa \equiv (m_e/m_p)^{3/4} \ll 1$--of the
order of the standard adiabatic parameter in quantum molecules
theory (see Section 4.2.2). As distinct from eq.(18), the small
term on the rhs of eq.(21) reveals the presence of entanglement
for the atomic $e+p$ structure. Hence quantum dynamics  for the
two structures ($e+p$ and $CM+R$) of the hydrogen atom are {\it
not} mutually equivalent, while eq.(18) is known to be in
accordance with experimental observations.

\bigskip

\noindent {\bf 3.1 Quantum entanglement relativity}

\bigskip

\noindent From eqs. (14)-(21) we can respectively write: %%
\begin{eqnarray}
&\nonumber&   \vert u \rangle_1 {1 \over \sqrt{2}} (\vert 0
\rangle_2 \vert 1 \rangle_3 - \vert 1 \rangle_2 \vert 0 \rangle_3)
= \vert \Psi\rangle_{123} =
\\&&
{1 \over 2 } [\vert \Phi\rangle_{12} \vert u_1\rangle_3 + \vert
\Phi\rangle_{12} \vert u_1\rangle_3 + \vert \Phi\rangle_{12} \vert
u_1\rangle_3 + \vert \Phi\rangle_{12} \vert u_1\rangle_3]
\end{eqnarray}

\noindent and:

\begin{equation}
\sqrt{1 - \kappa^2} \vert \phi_n\rangle_e \vert \vec
r^{\circ}\rangle_p + \vert O(\kappa)\rangle_{ep} = \vert
\psi\rangle_{atom} = \vert \chi\rangle_{CM} \vert n l m_l
\rangle_R.
\end{equation}

\noindent In the position representation,  eq.(23) [with a slight
change in notation] reads: %%
\begin{equation}
\sqrt{1 - \kappa^2} \phi_n(\vec r_e - \vec r_p^{\circ})
\delta(\vec r_p - \vec r_p^{\circ}) + \kappa \Phi_O(\vec r_e, \vec
r_p) = \chi(\vec r_{CM}) \varphi_{nlm_l}(\vec r_R).
\end{equation}

\noindent The expressions eq.(22) and eq.(24) are instances of the
general quantum mechanical {\it rule}--of the so-called {\it
Entanglement Relativity} (Vedral 2003, Caban et al 2005, Dugi\' c
and Jekni\' c 2006, Dugi\' c and Jekni\' c-Dugi\' c 2008,  Zanardi
2001, Ciancio et al 2006, De la Torre et al 2010, Harshman and
Wickramasekara 2007, Jekni\' c-Dugi\' c and Dugi\' c 2008, Terra
Cunha et al 2007). For a composite system $\mathcal{C}$ that can
be decomposed as $S+S'$ or as $A+B$, Entanglement Relativity (ER)
establishes for an instantaneous state, $\vert \Psi\rangle$, e.g.
\begin{equation}
\vert \xi \rangle_S{} \vert \phi\rangle_{S'} = \vert \Psi \rangle
= \sum_i c_i \vert i \rangle_A \vert i \rangle_B.
\end{equation}

\noindent For every set (e.g. an orthonormalized basis) of the
tensor-product states, there is infinitely many entangled
states--the number of entangled states in a Hilbert space is
incomparably larger than the number of the tensor-product states.
Hence the tensor product states are rare ("improbable") in the
Hilbert state space and for the most of the practical
purposes--but see eq.(3)--this possibility can be neglected. So,
ER implies that there is entanglement for practically every state
$\vert \Psi\rangle$ of a composite system. Of course, a state can
have entanglement for both structures of the composite system;
some subtleties regarding the finite-dimensional systems can be
found in (Harshman and Ranade 2011). Nevertheless, amount of
entanglement for a quantum state is not an LCTs invariant.

\smallskip

\noindent From eqs. (22)-(25), we directly realize: "quantum
entanglement" is not a feature of a composite system, or of a
system's state, but is a {\it feature of the composite system's}
structure.

\bigskip

\noindent {\bf 3.2 Quantum discord relativity}

\bigskip

\noindent "Quantum discord" is  a common name for a number of
mutually different measures of non-classical (quantum)
correlations in composite quantum systems (Olivier and Zurrek
2001, Henderson and Vedral 2001, Modi et al 2012, Xu and Li 2012).
For closed systems, "discord" coincides with "entanglement". For
 open quantum systems, there are quantum correlations that are
not identical with entanglement. Total amount of non-classical
correlations is measured by "discord".

\smallskip

\noindent If a composite (e.g. bipartite) quantum system carries
the total correlation  $I$, then the non-classical correlation can
be quantified by subtracting the classical correlation, denoted
$J$, from the total $I$.
 Of course, operationally,
information is acquired by performing a quantum measurement, e.g.
on one subsystem of the composite system. On this basis, one tries
to conclude about the amount of non-classical correlations in the
composite system.

\smallskip

\noindent Two systems, $S$ and $S'$, constitute a bipartite system
$\mathcal{C} = S+S'$. A quantum measurement performed on $S'$ and
defined by a projector, $\Pi_{S'i}$, provides the final state of
the composite system: $\rho_S\vert_{\Pi_{S'i}} = I_S \otimes
\Pi_{S'i} \rho I_S \otimes \Pi_{S'i}$. Then the maximum classical
correlations can be defined as: %%
\begin{equation}
J^{\leftarrow} (S\vert S') = \mathcal{S}_{vn}(S) -
\inf_{\{\Pi_{S'i}\}}\sum_i \vert c_i \vert^2
\mathcal{S}_{vn}(\rho_S\vert_{\Pi_{S'i}}) \ge 0,
\end{equation}

\noindent where $\mathcal{S}_{vn}$ represents von Neumann entropy,
 $\mathcal{S}_{vn}(\rho) = - tr \rho \ln \rho$, for instantanoeus state $\rho$.

 \smallskip

\noindent The total correlation in the system is defined as the
mutual information, $I(S:S') = \mathcal{S}_{vn}(S) +
\mathcal{S}_{vn}(S') - \mathcal{S}_{vn}(S,S') \ge 0$ that,
according to the above described idea, provides the following
measure, termed "one-way discord",  of quantum correlations in the
composite system $\mathcal{C}$ (Olivier and Zurek 2001): %%
\begin{equation}
D^{\leftarrow}(S|S') = I(S:S') - J^{\leftarrow}(S|S') \ge 0.
\end{equation}

\noindent The arrows appearing in eq.(27) emphasize that the
measurement is performed on $S'$. If the measurement is performed
on $S$, then the roles of $S$ and $S'$ in eq.(27) are mutually
exchanged. Then the one-way discord $D^{\leftarrow}(S'|S) =
D^{\rightarrow}(S|S')$ in full analogy with eq.(27). The closely
related measure is the "two-way discord",
$D^{\leftrightarrow}(S,S') = max \{D^{\leftarrow}(S|S'),$
$D^{\leftarrow}(S'|S)\}$, which tends to be larger than one-way
discord.

\smallskip

\noindent The one-way discord $D^{\leftarrow}(S|S')$ equals zero
if and only if the composite system's state, $\rho_C$, is of the
form (Modi et al 2012, Xu and Li 2012), (and the references
therein): %%
\begin{equation}
\rho_C = \sum_k p_k \vert k \rangle_S\langle k \vert \otimes
\rho_{S'k}, \quad _S\langle k \vert k' \rangle_S = \delta_{kk'},
\sum_k p_k = 1,
\end{equation}

\noindent while the two-way discord equals zero if and only if the
composite system's state is of the form (Modi et al 2012, Xu and
Li 2012), (and the references therein): %%
\begin{equation}
\rho'_C = \sum_{kl} p_{kl} \vert k\rangle_{S}\langle k \vert
\otimes \vert l \rangle_{S'}\langle l \vert, _S\langle k \vert k'
\rangle_S = \delta_{kk'}, _{S'}\langle l \vert l' \rangle_{S'} =
\delta_{ll'}, \sum_{k,l} p_{kl} =1.
\end{equation}

\noindent Apparently, the state $\rho'_C$, eq.(29), is a special
case of $\rho_C$ appearing in eq.(28): commutativity of all
$\rho_{S'k}$ in eq.(28) gives rise to $\rho'_C$ in eq.(29).

\smallskip

\noindent Let us now consider an alternate structure (Dugi\' c et
al 2013), $\mathcal{C} = A+B$.

\smallskip

\noindent By definition,  mixed states are "mixtures" of  pure
states: in eq.(29), the pure states $\vert k\rangle_S \vert
l\rangle_{S'}$ are "mixed" with the probability distribution
$\{p_{kl}\}$. So, we can use ER, Section 3.1. As it can be easily
shown, only if for every $\vert k \rangle_S \vert l\rangle_{S'}$
there exists some $\vert \alpha\rangle_A \vert \beta \rangle_B$
[such that $_A\langle \alpha \vert \alpha'\rangle_A = \delta
_{\alpha\alpha'}$ and $_B\langle \beta \vert \beta'\rangle_B =
\delta_{\beta, \beta'}$] the state eq.(29) obtains the form:
$\rho'_C = \sum_{\alpha, \beta} p'_{\alpha \beta} \vert
\alpha\rangle_{A}\langle \alpha \vert \otimes \vert \beta
\rangle_{B}\langle \beta \vert, \sum_{\alpha,\beta}
p_{\alpha\beta} =1$, for which the two-way discord for the $A+B$
structure $D^{\leftrightarrow}(A, B)=0$. In all other cases, the
state $\rho'_C$ is not of the form of eq.(29) for the alternate
($A+B$) structure. So, a change of the composite system's
structure induces a change in two-way discord: two-way discord
that equals zero for one structure (e.g. for $1+2$) becomes
nonzero for an alternate structure (for $A+B$).

\smallskip

\noindent Entanglement Relativity, Section 3.1, states: %%
\begin{equation}
\vert k \rangle_S \vert l \rangle_{S'} = \sum_{\alpha, \beta}
c_{\alpha\beta}^{kl} \vert \alpha\rangle_A \vert \beta \rangle_B.
\end{equation}

\noindent Substituting eq.(30) into eq.(29) gives: %%
\begin{eqnarray}
&\nonumber&  \rho'_C = \sum_{k, l, \alpha, \beta} p_{kl} \vert
C^{kl}_{\alpha\beta} \vert^2 \vert \alpha \rangle_A\langle \alpha
\vert \otimes \vert \beta \rangle_B\langle \beta \vert
\\&&
+ \sum_{k, l, \alpha\neq \alpha', \beta \neq \beta'} p_{kl}
C^{kl}_{\alpha\beta} C^{kl\ast}_{\alpha' \beta'} \vert \alpha
\rangle_A\langle \alpha' \vert \otimes \vert \beta
\rangle_B\langle \beta' \vert.
\end{eqnarray}

\noindent In order for eq.(31) takes the form of eq.(29) also for
the $A+B$ structure, the following conditions should be satisfied:
\begin{equation}
\sum_{k,l} p_{kl} C^{kl}_{\alpha\beta} C^{kl\ast}_{\alpha' \beta'}
= 0, \forall{\alpha \neq \alpha'}, \forall{\beta \neq \beta'},
\end{equation}

\noindent Analogous analysis of the state $\rho_C$, eq.(28), gives
rise to the following conclusion: in order for one-way discord
equals zero also for the alternate structure ($A+B$), the
following conditions must be fulfilled: %%
\begin{equation}
\sum_{k,l} p_k \omega^k_l C^{kl}_{\alpha\beta} C^{kl\ast}_{\alpha'
\beta'} = 0, \forall{\alpha \neq \alpha'}, \forall{\beta, \beta'},
\end{equation}

\noindent while $\quad p_k \ge 0, \omega_l^k \ge 0, \sum_k p_k =1
= \sum_l \omega_k^l, \forall{l}$. Both conditions, eq.(32) and
eq.(33), represent the sets of the simultaneously fulfilled
equalities. For the continuous variable systems, the number of
such equalities is infinite. However, this does not mean that
these conditions can never be fulfilled. Actually, the number of
the normalization and orthogonality conditions for the states
$\vert k\rangle_S \vert l \rangle_{S'}$ as well as for $\vert
\alpha\rangle_A \vert \beta\rangle_B$ is also infinite. So, we
cannot conclude that these conditions, or at least one of them, is
never fulfilled. Nevertheless, for every combination of the
coefficients $C_{\alpha\beta}^{kl}$ satisfying eq.(32) or eq.(33),
there is an infinite number of variations of the coefficients
$p_k$ and $ \omega^k_l$, or $p_{kl}$, respectively, that do not
satisfy eq.(32) and eq.(33). In practice, it means one may forget
about the states fulfilling  the conditions eq.(32) and/or
eq.(33).

\smallskip

\noindent In effect, virtually every change in structure of a
bipartite quantum system gives rise to a change in quantum
discord: e.g. if quantum discord is zero for one structure, it is
almost certainly non-zero for arbitrary alternative structure of
the composite system (Dugi\' c et al 2013) %%
\begin{equation}
\rho_S \otimes \rho_{S'} = \sum_i \lambda_i \rho_{Ai} \otimes
\rho_{Bi}.
\end{equation}

\noindent Eq.(34) exhibits quantum discord relativity (QDR).
Bearing in mind that "quantum discord" is more general than
"entanglement", we find Quantum Correlations Relativity (QCR):
{\it there is quantum correlation for practically every state of a
composite quantum system} (Dugi\' c et al 2013). Or the other way
around: "quantum correlation" does not concern of a composite
system, or of any of its possible states, {\it but of the
composite system's structure}. For an elaborated model see
Fel'dman and Zenchuk 2014b.

\bigskip

\noindent {\bf 3.3 Some mathematical remarks}

\bigskip

\noindent Considerations in Section 3.2 are based on ER, which is
a {\it universal} rule--a corollary of the universally valid
quantum mechanics--applicable for finite- as well as
infinite-dimensional, open or closed systems and for every kind of
LCTs.

\smallskip

\noindent The task of estimating  amount of non-classical
correlations in a composite  system for different structures can
be reduced to the following task:

\smallskip

\noindent {\bf T}. {\it Starting from a given form of a composite
system's state, provide a  form of the state for some alternate
structure of the composite system.}

\smallskip

\noindent Regarding the trivial LCTs (of grouping or decomposing
or permutations of the constituent particles), it is
straightforward (although probably sometimes tedious) to provide
the alternate forms of the state--cf. eqs. (14) and (15).

\smallskip

\noindent However, regarding the non-trivial LCTs,  and
particularly those distinguished in Section 2.2.2, for the {\it
closed} systems the task $\bf T$ coincides with providing the
Schmidt canonical form (for a fixed instant in time) of the state.
To this end, we are not aware of any {\it algebraic} recipe.
Typically, obtaining Schmidt form of the state assumes a
representation (e.g. the position representation). The methods are
proposed  that do not provide any analytical solutions: numerical
analysis may be helpful for Gaussian states of the continuous
variable systems, see e.g. (Ciancio et al 2006). For an example
regarding the spin-chain systems see (Fel'dman and Zenchuk 2012,
2014a,b).

\smallskip

\noindent Finally, regarding the mixed states (of interest for
{\it open} systems), the task {\bf T} is an instance of the
so-called "quantum separability (QUSEP)" problem, which is
thoroughly investigated for  the finite-dimensional systems. The
QUSEP problem is known to be computationally an NP-Hard problem
(Gharibian 2010). Also for the finite-dimensional systems, it
appears that calculating quantum discord is NP-Hard (Huang 2013).

\smallskip

\noindent So there is not a universal method for solving the task
{\bf T}. This is an open issue pointed out by our considerations.

\bigskip

\noindent {\bf 3.4 Some physical remarks}

\bigskip

\noindent It seems unavoidable to conclude, that QCR constitutes
the core of the most, if not all, of the problems (Zeh 1993, 2005,
Primas 1994) appearing in the "one-particle" methods in solid
state physics, nuclear physics and quantum chemistry.
Nevertheless, introducing quantum correlations into consideration
can be useful as we emphasize below.

\smallskip

\noindent There is a simple idea for {\it avoiding decoherence}
(Jekni\' c-Dugi\' c and Dugi\' c  2008). Decoherence is expected
to reduce amount of quantum correlations in a composite system. If
decoherence ruins quantum correlations for the $S+S'$ structure,
this is {\it not expected} to be the case for an alternate
structure $A+B$. So, quantum engineers  can avoid decoherence
(that is unfolding in the $S+S'$ structure) by simply targeting
the observables of the alternate, the $A+B$, structure.

\smallskip

\noindent QCR  implies: manipulating the composite system
$\mathcal{C}$ can be virtually independent of the $\mathcal{C}$'s
{\it initial preparation}. No matter of the initial state of
$\mathcal{C}$, there is always a possibility to use quantum
correlations. e.g. If the initial state is tensor product for the
$S+S'$ structure, one can use quantum correlations by
operationally targeting the observables of an alternate $A+B$
structure. Accessibility of a composite system's observables is a
subtle an issue to be discussed in Chapters 5 and 7. So, keeping
in mind this subtlety, in principle, one may forget about the
problems posed by the initial state preparation for bipartitions
of a composite quantum system.

\smallskip

\noindent QCR changes our {\it intuition} about the composite {\it
quantum} systems. The universally valid quantum mechanics does not
{\it a priori} select a preferred structure of a composite system.
All structures can be formally treated on the equal footing thus
challenging the classical prejudice of Section 2.1. The place of
the classical intuition, Section 2.1, in the  quantum mechanical
context is a subject of  Chapters 7 and 8, and particularly of
Section 8.2.

\smallskip

\noindent Finally, the following consequences of QCR will be
presented in the remainder of this book: the so-called parallel
occurrence of decoherence (Section 6.3.3), a limitation of the
Nakajima-Zwanzig projection method in open quantum systems theory
(Section 6.3.4), and the preferred structure of open system
(Chapter 7).

\newpage

\noindent {\bf Chapter 4}

\noindent\textbf{\large Quantum Molecule Structures}

\vspace*{55mm}

\noindent There are different ways of defining "molecule" as a
physical system. We distinguish (in somewhat simplified terms) a
few typical models of interest in physics and chemistry.

\smallskip

\noindent {\it Chemical model} (ChM). In chemistry, "molecule" is
often defined as "{\it An electrically neutral entity consisting
of more than one atom}"\footnote{\small IUPAC Recommendations
1994; doi:10.1351/pac199466051077.}. Physically, it is a set of
atoms mutually linked\footnote{\small Phenomenologically inspired
boundary conditions.}  by chemical bonds. If the atoms are
point-like (unstructured) particles, "molecule" is simply a chain
or a lattice of point-like oscillators. Spatial distribution of
atoms defines the molecule geometrical shape (molecule
configuration)--the very basic concept of stereochemistry.

\smallskip

\noindent {\it Solid state model} (SSM). If  internal atomic
excitations are added to every atomic position introduced in the
ChM, then one obtains another definition (model) of "molecule".
This "molecular excitons" model is typical for  solid state
physics and applications.

\smallskip

\noindent {\it Quantum chemistry model} (QC). In quantum
chemistry, "molecule" is defined as a set of the atomic nuclei and
the electrons in mutual interaction via the Coulomb electrostatic
field.

\smallskip

\noindent Taking the spin into consideration  complicates the
analysis. So, in order to exhibit the structural relations between
the different
 models, we will ignore the molecule spin.

\bigskip

\noindent {\bf 4.1 Mutual relations of the molecule structures}

\bigskip

\noindent The chemical model (ChM) is simply a set of point-like
atoms, $\mathcal{S}_{ChM} = \{1, 2, 3, ...\}$--internal atomic
degrees of freedom are ignored. Chemical bonds typically enter the
picture through a definite geometric shape (a configuration) of a
molecule. Hence it is reasonable to assume that the
$\mathcal{S}_{ChM}$ represents a set of harmonic oscillators--the
chemical bonds provide the effective harmonic field for every
oscillator (which is defined by its mass, spatial
equilibrium-position and frequency). The assumption of existence
of the atomic equilibrium-positions is classical in its spirit.
There is not a quantum mechanical reason to think so, for an
isolated molecule (Hund 1927). Therefore, in the quantum
mechanical context, the ChM model is not physically complete. For
large molecules (such as
 bio-polymers), the ChM model raises the following foundational
problems: why, and how,  large molecules obtain  different
configurations (Hund 1927, Giulini et al 1996), and how can be
described transitions between the molecule configurations (a
variant of the celebrated protein folding problem, see (Levinthal
1968)).

\smallskip

\noindent For the ChM, the molecule Hilbert state space is tensor
product of the individual oscillators Hilbert spaces: %%
\begin{equation}
\mathcal{H}_{ChM} = \otimes_i \mathcal{H}_i,
\end{equation}

\noindent while the Hamiltonian reads as: %%
\begin{equation}
H_{ChM} = \sum_i ({\vec p_i^2 \over 2m_i} + {1 \over 2} m_i
\omega_i^2 \vec r_i^2) + \sum_{i\neq j} V_{ij}.
\end{equation}

\noindent Non-harmonic corrections are neglected in eq.(36). For
non-interacting oscillators, $V_{ij} = 0, \forall{i,j}$

\smallskip

\noindent The solid-state model (SSM) enriches the ChM
description. If the excitations are denoted by "ex", then the
molecule structure reads as: $\mathcal{S}_{SSM} = \{1, 1ex, 2,
2ex, 3, 3ex,...\}$. Detailed physical nature of "exciton" is here
of secondary importance. Here we assume the excitations are well
spatially defined--joined with the respective oscillator
equilibrium-positions. "Coarse graining" of the
$\mathcal{S}_{SSM}$ structure can provide a bipartite structure,
$\mathcal{S'}_{SSM} = \{\mathcal{S}_{ChM}, Ex\}$, where the
exciton system $Ex = \{1ex, 2ex, 3ex,...\}$. Quantum mechanically,
the exciton system can be analysed independently of the lattice
vibrations: the excitation transfer from one to another cell of
the lattice is allowed and numerous interesting physical effects
may occur.

\smallskip

\noindent For the SSM structure, the Hilbert state space acquires
the form: %%
\begin{equation}
\mathcal{H}_{SSM} = \mathcal{H}_{ChM} \otimes_i \mathcal{F}_i,
\end{equation}

\noindent where $\mathcal{F}_i$ represents the Fock space for the
$i$th excitation. The Hamiltonian reads\footnote{\small Coupling
between the molecule vibrations and excitations is typically
provided by applying some external field to the molecule.}: %%
\begin{equation}
H_{SSM} = H_{ChM} + H_{ex}.
\end{equation}

\noindent As we emphasize below, the quantum chemistry  model
(QCM) is the most fundamental model of "molecule". As a kind of
generalization of the standard definition of "atom", in quantum
chemistry (Gribov and Mushtakova 1999, Atkins and Friedman 2005),
"molecule" is defined as a set of the atomic nuclei (denoted $n$)
and of the atomic electrons ($e$)--the $\mathcal{S}_{QCM} = \{1e,
1n, 2e, 2n,...\}$ structure; compare to the $\mathcal{S}$
structure in Example 2, Section 2.2.1. The $i$th atomic nucleus
brings some electrostatic charge, $Z_i e$, and there is the
electrostatic Coulomb interaction between the molecule's
constituents.

\smallskip

\noindent Let $\mathcal{H}_{ei}$ represents the Hilbert state
space of the $i$th electron, and  $\mathcal{H}_{n\alpha}$
represents the Hilbert state space of the $\alpha$th atomic
nucleus. Then the molecule Hilbert state space factorizes as: %%
\begin{equation}
\mathcal{H}_{QCM} = \otimes_i \mathcal{H}_{ei} \otimes_{\alpha}
\mathcal{H}_{n\alpha}.
\end{equation}

\noindent The molecule Hamiltonian reads as: %%
\begin{equation}
H_{QCM} = \sum_i {\vec p_{ei}^2 \over 2 m_e} + \sum_{i\neq j}
V_{ij} + \sum_{\alpha} {\vec p_{n\alpha}^2 \over 2 m_{\alpha}}
 + \sum_{\alpha \neq \alpha'}V_{\alpha\alpha'} + \sum_{i, \alpha}
 V_{i\alpha}.
\end{equation}

\noindent In eq.(40): the double-script terms refer to the Coulomb
interactions; the Latin indices refer to the electrons while the
Greek indices refer to the atomic nuclei.

\smallskip

\noindent A molecule of a given chemical kind is a unique entity
that, as a quantum system, is described by the unique Hilbert
state space, unique Hamiltonian and unique quantum state in every
instant in time. With this in mind, the indices that appear in
eqs. (35)-(40) emphasize the structures of a single molecule,  not
the different molecules. Below, we point out  relations between
the different structures of a single molecule.

\smallskip

\noindent By grouping the electrons one can obtain the structural
change, $\mathcal{S}_{QCM} \to \mathcal{S}_{SSM}$. This can be
achieved by joining $Z_i$ electrons with the $i$th atomic nucleus,
so as to obtain the $i$th electrically neutral atom. e.g. For the
hydrogen molecule, the structure $\mathcal{S}_{QCM} = \{1e, 1p,
2e, 2p\}$ can be transformed by grouping\footnote{\small Of
course, we assume the bound states--otherwise we have  free
particles. } to obtain $\{(1e, 1p), (2e, 2p)\} = \{1H, 2H\}$--cf.
the point (C) in Section 2.2.1. The next step may be to introduce
the atomic $CM$ and $R$ systems: $H = CM+R$. Then the hydrogen
molecule is described by the following structure: $\{(1CM, 1R),
(2CM, 2R)\}$. The $R$-system's excitations can be described in the
Fock-space representation, eq.(37)-(38). Finally, by neglecting
the atomic excitations, one obtains the ChM model (structure) of
the molecule.

\smallskip

\noindent This chain of transformations can be shortly presented
as follows: %%
\begin{equation}
\mathcal{S}_{QCM} \longrightarrow \mathcal{S}_{SSM}
\stackrel{negl-excit}{\longrightarrow} \mathcal{S}_{ChM}.
\end{equation}

\noindent Eq.(41) can be readily presented in the Hilbert state
space structure terms. Regarding the Hamiltonian, eq.(41) can be
presented as follows\footnote{\small See Supplement for some
details.}: %%
\begin{eqnarray}
&\nonumber&  H_{QCM} \stackrel{grouping}{\longrightarrow}
H_{group} = \sum_{\alpha} [ T_{n\alpha} +
\sum_{q_{\alpha}=1}^{Z_{\alpha}} (T_{eq_{\alpha}} + V^{en}_{\alpha
q_{\alpha}}) + \sum_{q_{\alpha}, q'_{\alpha}(\neq
q_{\alpha})=1}^{Z_{\alpha}} V^e_{q_{\alpha}q'_{\alpha}} ]
\\&& + H' \stackrel{nontriv}{\longrightarrow}
\sum_{\alpha} [ T_{CM\alpha} + m_{CM\alpha} \omega_{\alpha}^2
x_{CM\alpha}^2/2 + T_{R\alpha} + V_{R\alpha} ]
\stackrel{excit}{\longrightarrow} \nonumber \\&&
 H_{SSM} = \sum_{\alpha} [
T_{CM\alpha} + {m_{CM\alpha} \omega_{\alpha}^2 x_{CM\alpha}^2} +
H^{ex}_{\alpha} ]
 \stackrel{negl-excit}{\longrightarrow} H_{ChM}.
\end{eqnarray}

\noindent In eq.(42), "$\alpha$" enumerates the atoms, while $H'$
contains  interactions between the constituents of the different
atoms. There are  exactly decoupled $CM$ and $R$ systems for every
atom, while the $H'$ is the origin for the (effective) harmonic
potentials for the atoms. The terms in eq.(42) are simplified
since we do not take into account the electrons that are shared by
the neighbor atoms. Of course, there may be some corrections to
the exact harmonic potential that are not made explicit in
eq.(42). Stating eq.(42) in the more rigorous form does not alter
our main observations.

\smallskip

\noindent Compare the $H_{group}$ from eq.(42) with the $H_{QCM}$
eq.(40). For the QCM structure, all the electrons, by definition,
are subject to the Pauli exclusion principle. However,  for the
$H_{group}$, the Pauli exclusion principle applies exclusively to
the electrons belonging to the same atom.

\smallskip

\noindent Quantum state of $Z$ electrons in the QCM structure is
the following Slater determinant: %%
\begin{equation}
\vert \Psi\rangle_{molecule} = {1 \over \sqrt{Z!}} \left|
\begin{array}{ccc}
\vert\Psi_1\rangle_1 & \vert\Psi_2\rangle_1 &... \vert\Psi_Z\rangle_1\\
\vert\Psi_1\rangle_2 & \vert\Psi_2\rangle_2 &... \vert\Psi_Z\rangle_2\\
\\
\dots\\
 \vert\Psi_1\rangle_Z & \vert\Psi_2\rangle_Z &... \vert\Psi_Z\rangle_Z\\
\end{array}
\right|
\end{equation}

\noindent However, for the electrons-system  of the $\alpha$th
atom, the Slater determinant reads as: %%
\begin{equation}
\vert \Phi\rangle_{\alpha} = {1 \over \sqrt{Z_{\alpha }!}} \left|
\begin{array}{ccc}
\vert\Psi_1\rangle_1 & \vert\Psi_2\rangle_1 &... \vert\Psi_{Z_{\alpha}}\rangle_1\\
\vert\Psi_1\rangle_2 & \vert\Psi_2\rangle_2 &... \vert\Psi_{Z_{\alpha}}\rangle_2\\
\\
\dots\\
 \vert\Psi_1\rangle_{Z_{\alpha}} & \vert\Psi_2\rangle_{Z_{\alpha}} &... \vert\Psi_{Z_{\alpha}}\rangle_{Z_{\alpha}}\\
\end{array}
\right|
\end{equation}

\noindent For noninteracting atoms,  the total electrons system
state is simply tensor product: %%
\begin{equation}
\vert \Phi\rangle_{molecule} = \otimes_{\alpha} \vert
\Phi\rangle_{\alpha}, \quad Z = \sum_{\alpha} Z_{\alpha}.
\end{equation}

\noindent The point to be emphasized is that: %%
\begin{equation}
\vert \Psi\rangle_{molecule} \neq \vert \Phi\rangle_{molecule}.
\end{equation}

\noindent Correlations of  identical particles are so specific,
that
  $\sum_i c_i
\vert \Phi_i\rangle_{molecule} \neq$ $\vert
\Psi\rangle_{molecule}$, where $\vert \Phi_i\rangle_{molecule}$ is
of the form of eq.(45) for every index $i$.

\smallskip

\noindent Needless to say,  eq.(43) is  reducible to eq.(45). So,
$\mathcal{S}_{QCM}$ is the most general and the most fundamental
[non-relativistic] model of molecule.

\smallskip

\noindent Of course, this generality of the QCM structure does not
imply that the QCM structure is reducible onto the other
structures {\it in the sense} of the point (C) of Section 2.2.1.
To this end, only the SSM structure, eq.(37), is reducible to the
QCM structure via decomposing the atoms, and is also reducible to
the ChM structure by neglecting the electrons system.

\smallskip

\noindent It is important to stress: {\it a huge amount of
information is lost} in  transition from the QCM to the SSM
structure. Already at the first step of  grouping, certain
electrons correlations are lost, cf. eq.(46). By introducing the
atomic $CM$ and $R$ systems, correlation of the electrons and the
atomic nuclei is lost. Thereby, as distinct from the QCM
structure, for the SSM structure there is no hope for obtaining
the molecule configuration change (transformation) via influencing
the atomic internal degrees of freedom--some external action that
could couple the $CM$ and the $R$ degrees of freedom is needed.
So, the following question is in order: {\it to what extent the
conclusions obtained for one molecule structure can be applied to
another structure?} A partial answer will be given in Section 4.3.

\bigskip

\noindent {\bf 4.2 The protein folding problem}

\bigskip

\noindent Protein molecules are large--of the mass in the interval
$10^4-10^9$ a.m.u. (atomic mass units). There is really a huge
number of the possible geometric shapes  for proteins.
Interestingly,  biochemists claim that there exists a special,
globular shape, the so-called "native" shape of a molecule, that
is biologically active. Non-globular ("degenerate" shape) protein
configurations are biologically inactive, "dead", or even toxic.
Transition from the non-native to the native form (conformation)
is the protein folding problem (PFP).

\smallskip

\noindent In living organisms, protein molecules are suspended in
water. Therefore, it is expected that their dynamics is
non-trivially influenced by interaction with the solvent
molecules. For a single molecule, the structures considered in
Section 4.1 are global, while for a molecule suspended in a
solution, the structures are local (the transformations of
variables do not include the solvent-molecules degrees of
freedom).

\smallskip

\noindent PFP is fairly described by the so-called Levinthal
paradox (Levinthal 1968): If a single protein molecule is going to
be folded by sequentially sampling of all possible
conformations\footnote{\small For simplicity, we further
interchange the use of [molecule] "shape", "configuration" and
"conformation". Terminological subtlety is of no importance for
our considerations.}, it would take an astronomical amount of time
to do so, even if the conformations were sampled at a rapid rate
(on the nanosecond or picosecond scale). Based upon the
observation that proteins fold much faster than this, Levinthal
then proposed that a random conformational search does not occur,
and the protein must, therefore, fold through a series of
meta-stable intermediate states. This kinetic picture of the
protein folding is at the core of the modern approach (Dill and
Chan 1997).

\smallskip

\noindent PFP traditionally refers to the ChM molecule structure,
Section 4.1. A molecule is imagined as a random coil that should
fold in a sequence of some well defined conformation changes. The
quantum mechanical counterpart does not seem to be much more
useful for resolving the PFP. In the next section we briefly
review some classical approaches, and emphasize the kinetic nature
of the problem. In Section 4.2.2 we briefly describe a new,
quantum-decoherence based paradigm that refers to the quantum
chemistry molecule structure.

\bigskip

\noindent {\bf {\emph {4.2.1 The statistical-thermodynamic
approach}}}

\bigskip

\noindent The ChM structure is of interest. The point-like-atoms'
equilibrium-positions form a three-dimensional lattice. Regarding
the large molecules, "conformation" is a lattice with the fixed
(average) distance between the adjacent atoms and the fixed
(average) angles between the adjacent lattice segments. For every
change of conformation it is assumed {\it not} to change the
distances and the angles--conformal transformations.

\smallskip

\noindent Protein folding is defined as a series of local
rotations that sequentially change the molecule's shape. Even for
small protein molecules, the number of  combinations of  local
rotations is huge. It is not expectable that a molecule quickly
find the native conformation--the Levinthal paradox.

\smallskip

\noindent The most of the current research on protein folding
considers an ensemble of molecules suspended in a solution at
fixed temperature. Related methods provide powerful means for
determining conformations even for the very large protein
molecules. Some computational methods are based on the assumption
that the native state is very stable--it can be imagined as a
minimum of the configuration-energy landscape. While all of these
methods can provide {\it existence} of the native conformation (as
well as, in general, some metastable conformations), the PFP
problem, as stated by Levinthal, is more subtle--it's {\it
kinematic}. As (Dill and Chan 1997) strongly emphasize: the PFP is
{\it not} merely about existence of the (meta)stable
conformations, but rather about the possible configuration
transitions within the classical configuration space of a {\it
single} molecule.

\smallskip

\noindent Even the kinetic approach does not challenge the basic
strategy stemming from the classical ChM molecule structure: for a
single molecule it is supposed that there is a {\it pathway}
("trajectory") in the molecule conformation space, while in every
instant in time a molecule has a definite geometrical form. In
addition,  internal degrees of freedom of molecule are neglected,
and are sometimes treated as  un-necessary complication. The
approach (Dill and Chen 1997) is still in a purely qualitative
form.

\bigskip

\noindent {\bf {\emph {4.2.2 A quantum decoherence approach}}}

\bigskip

\noindent Quantum mechanical approach introduces the atomic
internal degrees of freedom into consideration. To this end, the
results that can be obtained for the SSM or QCM structure are in
principle not achievable on the basis of the ChM structure. So,
partial answer to the above question (cf. Section 4.1) reads as:
there is a lot of information about the folding mechanism that are
inaccessible within the ChM-based approach.

\smallskip

\noindent To see  richness of the SSM and QCM structures compared
to the ChM structure, we recall and extend what is told in Section
4.1. Both the SSM and especially the QCM structure provide a basis
for the electrons-system-mediated change of conformation. To this
end, it is well known that the molecule fluorescence and
phosphorescence are phenomena closely related to protein folding.
On the other hand, quantum mechanical approach,  in principle,
does not allow a definite pathway in the configuration space of
the molecule. This may be a hint for avoiding the Levinthal
paradox.

\smallskip

\noindent However, there is more subtlety to the quantum
mechanical approach to PFP. Ever since Hund's  remark (Hund 1927),
it is a foundational issue of the whole of chemistry: how do the
definite, the classical-like, stable molecule configurations
appear from the quantum mechanical substrate? Furthermore, if
quantum mechanics can provide protein conformation as a
classical-like stable characteristic, one can wonder if it may
 happen that, after all, the configuration transitions are
inevitably classical--i.e. that follow some special pathways in
the configuration space?

\smallskip

\noindent In the remainder of this section we offer answers to
both questions, {\it in} the context of the QCM molecule
structure. The answers are due to the process of quantum
decoherence: {\it both} configuration stability and transitions
can be naturally [but purely qualitatively] described (Jekni\'
c-Dugi\' c 2009a). This way both, the Hund's  and the Levinthal
paradox, are resolved; see also Rakovi\' c et al 2014.

\smallskip

\noindent The QCM structure is defined by eq.(40). Here we apply
the standard adiabatic approximation to the following variation of
the QCM structure: $\mathcal{S'}_{QCM} = \{E, N\}$, where  $E$
stands for the electrons, and the $N$ for the system of
atomic-nuclei. This bipartition helps us straightforwardly to
introduce the $CM$ and $R$ degrees of freedom for the later: $N =
CM_N + R_N$. The set of the relative atomic-nuclei positions,
$\vec \rho_{ij} = \vec r_i - \vec r_j$ can be further decomposed.
Actually, the set $\{\vec \rho_{ij}\}$ can be divided into two
subsets, which define the rotational ($Rot_N$) degrees of freedom
and the internal, the conformation ($K_N$), degrees of freedom.

\smallskip

\noindent So, a molecule is defined by the following factorization
of the Hilbert space: %%
\begin{equation}
\mathcal{H} = \mathcal{H}_E  \otimes \mathcal{H}_{CM_N} \otimes
\mathcal{H}_{Rot_N} \otimes \mathcal{H}_{K_N}.
\end{equation}

\noindent The related form of the molecule
Hamiltonian\footnote{\small See Supplement for details.}: %%
\begin{eqnarray}
&\nonumber&  H_{QCM} = \sum_i {\vec p_{ei}^2 \over 2 m_e} +
\sum_{i\neq j} V_{ij} + T_{CM_N} + T_{Rot_N} + T_{K_N} +
\\&&
V_{E,CM_N} + V_{E,Rot_N} + V_{E,K_N} + V_{Rot_N, K_N}.
\end{eqnarray}

\noindent where, as usually, the double subscripts distinguish the
interaction terms; the $T$ denoting the kinetic energy terms and
the index $E$ standing for the total electrons system.

\smallskip

\noindent Compared with the standard QCM structure, $E+N$, our
approach has the following virtues. First, cf. Section 4.3 below,
it can be directly compared with the ChM and the SSM structures.
Second, there are at least three different channels of the
environmental (the solvent) influence on $K_N$--the last two
interaction terms in eq.(48) provide the possible "channels" for
influencing the molecule conformation. Third, one can easily apply
the standard adiabatic approximation to the structure eq.(47).
Fourth, on this basis, one can define the electrons positions to
be measured from the  $CM_N$ system as the reference system.
Classically, the $\vec r_{CM_N}$ is a $c$-number, not a dynamical
variable; so, rigorously, the electrons variables presented
formally as the operators acquire the form: $\hat{\vec{r}}_{ei} -
\vec r_{CM_N}\hat I$. The quantum mechanical reference frames will
be considered in Chapter 8. Due to the presence of the electrons,
the  $CM_N$ and $R_N$ systems are in mutual
interaction\footnote{\small This coupling is absent for the atomic
$CM$ and $R$ of the {\it total} atom.}.

\smallskip

\noindent The following masses are  implicit in  the
kinetic-energy terms in eq.(48): the electron mass $m_e$, the
$CM_N$ mass $M$, the rotational moment of inertia
$I_{Rot_N}$\footnote{\small Properly expressed in the mass
units.}, and the "reduced masses" $\mu_{K_Ni}$. Now it is easy to
show that the adiabatic parameter $\kappa^{4/3} = max \{m_e/M,
m_e/I_{Rot_N}, m_e/\mu_{min}\} \sim m_e/\mu_{min} < 10^{-3}$,
where $\mu_{min} $ is the minimum "reduced mass"--of the order of
the minimum nucleus-mass. Physically it means that, like for the
standard QCM structure, $E+N$, one can adiabatically cut off the
electrons system  from the rest of the molecule.

\smallskip

\noindent In the zeroth order of approximation, when dynamics of
the non-electronic degrees of freedom is "frozen", the Hamiltonian
eq.(48) reduces to the electrons-system's Hamiltonian: %%
\begin{eqnarray}
&\nonumber&  H_E = \sum_i {\vec p_{ei}^2 \over 2 m_e} +
\sum_{i\neq j} V_{ij} + V(CM_N) + V(E,Rot_N) + V(E,K_N) +\\&&
V(Rot_N, K_N) \approx \sum_i {\vec p_{ei}^2 \over 2 m_e} +
\sum_{i\neq j} V_{ij}  + V(Rot_N) + V(K_N).
\end{eqnarray}

\noindent The terms on the rhs of eq.(49) represent the effective
external classical fields  for the electrons
system\footnote{\small  The fixed atomic nuclei positions enter as
the fixed parameters in eq.(49).}. We assume that the $CM_N$
system will not affect the electrons, at least as long as
adiabatic approximation is satisfied.

\noindent The electrons-system Hamiltonian eq.(49) gives rise to
the zeroth order Schr\" o\-digner equation: %%
\begin{equation}
H_E \vert \phi_n(K_N)\rangle_E = E_{n}(K_N) \vert
\phi(K_N)\rangle_E.
\end{equation}

\noindent In eq.(50), we keep only  dependence on the
conformation, $K_N$, since it  defines  spatial configuration of
the positively charged atomic nuclei. Tentatively neglecting the
$Rot_N$ and $CM_N$ systems, the zeroth order form of the molecule
quantum state reads as: %%
\begin{equation}
\vert \phi_n(K_N)\rangle_E \otimes \vert K\rangle_N + \vert
O(\kappa) \rangle,
\end{equation}

 \noindent Just like in eq.(21), there is
entanglement--for the  $E$ and $K_N$ systems.

\smallskip

\noindent Now dynamics of the $K_N$ system is adiabatically
defined by the effective ("average") Hamiltonian (Gribov and
Mushtakova 1999): %%
\begin{equation}
H_{K_N} = _e\langle \phi_n(E_n(K_N)) \vert H_{QCM} \vert
\phi_n(E_n(K_N))\rangle_e.
\end{equation}

\noindent Intuitively, eq.(52) describes what the $K_N$ system can
"see" of the fast electrons-system dynamics. Without delving  into
details, we emphasize: the energy eigenvalue for the fixed quantum
number $n$, $E_n(K_N)$, represents an effective potential-energy
(hyper)surface for the configuration system. It is usually assumed
that there are certain depressions (the local minimums) in the
potential energy landscape that would correspond to the
phenomenologically observed stable conformations (Gribov and
Mushtakova 1999, Atkins and Friedman 2005). Of course, the number
of such local minimums (the stable conformations) is enumerable
($K_{N1}, K_{N2},...$) for every $n$.

\smallskip

\noindent Conformation dynamics generated by the  Hamiltonian
eq.(52) is usually imagined as conformational {\it vibrations}
(oscillations) in the vicinity of a local minimum of the
conformation-energy hypersurface. Given that the weak $V_{K_N,
Rot_N}$ interaction can be considered as a perturbation, the {\it
exact} form of the molecule state  is of the form\footnote{\small
Integrating over the electronic degrees of freedom in eq.(52)
turns all the electrons-system's couplings [appearing in eq.(48)]
with the rest into the external fields, and the only remaining
interaction term is the $V_{K_N, Rot_N}$ term.}: %%
\begin{equation}
\vert \Phi\rangle_{molecule} = \vert \phi_n(K_N)\rangle_E \vert
K\rangle_N \vert \chi \rangle_{Rot_N} \vert \phi\rangle_{CM_N} +
\vert O(\kappa)\rangle_{E,K_N,Rot_N,CM_N}.
\end{equation}

\noindent So, if one deals with the first (the dominant) term on
the rhs of eq.(53), the adiabatic method guarantees that he will
obtain the results with an error not larger than $\kappa \ll 1$.

\smallskip

\noindent But this is strange, since neither the exact state
eq.(53) nor its dominant term are eigenstates of the molecule
Hamiltonian eq.(48). In the dominant term, the conformation state
$\vert K\rangle_N$ is the molecule-conformation eigenstate.
Bearing in mind that $[H_{QCM}, K_N] \neq 0$, one may say that the
adiabatic approximation provides a partial answer to the  Hund's
paradox (Hund 1927). Nevertheless, the adiabatic "mechanism" is
not sufficient for this purpose (Gribov and Magarshak 2008, Dugi\'
c and Jekni\' c-Dugi\' c 2009).

\smallskip

\noindent On the other hand, for the stable classical-like
conformations, one may wonder if {\it any quantum-mechanical
mechanism can provide the finite-time conformational
transitions}--the Levinthal paradox (Levinthal 1968)?

\smallskip

\noindent Interestingly enough, the following plausible
stipulations provide a coherent and rather general background for
answering both   questions. The stipulations are
phenomenologically inspired:  typical experimental investigations
are performed on an ensemble of molecules in a solution (e.g. in
water). In this new context both the Hund's and the Levinthal
paradoxes are resolved (Jekni\' c-Dugi\' c 2009a)\footnote{\small
http://www.verticalnews.com/premium-newsletters/Journal-of-Physics-Research-/2009-03-31/71094PR.html.}.

\smallskip

\noindent {\bf Stipulation 1}. {\it For every} stationary {\it
state of the composite system "[ensemble of] molecules +
solution", the solution acts as a decoherence-inducing environment
for the molecule conformation system $K_N$.}

\smallskip

\noindent {\bf Stipulation 2}. Non-stationary {\it state of the
"[ensemble of] molecules + solution" system does not preserve
molecular conformations. Every non-stationary state terminates by
a stationary state.}

\smallskip

\noindent If there is not any severe external influence on the
"molecules+environment" system, then we say the system is in
stationary state. "Non-stationary" means the opposite, i.e. the
different ways the composite system can be disturbed, e.g., by
heating, by intense illuminating, by adding new solvent (this can
change the solution pH value) etc. Both "stationary" and
"non-stationary" are {\it phenomenologically} inspired. By
definition, "stationary state" is a state (or, physically, a set
of states) that follows from some kind of the environment
relaxation. For large environment, this can be thermodynamic
equilibration.

\smallskip

\noindent The stipulations do not prejudice either the decoherence
mechanism or the asymptotic ($t \to \infty$) relaxation into a
(possibly unique) stationary state of the open system, $K_N$. The
initial and the final (non-asymptotic) stationary states may be
physically totally different, {\it except} in that that they
should provide the occurrence of decoherence for the $K_N$ system.

\smallskip

\noindent Stipulation 1 establishes:  arbitrary initial state of
the molecule conformation quickly becomes a mixture of
conformations: %%
\begin{equation}
\rho_{K_N} = tr_{E,CM_N,Rot_N} \rho_{molecule} = \sum_i p_i \vert
k_i\rangle_{K_N}\langle k_i \vert.
\end{equation}

\noindent {\it Every} "stationary state" is described by eq.(54).
Of course, the sum in eq.(54) can sample different sets of
conformations for  different stationary states.

\smallskip

\noindent Now, according to Stipulation 2,  external influence
does {\it not} preserve the states in the form of eq.(54). Even
more, one can expect that  external influence (giving rise to
"non-stationary state") provides a time dependent state
$\rho'_{K_N}(t)$, such that: %%
\begin{equation}
[\rho'_{K_N}(t), \rho'_{K_N}(t')] \neq 0, \quad t \neq t'.
\end{equation}

\noindent Totally independently of the non-stationary state
dynamics\footnote{\small Except if one assumes nonrealistic
scenario that the external influence preserves the
conformation-system state.},
 relaxation  into another
stationary state provides [Stipulation 1] the final
conformation-system state, $\rho''_{K_N}$, of the general form of
eq.(54)\footnote{\small This is a direct consequence of the fact
that decoherence is  a quantum measurement performed by the
environment on the open system. The final state is a mixture of
the measured-observable eigenstates. Here it's the molecule
conformation that is measured.}. Therefore, the total dynamics of
the open system can be described as follows: %%
\begin{equation}
\rho_{K_N} = \sum_i p_i \vert k_i\rangle_{K_N}\langle k_i \vert
\to \rho'_{K_N} = \sum_i \pi_i \vert \chi_i\rangle_{K_N}\langle
\chi_i \vert \to \rho''_{K_N} = \sum_i q_i \vert
k'_i\rangle_{K_N}\langle k'_i \vert,
\end{equation}

\noindent where: $\sum_i p_i = 1 = \sum_i \pi_i = \sum_i q_i$,
$\vert \chi_i \rangle = \sum_j c_{ij} \vert k_j\rangle$, and
$[\rho, \rho'] \neq 0 \neq [\rho', \rho'']$, but $[\rho, \rho''] =
0$. The final state ($\rho''$) can mix the conformations that are
not present in the initial state ($\rho$), while the statistical
weights for the common conformations need not be equal for the
initial ($\rho$) and the final ($\rho''$) state; i.e. $p_i \neq
q_i$ for at least some index $i$.

\smallskip

\noindent This possibility of appearance of the new conformations,
as well as of the different probabilities for the common
conformations for the initial and the final conformation-system
state, is the {\it decoherence-based model of the conformation
transitions} in large molecules.

\smallskip

\noindent So, Stipulation 1 provides an answer to the Hund's
paradox. On the other hand, both Stipulation 1 and Stipulation 2
provide a general basis for the conformational transitions,
eq.(56). The time needed for such transitions is of the order of
the "decoherence time" thus not leaving room for the Levinthal
paradox.

\pagebreak

\noindent {\bf {\emph {4.2.3 Overview}}}

\bigskip

\noindent The approach presented in Section 4.2.2 is purely
qualitative. But this is, as yet, unavoidable--the composite
system of interest is too complicated. This is also the case with
the classical-physics approach (Dill and Chen 1997).

\smallskip

\noindent As distinct from the classical models, the model of
Section 4.2.2 does not leave room for the classical pathways in
the molecule configuration space, while there are different
"channels" for the conformation transitions. The adiabatic
approximation provides the local minimums on the energy
hypersurface as the preferred, the decoherence-distinguished
stable conformations--an answer to the Hund's paradox. The fast
decoherence process dissolves the Levinthal paradox.

\smallskip

\noindent From the quantum mechanical point of view, while the
occurrence of decoherence is expected for the conformation system
(Stipulation 1), a rigorous proof of this expectation is virtually
intractable. To this end, a part of the
difficulties\footnote{\small The main difficulty is the fact that
the protein molecules in the living biological cells are far from
the thermodynamic equilibrium.} will be presented and discussed in
Chapters 6 and 7. Here we finish our considerations by comparing
the approach of Section 4.2.2 with some similar models/approaches
in the literature.

\smallskip

\noindent (Gelin et al 2011) derive a master equation for a
molecular aggregate in contact with the heat bath. Their model is
similar to the model of Section 4.2.2, yet with simplification of
identical constituents of the aggregate. Adiabatic approximation
is implicit to their model, which  couples the aggregate's $CM$
system  (but not the conformational system) with the environment.
In effect, they obtain a basis for the $CM$-system's quantum
Brownian-like motion, while the internal degrees of freedom remain
purely quantum mechanical. This result is a consequence of a
number of approximations, notably of the assumption that all the
constituents are mutually identical (chemically and physically).
Due to the absence of solutions to the master equation, they do
not tackle either the Hund's or the Levinthal paradox.

\smallskip

\noindent In a recent paper (Luo and Lu 2011), the authors
consider the quantum mechanical transitions of the protein
conformations for the different temperature regimes. The QCM
molecule structure is of interest. This approach regards the
thermodynamic description (Section 4.2.1) while assuming existence
of the definite (the initial and the final) conformation. So, they
don't even  tempt to answer the Hund's paradox.

\smallskip

\noindent On the other hand, recent papers (Trost and Hornberger
2009, Bahrami et al 2012) consider the Hund's paradox for the
quantum-mechanical counterpart of the ChM structure, but
exclusively for the {\it small}-molecules chirality, while leaving
the configuration transitions issue (and the Levinthal paradox)
intact. Complexity of the occurrence of decoherence for small
molecules suggests virtual intractability of the same issue for
the large molecules (Stipulation 1). A similar quantum mechanical
approach to Hund's problem can be found in (Jona-Lasinio and
Claverie 1986, Amann 1991). Therein, interaction with the
environment is {\it designed} so as to provide decoherence, while
the microscopic and structural considerations are completely left
out. The authors don't even try to describe the configuration
transitions.

\bigskip

\noindent {\bf 4.3 Quantum structures in context}

\bigskip

\noindent Probably the main lesson of Section 4.2.2 is a need for
a proper selection of the degrees of freedom (of a subsystem of a
composite system)--the conformation $K_N$ system can be compared
with the molecules conformation-systems for the ChM and SSM
models. This is achieved by performing the proper LCTs
 in conjunction with adiabatic approximation. Notice that the
LCTs are applied {\it locally} to the system of atomic nuclei by
introducing the $CM_N$ and $R_N$ subsystems, and then by the "fine
graining" of the $R_N$ system to introduce the $Rot_N$ and $K_N$
subsystems: $\{1n,2n,...\} \to \{CM_N, R_N\} \to \{CM_N, Rot_N,
K_N\}$.

\smallskip

\noindent There is a chain of the molecule-structure
transformations (compare to eqs. (41), (42))\footnote{\small Of
course, $1E+1n=1CM+1R=1CM+1ex$ are the different decompositions of
the one and the same atom denoted $1$.}: %%
\begin{eqnarray}
&\nonumber&  \mathcal{S}_{QCM} = \{1e,1n, 2e, 2n,...\}
\stackrel{electrons-grouping}{\longrightarrow} \{1E, 1n, 2E,
2n,...\} \stackrel{nontrivial}{\longrightarrow}\\&&  \{1CM,1R,
2CM, 2R,...\} \stackrel{introd-excit}{\longrightarrow}
\mathcal{S}_{SSM} = \{1CM, 1ex, 2CM, 2ex,...\}\nonumber \\&&
\stackrel{negl-excit}{\longrightarrow} \mathcal{S}_{ChM} =\{1,
2,...\}.
\end{eqnarray}

\noindent Every step in eq.(57) is subject to quantum correlations
relativity, Section 3.2. So, there is not direct transition of
conclusions from one to another structure. Nevertheless, due to
the small mass ratio $m_e/m_n$, the atomic center of mass is close
to the atomic nucleus position (e.g. $1n\approx$ 1CM $ \approx1$
for the structures appearing in eq.(57)). Bearing this (i.e.
eq.(53)) in mind, we can hope, that Section 4.2.2 provides a
qualitatively useful description of the conformation stability and
transitions also for the  ChM and SSM structures. However, this
conclusion does not directly apply to the electrons system--cf.
eq.(46)--as well as to channelling the conformation transitions.

\smallskip

\noindent From eq.(53) we can see, that the external influence
exerted on the electrons system $E$, or on  the rotational degrees
of freedom $Rot_N$, can also influence the molecule conformation
$K_N$ system. The details regarding the preferred configuration
states (the configuration "pointer basis") as well as a scenario
regarding the electrons-system mediated configuration transitions
 can be found in (Jekni\' c-Dugi\' c
2009a,b).

\smallskip

\noindent On the other hand, from eq.(42), the only way indirectly
to influence conformation for the SSM  structure, is, to
externally induce interaction between the $\{CM_{\alpha}\}$ and
the excitation systems (Caspi and Ben-Jacob 2000). Even this
possibility is absent for the ChM model, for which the only way to
change conformation is directly to target the conformation system
(Jona-Lasinio and Claverie 1986, Amann 1991).

\smallskip

\noindent Our considerations do not exhaust the list of the
possible molecule structures\footnote{\small Subtlety of the
molecular structures are also presented in (Michal Svr\v cek,
2012). Regarding the foundations and limits of the adiabatic
approximation, see (Gribov and Magarshak 2008, Dugi\' c and
Jekni\' c-Dugi\' c 2009a).}, neither the list of the possible ways
to manipulate the molecule degrees of freedom. Similarity of the
effects as well as the quantum correlations relativity, Section
3.2, suggest that the realistic experimental situations are hardly
structurally as clear and neat as our (idealized) theoretical
formulations.

\newpage

\noindent {\bf Chapter 5}

\noindent\textbf{\large Realistic Physical Structures}

\vspace*{55mm}

\noindent  It is a universal physical fact: of a composite system,
only a fraction of the degrees of freedom is practically
accessible (Giulini et al 1996, Zurek 2003, Schlosshauer 2004,
Nielsen and Chuang 2000). The classical, macroscopic bodies are
described by their spatial shape and orientation. In formalism,
those are the "collective" variables of the center of mass and the
Euler angles. Internal degrees of freedom are not directly
observable and provide a basis for the macroscopic-bodies
temperature and radiation.

\smallskip

\noindent Quantum mechanical systems (atoms, molecules etc.) are
also described by the center-of-mass and the relative-positions
degrees of freedom. These degrees of freedom are  presented by
eq.(4) and by the rhs of eq.(5). Manipulating these degrees of
freedom makes them realistic in the operational physical sense.
Bearing Chapters 2 and 3 in mind, in this chapter, we provide a
fresh view of some well known experimental situations and we
highlight {\it operational reality} of the $CM$ and $R$ degrees of
freedom.

\bigskip

\noindent {\bf 5.1 Relativity of "local operations" }

\bigskip

\noindent The concept of "structure" assumes locality of the
subsystems degrees of freedom.  Manipulating the subsystems
degrees of freedom assumes their (local) accessibility.

\smallskip

\noindent "Locality" is structure dependent, Section 2.4 (not
necessarily incorporating the relativistic locality).  "Local
operation" assumes {\it non-disturbance of the rest}, which is a
part of the same structure of the composite system. In quantum
information science, "local operations" are presented e.g. by the
so-called "local operations and classical communication" (LOCC)
procedures. Formally, local operations are  defined by the
"single-particle" operators of the form $A \otimes I$.  e.g. The
center of mass position, $X_{CM}$, eq.(4), takes the form $X_{CM}
\otimes I_R$ relative to the $CM+R$ structure, {\it but} for the
$1+2$ structure, it takes the form $(m_1/M) x_1 \otimes I_2 +
(m_2/M) I_1 \otimes x_2$. So for the $CM+R$ structure, $X_{CM}$ is
a local, while for the $1+2$ structure, it is a "collective"
("composite") observable. Direct measurement of $X_{CM}$ is
supposed not to disturb the $R$ system, while partially disturbing
both the $1$ and $2$ systems. On the other hand, $X_{CM}$ can be
indirectly measured by directly measuring $x_1$ and $x_2$, and
then, according to eq.(4), to calculate $X_{CM}$. However, due to
eq.(4), such measurement provides information also about the $R$
system and is therefore not local. The told equally refers to
arbitrary observable of the composite system. For instance, $x_1$
can be indirectly measured by directly measuring $X_{CM}$ and
$r_R$. Therefore, the concept of "local observable/operation" as
well as of the "composite observable/measurement" is structure
dependent--the electron's position $\vec{r}_e$ in the hydrogen
atom is a collective observable {\it relative} to the atomic
$CM+R$ structure. Formally, there is nothing "more local"
regarding the electron's position, $\vec{r}_e$, than regarding the
$CM$ position, $\vec{R}_{CM}$.

\smallskip

\noindent Whether an observable is accessible to measurement in a
given physical situation is a separate question (Zanardi 2001,
Harshman 2012a). Here we adopt the following:

\smallskip

\noindent {\bf Def.5.1} "Accessibility" {\it of an observable of a
system assumes a measurement procedure, which does not make use of
any indirect measurement, i.e. measurement of other observables of
the system.}

\smallskip

\noindent Inaccessibility of a macroscopic-system center-of-mass
is at the root of the classical prejudice on the transformations
of variables, Section 2.1. e.g. For the classical systems, the
formal $CM$ system pertains to an empty point in space, not to a
physical object\footnote{\small You cannot move a pair of apples
by hitting their center of mass. In order to measure the apples
$CM$ position, you need to perform measurement of the apples'
positions, and then to calculate the apples $CM$  position.}.

\smallskip

\noindent The most of the realistic quantum measurements employ
detection of quantum particles. e.g. In atomic and  molecule
spectroscopy, the photon field  is accessible (directly measured,
detected). This detection provides an (indirectly acquired)
information about the atomic (molecule) internal energy and state.
This is a local operation relative to the atomic $CM+R$ structure,
but is a global operation relative to the $e+p$ structure.
Mechanism of quantum measurements, even the simplest ones, is not
yet known. So Def.5.1 does {\it not} refer to such details.
Rather, Def.5.1 assumes, that measurement of an observable does
not assume or reveal the values of the observables\footnote{\small
All but those that can be trivially linked with the measured one.}
of any other system. Hence "accessibility" requires locality of
measurement but is  more stringent: it also requires absence of
information about any other observable.

\smallskip

\noindent Accessibility (direct measurement) of the
hydrogen-atom's electron's and the proton's positions, $\vec r_e$
and $\vec r_p$, provides indirect measurement of the atomic $CM$
and $R$ positions. This is, of course, a local operation relative
to the atomic $e+p$ structure, but is a global operation relative
to the $CM+R$ structure. So we emphasize the following universal
physical fact: accessibility of an observable of a quantum system
is a matter of a {\it specific physical situation}, which is
defined by the choice of the "apparatus" and of its initial state.
An example of accessibility, which is determined by the
environment characteristics, can be found in Section 7.3.

\smallskip

\noindent The concept of locality now emphasizes subtlety of the
concept of "multi-particle entanglement (correlation)" (Brus 2002,
Facchi et al 2006, Wichterich 2011, Bellomo et al 2011). Consider
a system $\mathcal{C}$ of $N$ non-identical particles. Its Hilbert
state space $\mathcal{H} = \otimes_i^N \mathcal{H}_i$ and the
state $\vert \Psi\rangle = \sum_{i_1, i_2,...i_N} C_{i_1i_2...i_N}
\otimes_{j=1}^N \vert \phi_{i_j}\rangle_j$; $\vert
\phi_{i_j}\rangle_j$ is the $i$th state of the $j$th particle. A
bipartition $\mathcal{C} = A+B$ determines the factorization
$\mathcal{H}_A \otimes \mathcal{H}_B$ and the state $\vert
\Psi\rangle =  \sum_i c_i \vert i\rangle_A \vert i \rangle_B$,
which is given in the Schmidt canonical form. The point to be
emphasized: bipartition $A+B$ is comparable with a pair of
unstructured particles. In other words: there is no {\it a priori}
more entanglement in the $A+B$ structure than in a state $\vert
\Psi\rangle = \sum_i c_i \vert i\rangle_1 \vert i \rangle_2$ for a
pair of unstructured particles $1$ and $2$. However, this
similarity fades if the $A$'s and $B$'s structures are taken into
consideration. In this case, the task of "multi-particle
correlations" refers to the correlations of a numerous set of
particles belonging to the different partitions ($A$ and $B$,
respectively).

\smallskip

\noindent Measurements of  observables that are local {\it
relative} to the structure of interest may reveal quantum
correlations in {\it that} structure. So, e.g., measurement of an
obsevable $\hat A$ that is sensitive to the $A$-block's  structure
is not necessarily useful for detecting entanglement for the $A+B$
partition (regarding the above Schmidt form, $\vert \Psi\rangle =
\sum_i c_i \vert i\rangle_A \vert i \rangle_B$). Detecting
"multi-particle" entanglement refers to entanglement between the
{\it pairs of particles}, $1$ and $2$, which belong to $A$ and
$B$, respectively.

\smallskip

\noindent Hence, relativity of the concept of locality calls for
caution: in order for the observable $\hat A$ be insensitive to
the $A$'s structure, it must be "collective observable" {\it
relative} to the {\it constituent particles} of the $A$ system.
Whether the $\hat A$ observable pertains to another structure of
the composite system $\mathcal{C}$ is irrelevant\footnote{\small
Plenty of the observables, e.g. the Hamiltonian, are "absolutely
collective (non-local)", in the sense that they cannot be local
for any partition. However, "insensitivity to structure" is subtle
and poses the following question: are there "intensive" quantum
observables, which are both "absolutely non-local" {\it and}
structure insensitive?}.

\bigskip

\noindent {\bf 5.2 Manipulating the center of mass}

\bigskip

\noindent The most of the realistic manipulations of the atomic
species refer to the atomic $CM+R$ structure, see e.g. the rhs of
eq.(58). In this section we are interested in actions that can be
clearly expressed in terms of the atomic/molecule $CM$
system--which includes the actions not affecting the atomic $R$
system. Of course, such actions are precluded in the classical
physics realm [see Footnote 27].

\smallskip

\noindent "{\it Two types of degrees of freedom have to be
considered for an atom: (i) the internal degrees of freedom, such
as the electronic configuration or the spin polarization, in the
center-of-mass reference frame; and (ii) the external degrees of
freedom, i.e. the position and momentum of the center of mass of
the atom}." (Cohen-Tanoudji and Dalibard 2006).

\smallskip

\noindent Electromagnetic forces and trapping (of charged or
neutral particles), atomic laser and lithography, atomic
interferometry,  refer to the atomic $CM$ system. For certain
purposes, one can forget\footnote{\small Of course, ignorance
about some degrees of freedom is not equivalent with the locality
of measurement, Section 5.1.} about the internal degrees of
freedom and consider an atom as a point-like particle with the
total (center-of-mass) atomic mass $M$\footnote{\small See the ChM
molecule model in Section 4.1.}. Temperature of an atomic gas (in
thermal equilibrium) is defined by statistical distribution of the
atomic-$CM$ momentums (velocities)--the internal atomic structure
is not of interest. Nevertheless, this picture is strict only for
a gas of atoms on sufficiently high (e.g. room) temperature. For
lower temperatures, the atomic $CM$ system can be described e.g.
by a wave packet (rather than by a point-like particle)--the
quantum effects become relevant. This is still a particle-like
description of the atoms in a gas. Temperature of the gas defines
the average the de Broglie wavelength of the atomic $CM$
systems--thus providing a quantitative criterion for the
particle-like versus the wave-like behavior of the atomic $CM$
systems. At  sufficiently low temperature one can no longer
distinguish the individual-atoms' center-of-mass systems from each
other--e.g., in the Bose condensate, the $CM$ systems of all atoms
have the same wave-function.

\smallskip

\noindent Conceptually the same physical basis apply to the
cooling of molecules: a molecule gas is cooled if the molecules
center-of-mass systems are sufficiently slow in the laboratory
reference frame (M. Zeppenfeld et al 2012)--there are no specific
conditions that are imposed for the atomic internal degrees of
freedom.

\smallskip

\noindent Experimental observation of diffraction and decoherence
of {\it large molecules} is striking an effect (Hackermuller et al
2003, Hackermuller et al 2004).
 An obvious motivation for doing experiments with matter waves is the everyday
 experience that physical bodies do not at all spread out like waves; rather they have
 a well-defined position whenever they are observed. Of course,
 the  center-of-mass position is of interest.

 \smallskip

\noindent {\it Micro- and nano-mechanical resonators} are
macroscopic systems--they can be seen with the naked eye.
Nevertheless, center of mass of these systems can be modelled as a
harmonic oscillator, which can undergo the quantum Brownian motion
dynamics (Gr\" oblacher et al 2013)\footnote{\small See Section
6.3.2 for some technical details.}. Indirect observation of the
$CM$ dynamics can reveal non-Markovian characteristics of the
environment, which monitors the $CM$ system. Although the  study
is performed at room temperature, it can be directly applied to
other mechanical resonators that operate close to the "quantum
regime".

\bigskip

\noindent {\bf 5.3 Manipulating the relative positions}

\bigskip

\noindent In this section we  consider  the physical situations
that can be clearly described in terms of the atomic/molecule $R$
system--which includes the situations in which the corresponding
$CM$ system is not affected. Direct manipulating the internal
degrees of freedom cannot be even defined in classical mechanics
[see Footnote 27].

\smallskip

\noindent Typically,  internal degrees of freedom are indirectly
observed--e.g. by detecting the emitted radiation. This detection
is all about the atomic/molecule {\it spectroscopy}.  Atomic
(molecule) excitation and de-excitation can be considered without
taking the $CM$ system into account. This, however, is not the
only possibility. Accessibility (direct measurement, Def.5.1) of
the "relative" degrees of freedom has recently been theoretically
(Rau et al 2003, Dunningham et al 2004) and also experimentally
(Maeda et al 2005) considered.

\smallskip

\noindent In  (Rau et al 2003), the authors come to the following
conclusion:

\smallskip

\noindent "{\it Thus, we have a consistent definition of relative
position that implies that relationships between objects, rather
than coordinates and absolute variables, are fundamental in the
quantum world.}"

\smallskip

\noindent Furthermore, they extend their observation for every
pair of mutually conjugate observables: "{\it This suggests that
some form of entanglement-driven localization might occur for any
pair of relative conjugate observables.}".  In (Dunningham et al
2004), physical reality of the relative position is claimed:

\smallskip

\noindent"{\it We have discussed how light scattering from
delocalized quantum particles can lead to the emergence of
'classical' relative positions. This process occurs even though
the absolute positions of the particles remain undefined and
suggests that the natural spatial framework for such a system is
relative position.}"

\smallskip

\noindent Furthermore, the authors (Dunningham et al 2004) claim
non-disturbance of the particles $CM$ system. That is, they
consider a purely local action exerted on the $R$ system.

\smallskip

\noindent In an experiment targeting the "atomic electron orbit",
(Maeda et al 2005), the authors say:

\smallskip

\noindent "{\it Nonetheless, our intuitive picture of an atom is
an electron moving in a Kepler orbit about an ionic core.... Using
picosecond or femtosecond laser pulses, it is now straightforward
to create wave packets of atoms of high principal quantum number
n, in which the electrons move in Kepler orbits ... However,
adding a small oscillating field at the orbital frequency can
phase-lock the motion of the electron to the oscillatory field
(5-11), such that the localization of the electron persists at
least for thousands of orbits (11)--perhaps long enough to
actually use these classical atoms in applications such as
information processing (12).}"

\smallskip

\noindent While the phrases that describe the experimental
findings are in terms of "electron orbits", the theoretical basis
is clearly presented in terms of the atomic $R$ degrees of
freedom. Bearing in mind that, in the experiment, the atomic $CM$
degrees of freedom are assumed to remain intact, with the aid of
Section 5.1, we realize that the experiment is another instance of
 direct manipulation with the atomic $R$ system.

\smallskip

\noindent Regarding the large-molecules species, the conformation
$K_N$, Section 4.2.2, represents the internal degrees of freedom.
These degrees of freedom are at the core of stereochemistry as
well as of the foundations of the biopolymer dynamics, e.g., in
the protein (un)folding and molecular recognition. Manipulation of
the large-molecules conformation with light is by now a routine
(Lendlein et al 2005).

\bigskip

\noindent {\bf 5.4 Quantum correlations relativity in use}

\bigskip

\noindent Entanglement relativity (or the more general quantum
correlations relativity) is in direct use via "entanglement
swapping"  (Bennett et al 1993, Ma et al 2012), (and the
references therein), and  via "coarse graining" (Ragy and Adesso
2012) structural transformations. Entanglement swapping is
formally a trivial kind of LCTs--regrouping of subsystems, Section
2.2.1, point (A). It is global, in the sense of Section 2.2.1,
point (B). On the other hand, grouping ("coarse graining") or
decomposing ("fine graining") the subsystems are also trivial but
local kinds of LCTs.

\smallskip

\noindent In (Ma et al 2012), the authors consider an
entanglement-based variant (Peres 2000) of the gedanken "delayed
choice" experiment (Wheeler 1978). At first sight, it may seem
that this is a delayed choice in the original Wheeler's spirit.
However, this is not the case. The theoretical proposal (Peres
2000) as well as the experimental realization (Ma et al 2012)
target entanglement [via entanglement swapping], rather than the
individual qubits, in a system of four qubits. In the experiment
it is clearly demonstrated: there is not individuality of the
single qubits or of the pairs of qubits; see also (Dugi\' c 2012).
Rather, the effects due to entanglement of the different
bipartitions of the system of four qubits are experimentally
observed. In other words: the object of investigation is
entanglement of different pairs of qubits, not the individual
qubits. Depending on the choice of the pair of qubits to be
measured, the remaining pair of qubits appears in entangled or in
a separable pure state. Such a measurement in a later instant
apparently changes the initially obtained record on entanglement
of the pair of qubits. This intuitively paradoxical situation is
described in the theoretical proposal (Peres 2000):

\smallskip

\noindent "{\it The point is that it is meaningless to assert that
two particles are entangled without specifying in which state they
are entangled, just as it is meaningless to assert that a quantum
system is in a pure state without specifying that state [9]. If
this simple rule is forgotten, or if we attempt to attribute an
objective meaning to the quantum state of a single system, curious
paradoxes appear: quantum effects mimic not only instantaneous
action-at-a-distance but also, as seen here, influence of future
actions on past events, even after these events have been
irrevocably recorded.}"

\smallskip

\noindent However, once Entanglement Relativity is properly
understood, the above quote can be re-phrased as follows: The
point is that it is meaningless to assert that two particles are
entangled without specifying the structure of interest. Even for a
specified (pure) state of the system of qubits, entanglement may
or may not be observed depending on the structure distinguished by
the chosen (local to that structure) measurement to be performed
in a later instant in time. Once again, we can say: entanglement
is a structure-dependent, i.e., a relative notion.
\smallskip

\noindent "Coarse graining" of a composite system's structure is
the "particles grouping" kind of LCTs. Recently, a "coarse
grained" picture of the "ghost imaging" technique has been
analyzed (Ragy and Adesso 2012). The authors analyzed the nature
of correlations in Gaussian light sources used for ghost imaging
from a quantum informational perspective, combining a microscopic
with an effective coarse-grained description. A transition from
the microscopic modes $a_i$ to the coarse grained two-mode boson
operators, $c_1, c_2$, provides a striking observation. The
findings are described as follows:

\smallskip

\noindent "{\it This reveals an interesting feature associated to
the coarse-grained formalism put forward here. It actually
indicates how the quantum nature of the light source becomes
quenched as we diverge from the photon-counting regime and enter
the classical limit of intensity correlations. For these high
illuminations, the quantum correlations available for detection by
our scheme tend to zero, and the physical model of the scheme does
not require a quantum description of the light to be accurate.}"

\smallskip

\noindent So, the authors observe a transition from  quantum to
classical regime in  the ghost imaging technique as a consequence
of averaging of  the coarse-grained structure of the light source.
The authors properly interpret their finding--not yet emphasizing
the quantum correlations relativity--, while not discussing a need
to perform averaging of the field modes. So, this is not a
solution to the problem of the transition from quantum to
classical. Nevertheless, this is a very important contribution to
this long-standing issue (Giulini et al 1996, Zurek 2003,
Schlosshauer 2004): the observed transition is {\it not known for
the original} (non-coarse-grained) degrees of freedom.

\smallskip

\noindent Recently, it has been understood that even if an open
system is equilibrated, i.e. is in thermal equilibrium in regard
of its thermal bath, the system need not act as a thermal bath
towards all reference systems (e.g. observers) (del Rio et al
2014). The very concept of thermal equilibrium is {\it relative}:
a system $S$ in thermal state due to its environment $E$ need not
be in thermal state {\it relative} to some other $S'$ system. In
order to have two systems, $S$ and $S'$, to act as thermal baths
relative to each other, absence of quantum correlations is
required. Now, bearing QCR in mind, it becomes clear that, in a
sense, "relative thermalization" is relative. That is, there is no
sense in saying that a pair of systems can act as thermal baths
towards each other unless the structure is defined  for the
systems. In other words: reative thermalization is a direct
corollary of quantum correlations relativity.

\smallskip

\noindent All kinds of the LCTs considered in this section are
classical in spirit. In classical physics, the subsystems
("particles") are assumed to have individuality that is {\it not}
jeopardized by the formally trivial LCTs of
decomposing/(re)grouping the particles. Some groups of particles
may be additionally charged by certain (local) boundary conditions
to appear in bound states, like in the Chemical Model of
molecules, Section 4.1. However,  quantum correlations relativity,
Section 3.2, substantially changes the picture as it is emphasized
throughout this book. Bearing in mind the classical spirit of
decomposing/(re)grouping systems, it is not surprising that this
kind of  LCTs is the main kind of structure transformations that
are considered in the literature so far. Physical relevance of the
more general ones (such as those to be presented in Chapters 6 and
7) is yet to be appreciated.

\bigskip

\noindent {\bf 5.5 Outlook}

\bigskip

\noindent The classically artificial, "collective", $CM+R$
structure proves itself operationally realistic in the quantum
realm\footnote{\small Those subsystems are of general use, for the
particles in bound states, as well as for the free particles.}.
Laboratory manipulation makes these degrees of freedom {\it at
least as realistic as the "fundamental" degrees of freedom of a
composite quantum system}. If it were not so, we would have
already been able more-or-less directly to observe the fundamental
constituents of the matter.

\smallskip

\noindent Loss of individuality of quantum subsystems as well as
quantum correlations relativity and relativity of locality,
Section 2.4, provide a consistent view of this phenomenological
fact. There is nothing "artificial", "collective" or "emergent" in
the quantum center-of-mass and internal-degrees of freedom.
Rather, physical situation defines a specific set of local degrees
of freedom (and observables) that are operationally accessible. A
set of such observables defines the local subsystems and  related
(composite system's) {\it operationally preferred}
structure\footnote{\small As stated in Chapter 2, we are
exclusively interested in the composite systems allowing  the
tensor-product-structure variations.} (Zanardi 2001, Zanardi et al
2004, Harshman and Ranade 2011). Foundational issues on the
operationally preferred structure of an open quantum system are
subject of Chapter  7. Experimental confirmation of entanglement
relativity gives rise to: (i) it changes our intuition on
"structure", and (ii) it opens practical applications of quantum
phenomena that are traditionally considered to be impossible in
the classical physics realm.

\newpage

\noindent {\bf Chapter 6}

\noindent\textbf{\large Parallel Occurrence of Decoherence}

\vspace*{55mm}

\noindent Quantum mechanics offers a stunning observation: a
quantum whole carries less uncertainty than its parts. In favor of
this observation is relativity of quantum locality and system as
well as of quantum correlations. This is in sharp contrast with
the classical intuition, which knows of "systems", their
individuality and distinguishability, separability which became
not only the goal but also a {\it means} for solving the
quantum-to-classical-transition problem.

\smallskip

\noindent Quantum decoherence is currently the main candidate for
establishing the quantum-to-classical transition (Giulini et al
1996, Zurek 2003, Schlosshauer 2004). The general task of the
decoherence program (Schlosshauer 2004) starts as follows: "There
is a system $S$ that is in (unavoidable) interaction with its
environment $E$. The composite system, $S+E$, is  subject to the
Schr\" odinger law."

\smallskip

\noindent This assumption on the pre-defined structure is the very
basis of the standard, actually a {\it bottom-up}, approach to
decoherence which is fairly presented by Zurek's (Zurek 2003):

\smallskip

\noindent "{\it In the absence of systems, the problem of
interpretation seems to disappear. There is simply no need for
'collapse' in a universe with no systems. Our experience of the
classical reality does not apply to the universe as a whole, seen
from the outside, but to the systems within it.}"

\smallskip

\noindent The idea on predefined structure is classical in its
spirit. Furthermore, in the decoherence context, it leads to a
circular reasoning: stipulate a structure, and then use
decoherence to justify the stipulation. However, bearing in mind
relativity of "system" and "locality", Section 5.1, the following
question appears:

\smallskip

\noindent ({\bf Q}) What might be the physical consequences of the
linear canonical transformations on the occurrence of decoherence?

\smallskip

\noindent Importance of this [as yet poorly posed] question can be
seen from the following quote (Zurek 1998):

\smallskip

\noindent "{\it In particular, one issue which has been often
taken for granted is looming big, as a foundation of the whole
decoherence program. It is the question of what are the
�systems� which play such a crucial role in all the
discussions of the emergent classicality. (. . . ) [A] compelling
explanation of what are the systems--how to define them given,
say, the overall Hamiltonian in some suitably large Hilbert
space--would be undoubtedly most useful.}"

\smallskip

\noindent Intuitively, the LCT-induced change of structure may
reveal non-trivial observations regarding the question of "what is
'system'?" (Dugi\' c and Jekni\' c 2006, Dugi\' c and Jekni\'
c-Dugi\' c 2008). In a sense, the  question ({\bf Q}) promotes a
new {\it top-down} approach to describing composite quantum
systems and decoherence.

\smallskip

\noindent However, the question ({\bf Q}) is too general and
imprecise. Not surprisingly, answer may depend on the number of
the underlying assumptions and/or variations. To this end, we
distinguish the following {contexts} of the question:

\smallskip

\noindent (1) LCTs refer to the closed, or to the open system;

\smallskip

\noindent (2) LCTs refer to a few- or to a many-particle system;

\smallskip

\noindent (3) Specific choice of the kind and/or of quantum state
of the environment;

\smallskip

\noindent (4) Interpretation of quantum theory.

\smallskip

\noindent "Interpretation" is a subject of Chapter 8. In this
Chapter we consider specific models that refer to the total
(closed) system $\mathcal{C}$ by mainly following (Dugi\' c and
Jekni\' c-Dugi\' c 2012). In Chapter 7, we will consider specific
models of  open bipartite systems.

\bigskip

\noindent {\bf 6.1 The task}

\bigskip

\noindent We are interested in global LCTs. More specifically: we
are interested in a pair of mutually global and  irreducible
structures, Section 2.2.2.

\smallskip

\noindent This kind of structure variation means that  LCTs
intertwined degrees of freedom of the $S$ and $E$ system. Instead
of the pair $S+E$, there appears a new structure $S'+E'$. We are
interested in the continuous variable (CV) systems; a similar
analysis regarding a finite-dimensional system can be found e.g.
in (Felï'dman and Zenchuk, 2012, 2014a,b), which is not devoted to
the occurrence of decoherence for the different structures.

\smallskip

\noindent Since  LCTs preserve the number of the degrees of
freedom, equal dimensionality of $S$ and $S'$ implies equal
dimensionality (complexity) of the respective environments, $E$
and $E'$.

\smallskip

\noindent For the considered structures, there is a number of
features of interest. To this end, we re-phrase the contents of
Section 2.2.2: (i) the structures irreducibility implies that
every structure is endowed by its own "elementary particles" and
their interactions; (ii) the subsystems (e.g. $S$ and $S'$)
belonging to different structures are information theoretically
separated; (iii) There is neither correlation nor any information
flow between the subsystems of the two structures. In this sense,
the two structures appear autonomous relative to each other.
Common for the two structures is the composite system's Hilbert
space, the Hamiltonian and the unique quantum state in every
instant in time.

\smallskip

\noindent Our task now reads as: for the proper LCTs, to
investigate the occurrence of decoherence in the alternate
structure $S'+E'$.

\bigskip

\noindent {\bf 6.2 The obstacles}

\bigskip

\noindent The above posed task faces some obstacles.

\noindent The first obstacle comes from Section 3: every change in
the degrees of freedom typically gives rise to a change in
correlation between  subsystems--presence of correlations
complicates analysis.

\smallskip

\noindent Derivation of the master equations for an open system
$S$ typically assumes (Breuer and Petruccione 2002, Rivas and
Huelga 2011) both  the initial tensor product state, $\rho_S
\otimes \rho_E$, as well as that the environment is a thermal
bath, i.e. $\rho_E = \rho_{th} = \exp(- \beta H_E)/ Tr_B (\exp(-
\beta H_E))$; $H_E$ is the environment's Hamiltonian and $\beta =
(k_BT)^{-1}$ is the standard "inverse temperature".

\smallskip

\noindent However, there is a direct consequence of the quantum
correlations relativity, Section 3.2: the tensor-product state
$\rho_S \otimes \rho_E$ bears correlations for the new structure,
i.e. $\rho_S \otimes \rho_E \neq \rho_{S'} \otimes \rho_{E'}$. As
a consequence, one directly observes: the new environment $E'$
need not be in thermal state--worse, it's state need not be even
stationary--see Lemma 6.1 below. Worse, non-factorized initial
state for the "system+environment" (here: $S'+E'$) challenges both
Markovianity as well as complete positivity of the open system's
dynamics (Breuer and Petruccione 2002, Rivas and Huelga 2011,
Rodriguez-Rosario and Sudarshan 2011, Brodutch et al
2012)\footnote{\small Interestingly enough, the basic method in
the field, the so-called Nakajima-Zwanzig projection method, is
inapplicable for the structural considerations, see Section
6.3.4.}.

\smallskip

\noindent On the other hand,  LCTs typically introduce the new
interaction terms. So, one can expect interactions of the
constituent particles of the new environment $E'$. This, in
general, poses significant technical difficulties in deriving
master equation for the open system (Breuer and Petruccione 2002,
Rivas and Huelga 2011, Breuer et al 2009, Laine et al 2010, Rivas
et al 2010a, Rodriguez-Rosario and Sudarshan 2011, Haikka et al
2011, Brodutch et al 2012).

\smallskip

\noindent Thus having in mind the  foundations of the Markovian
open systems theory (Rivas and Huelga 2011), the transition $\{S,
E\} \to \{S', E'\}$ can, in general, pose insurmountable obstacles
to solving the task.

\smallskip

\noindent However, there is a class of the open-systems models
that are immune to these obstacles--the so-called linear models.
This is the subject of the next section. \\

\bigskip

\noindent {\bf 6.3 Quantum Brownian motion}

\bigskip

\noindent"Brownian motion" is a realistic physical effect for the
center of mass of "Brownian particle" (BP). Internal structure of
the particle does not contribute to the Brownian motion effect.
For this reason it is legitimate to forget about the internal BP
degrees of freedom, and, for simplicity (without any loss of
generality), to investigate the one-dimensional system, $S$, which
can be modelled as a free particle or as a harmonic oscillator.

\smallskip

\noindent The particle's environment is usually modelled as a set
of non-interacting linear harmonic oscillators in thermal
equilibrium.

\bigskip

\noindent {\bf {\emph {6.3.1 The LCTs and the structures of
interest}}}

\bigskip

\noindent Let us consider a set of three-dimensional particles,
which are defined by their respective position and momentum
observables, ${\vec r}_i$, ${\vec p}_j$, where $i,j=1,2,..N$
enumerates the particles, and $[x_{i\alpha}, p_{j\beta}] =
\delta_{ij} \delta_{\alpha \beta}$, $\alpha, \beta = 1,2,3$.

\smallskip

\noindent We introduce the total system's center of mass and the
relative positions, denoted $CM$ and $R$,
respectively\footnote{\small A generalization of eqs. (4), (5).}:
\begin{equation}
\vec R_{CM} = \sum_i m_i \vec r_i/M, \quad \vec \rho_{Rl} = \vec
r_i - \vec r_j, \quad l(\equiv \{i,j\}) =1,2,...N-1.
\end{equation}

\noindent The inverse to eq.(58) reads as: %%
\begin{equation}
\vec r_i = \vec R_{CM} + \sum_{l=1}^{N-1} \omega_{li} \vec
\rho_{Rl},
\end{equation}

\noindent with the real coefficients $\omega$.

\smallskip

\noindent Regarding the system's Hamiltonian, there appears the
so-called "mass polarization" term [see Supplement]: %%
\begin{equation}
M_{Rll'} = {m_{l+1}m_{l'+1} \dot{\vec \rho}_{Rl} \cdot \dot{\vec
\rho}_{Rl'}\over M} = {m_{l+1}m_{l'+1} \vec p_{Rl} \cdot \vec
p_{Rl'}\over \mu_l \mu_{l'} M},
\end{equation}

\noindent where appear the time derivatives of the relative
positions and their scalar product; $[\vec \rho_{Rl}, \vec
p_{Rl'}] = \imath \hbar \delta_{ll'}$. The set of the "reduced
masses": %%
\begin{equation}
\mu_l = {m_{l+1}(M-m_{l+1}) \over M}.
\end{equation}

\noindent The kinetic term for  every constituent "particle"
preserves the standard form, e.g.: %%
\begin{equation}
T_{CM} = {\vec P_{CM}^2 \over 2M}, \quad T_{Rl} = {\vec p_{Rl}^2
\over 2 \mu_l}.
\end{equation}

\noindent As elaborated in Supplement,  external fields for the
original particles become interactions for the $CM$ and $R$
systems, while the distance-dependent interactions of the original
particles become external fields for the $R$ system. Hence the
form of the composite system's Hamiltonian completely
changes--except the kinetic terms. Nevertheless,  the composite
system's Hamiltonian, $H$, preserves its general form (compare to
eq.(5))
\begin{equation}
H_S + H_E + H_{SE} = H = H_{S'} + H_{E'} + H_{S'E'},
\end{equation}

\noindent where we assume that the open system $S$ is one of the
"original particles" while the rest constitutes the environment
$E$, and we identify the systems, $CM \equiv S'$ and $R \equiv
E'$. To this end [as emphasized above]--since the original open
system $S$ and the new one $S'$ are of the same number of the
degrees of freedom--the respective environments, $E$ and $E'$, are
 of the same number of the degrees of freedom.

 \smallskip

\noindent In the terms of Chapter 2, the considered
transformations
 are structurally described as follows:
\begin{equation}
\mathcal{S} = \{\vec r_S, \vec r_{Ei}\} \to \mathcal{S}' = \{\vec
r_{S'}, \vec \rho_{E'i}\}.
\end{equation}

\noindent So our task (Section 6.1) reads as: to investigate the
occurrence of decoherence for some bipartitions of certain models
described by the general expressions eqs.(58)-(64).

\bigskip

\noindent {\bf {\emph {6.3.2 The Caldeira-Leggett model}}}

\bigskip

\noindent We are interested in the Caldeira-Leggett model
(Caldeira and Leggett 1983) defined by the following Hamiltonian
for the "original" structure $S+E$: %%
\begin{equation}
H = {p_S^2 \over 2m_S} + V(x_S) + \sum_i \left( {p_{Ei}^2 \over
2m_{Ei}} + {m_{Ei} \omega_{Ei}^2 x_{Ei}^2\over 2}  \right) \pm x_S
\sum_i \kappa_i x_{Ei} \equiv H_S + H_E + H_{SE}
\end{equation}

\noindent Physically, this is a model of a one-dimensional system
$S$ immersed in a thermal bath of mutually non-interacting
harmonic oscillators (that are collectively denoted as the
environment $E$). The interaction %%
\begin{equation}
H_{SE} = \pm x_S \sum_i \kappa_i x_{Ei}
\end{equation}

\noindent is bilinear and with the strength determined by the
coefficients $\kappa$; both signs, $\pm$, appear in the literature
without making any substantial change regarding the open-system's
(the $S$'s) dynamics.

\smallskip

\noindent The open system's dynamics can be described in terms of
the open-system's state dynamics (the Schr\" odinger picture),
$\rho_S(t)$, or in terms of the open system's-variables dynamics
(the Heisenberg picture), $x_S(t)$, $p_S(t)$.

\smallskip

\noindent Typically\footnote{\small For the sake of Markovianity
of the particle's dynamics.}, the initial state for the composite
system is assumed to be tensor product, $\rho_S(t=0) \rho_E(t=0)$.
Furthermore, the following {\it ansatz} is typically used (Breuer
and Petruccione 2002): the total system's state in every instant
in time reads %%
\begin{equation}
\rho_S(t) \rho_E, \quad \rho_E = \exp(-\beta H_E)/Tr_B
(\exp(-\beta H_E)), \forall{t}.
\end{equation}

\noindent It is also assumed: the environment oscillators are
mutually uncoupled (non-interacting), while  interaction of  $S$
and $E$ is "weak". These simplifications come from the general
open-systems theory: without these simplifications, the search for
the general form of the Markovian master equations becomes
practically intractable (Rivas and Huelga 2011).

\smallskip

\noindent For the phenomenologically inspired choice of the
environment "spectral density", one obtains the  following (high
temperature) master equation [in the Schr\" odinger picture] for
the quantum Brownian motion (QBM): %%
\begin{equation}
{d\rho_S(t) \over dt} = - {\imath \over \hbar} [H_S, \rho_S(t)] -
{\imath \gamma \over \hbar} [x_S, \{p_S, \rho_S(t)\}] - {2m_S
\gamma k_B T \over \hbar^2} [x_S, [x_S, \rho_S(t)]];
\end{equation}

\noindent the curly brackets denoting the anticommutator, $\{x_S,
\rho_S(t)\} = x_S \rho_S(t) + \rho_S(t) x_S$, and $\gamma$
representing the phenomenological "friction"
parameter\footnote{\small As a consequence of the choice of the
spectral density.}.

\smallskip

\noindent The last term in eq.(68) is the decoherence term. The
approximate "pointer basis" (i.e. the "preferred") states are
Gaussian states\footnote{\small Not of the minimal uncertainty.}.

\smallskip

\noindent Physically, Brownian particle undergoes the decoherence
process and dissipation that become obvious in the
Heisenberg-picture for the particle's position and momentum
observables.

\smallskip

\noindent Let us now consider the transformations of variables
distinguished in Section 6.3.1.

\smallskip

\noindent Placing the expressions eq.(58), (59) into the
Hamiltonian eq.(65) for one-dimensional system, one obtains for
the alternate structure $S'+E'$ [while bearing in mind $S' \equiv
CM$ and $E' \equiv R$]: %%
\begin{equation}
H = H_{S'} + H_{E'} + H_{S'E'},
\end{equation}

\noindent with the following terms:
\begin{eqnarray}
&\nonumber& H_{S'} = {P_{S'}^2\over 2M} + {M \Omega_{S'}^2
x_{S'}^2\over 2} \\&& H_{E'} = \sum_i \left( {p_{E'i}^2\over
2\mu_i} + {\mu_i \nu_{E'i}^2 \rho_{E'i}^2\over 2} + V_{E'} \right)
\nonumber\\&& H_{S'E'} = \pm x_{S'} \sum_i \sigma_i \rho_{E'i}.
\end{eqnarray}

\noindent Formally, eq.(70) is similar to eq.(65). Physically,
there is another one-dimensional system $S'$ in interaction with a
set of the linear harmonic oscillators. The only formal
distinction lies in the appearance of the interaction term
$V_{E'}$ for the constituents of new environment $E'$. Interaction
in the new structure, $S'+E'$, is of the same, bilinear form of
eq.(66).

\smallskip

\noindent Delving into details, we obtain (Dugi\' c and Jekni\'
c-Dugi\' c 2012) precise definitions for the terms in eq.(70) for
the two cases: the $S$ system as the free particle, and the $S$
system as a harmonic oscillator\footnote{\small  By $\omega_{iS}
\equiv \omega_{Si}$ we assume the real parameters appearing in
eq.(59).}.

\smallskip

\noindent {\it Free particle} ($V_{S} = 0$): $M\Omega_{S'}^2 /2 =
\sum_i (\pm \kappa_i + m_{Ei} \omega_{Ei}^2/2)$, $\mu_i
\nu_{Ei}^2/2 = \pm \omega_{Si}$ $\sum_j \kappa_j \omega_{ij} +
\sum_j m_{Ej} \omega_{Ej}^2 \omega_{ij}^2/2$, and $\sigma_i =
\sum_j (\kappa_j \omega_{ij} + \kappa_j \omega_{Si} + m_{Ej}
\omega_{Ej}^2 \omega_{ij})$. The internal interaction term $V_{E'}
= \sum_{i\neq j} \left[C_{ij} p_{E'i} p_{E'j}/\mu_i \mu_j +
(\Omega_{ij} + \omega_{Si} \Omega_j) \rho_{E'i} \rho_{E'j}
\right]$; the "mass polarization terms" $C_{ij} = m_{E(i+1)}
m_{E(j+1)}/M$ and $\Omega_i = \sum_j \kappa_j \omega_{ij}$, while
$\Omega_{ij} = \sum_k m_{Ek} \omega_{Ek}^2 \omega_{ik}
\omega_{jk}/2$.

\smallskip

\noindent {\it Harmonic oscillator} ($V_{S} = m_S \omega_{S}^2
x_S^2 /2$): As distinct from the free particle case, the $S$
system as a harmonic oscillator provides the harmonic term, which
should be simply added to the free-particle Hamiltonian. So the
Hamiltonian for the harmonic oscillator follows  from adding the
following term to the free-particle  Hamiltonian: $m_S \omega_S^2
x_{S'}^2/2 + \sum_i m_{S} \omega_{S}^2 \omega_{iS}^2
\rho_{E'i}^2/2 + \sum_{i \neq j} m_S \omega_S^2 \omega_{iS}
\omega_{jS} \rho_{E'i} \rho_{E'j}/2 + x_{S'} \sum_i m_S \omega_S^2
\omega_{iS} \rho_{E'i}$.

\smallskip

\noindent The new open system, $S'$, is a harmonic oscillator even
if the original $S$ system is a free particle. LCTs inevitably
introduce the internal interaction $V_{E'}$, which couples the
$E'$-system oscillators; there are the linear momentum-momentum
and the position-position coupling for the constituents of the new
environment $E'$; see Supplement for details.

\bigskip

\noindent {\bf {\emph {6.3.3 $S'$ is a Brownian particle}}}

\bigskip

\noindent The two forms of the composite system's Hamiltonian,
eq.(65) and eq.(70), are almost isomorphic. However, this does not
{\it per se} imply that the dynamics of the open systems, $S$ and
$S'$, are mutually equal.

\smallskip

\noindent As emphasized in Section 6.2, in general, there is  a
number of obstacles for providing master equation for the new open
system $S'$ that is worth repeating.

\smallskip

\noindent First, if the initial state for the $S+E$ structure is
tensor product, $\rho_S(t=0) \rho_E$,  then there are initial
correlations regarding the alternate structure $S'+E'$ (in
general, see Section 3.2, there are quantum correlations with
non-zero one-way discord\footnote{\small Remind: the two-way
discord tends to be larger than one-way discord, Section 3.2.}).
As a consequence, the $S'$ system's dynamics may be non-Markovian
(and also non-completely positive). Second, if the original
environment $E$ is initially in thermal equilibrium, this is not
the case for the new one, $E'$ [see Lemma 6.1 below]. Third, the
"new" oscillators (subsystems of the new environment $E'$) are
mutually coupled. For interacting oscillators, the master equation
eq.(68) is not necessarily valid--there may be both memory effects
for the environment as well as a  change in the spectral density.

\smallskip

\noindent Nevertheless, as we show below, the open system $S'$
also undergoes Brownian dynamics.

\smallskip

\noindent Let us first emphasize irrelevance of the $V_{E'}$ term
in eq.(70).

\smallskip

\noindent For the new environment $E'$, we introduce the "normal
coordinates", $Q_{E'i}$, and the conjugate momentums, $P_{E'i}$,
as the new canonical variables: %%
\begin{equation}
Q_{E'i} = \sum_m \alpha_{mi} x_{E'm}, \quad P_{E'i} = \sum_n
\beta_{ni} p_{E'n}, \quad [Q_{E'i}, P_{E'j}] = \imath \hbar
\delta_{ij}.
\end{equation}

\noindent The choice of the new variables is constrained by the
requirement of non-coupling of the new variables as well as by the
commutator in eq.(71). For the potential $V_{E'}$, which is of the
bilinear form regarding the position and momentum observables,
this is always possible to do. So, the $E'$ system can be
considered as a set of mutually noninteracting linear harmonic
oscillators, which are described by the normal coordinates,
$Q_{E'i}$ in eq.(71), as the oscillators position-observables.
Then instead of eq.(70) one obtains: %%
\begin{eqnarray}
&\nonumber& H_{S'} = {P_{S'}^2\over 2M} + {M \Omega_{S'}^2
x_{S'}^2\over 2} \\&& H_{E'} = \sum_i \left( {P_{E'i}^2\over 2} +
{ \lambda_i^2 Q_{E'i}^2\over 2}  \right) \nonumber\\&& H_{S'E'} =
\pm x_{S'} \sum_i \sigma'_i Q_{E'i}.
\end{eqnarray}

\noindent where the new coupling constants $\sigma'_i = \sum_j
\alpha'_{ij} \sigma_j$; here we use the inverse to eq.(71),
$x_{E'i} = \sum_j \alpha'_{ij} Q_{E'j}$.

\smallskip

\noindent Thus we have performed the following
global\footnote{\small If we bear the $S$ system in  mind, the
transformation is, of course, local.} (non-trivial,
irreducible\footnote{\small See Section 2.2.}) change in the
new-environment's structure: %%
\begin{equation}
\mathcal{S}_{E'} = \{x_{E'i}\} \to \mathcal{S'}_{E'} =
\{Q_{E'i}\},
\end{equation}

\noindent while the environment Hilbert state space now obtains
new factorization, $\mathcal{H}_{E'} = \otimes_i
\mathcal{H}_{E'i}^{(Q)}$, and the environment a set of mutually
noninteracting oscillators, eq.(72); for a similar procedure see
Lim et al 2014. Thereby, in this step, the total system had
undergone the following local structure transformation: %%
\begin{equation}
\mathcal{S} = \{x_{S'}, x_{E'i}\} \to \mathcal{S'} = \{x_{S'},
Q_{E'i}\},
\end{equation}

\noindent which now makes the two forms of the Hamiltonian,
eq.(65) and eq.(72), {\it fully isomorphic}.

\smallskip

\noindent The open system's state, $\rho_{S'}$, is defined by the
tracing out operation, $\rho_{S'}(t) = tr_{E'}
\rho_{S'+E'}(t)$--the "$tr_{E'}$" operation is taken over the
whole Hilbert space of the $E'$ system. The basis-independence of
the tracing out operation can be represented e.g. %%
\begin{equation}
tr_{E'} A_{E'} = \int \langle \{x_{E'i}\} \vert A_{E'} \vert
\{x_{E'i}\}\rangle \Pi_i dx_{E'i} = \int \langle \{Q_{E'i}\} \vert
A_{E'} \vert \{Q_{E'i}\}\rangle \Pi_i dQ_{E'i}
\end{equation}

\noindent for the different structures of $E'$; $\vert
\{x_{E'i}\}\rangle = \otimes_i \vert x_i\rangle_{E'}$, $\vert
\{Q_{E'i}\}\rangle = \otimes_i \vert Q_i\rangle_{E'}$. So, eq.(75)
clearly states: the open system's state, $\rho_{S'}(t)$, is {\it
unique}, i.e. is {\it not}  environment-structure dependent.
Therefore dynamics of the open system $S'$ [e.g. derivation of the
related master equation] is not conditioned by the choice of the
$E'$-system's structure.

\smallskip

\noindent Now we return to the consequences of  QCR, Section 3.2.
As emphasized above, a change of the open system's structure will
in general lead to a change in form of the system's quantum state,
as well as in the amount of quantum correlations carried by the
state. However, there are the following special cases: (a) the
composite system is at zero temperature ($T=0$), and (b) the
composite system is at nonzero temperature $T$. These cases are
actually known and investigated, e.g. in (Paz 1996, Bellomo et al
2005, Anglin et al 1997).

\smallskip

\noindent Exact solutions to the Schr\" odinger equation for the
Hamiltonian eq.(65), i.e. eq.(72), are not yet known.
Nevertheless, for the case (a), it is known that the ground energy
state is non-degenerate (i.e. is unique) and entangled for the
bipartition, "system+environment" system. Regarding the case (b):
at non-zero temperature, the total system's state %%
\begin{equation}
\rho = {e^{-\beta H}\over Z}, \quad Z  = tr e^{-\beta H}
\end{equation}

\noindent where $H$ is the total system's Hamiltonian, for the
canonical ensemble;  $\beta$ is the "inverse temperature". Of
course, the state eq.(76) is non-factorized.

\smallskip

\noindent The following point should be strongly emphasized: all
the conclusions referring to the cases (a) and (b) equally concern
to both structures, $S+E$ and $S'+E'$. e.g.  The Hamiltonian $H$
appearing in eq.(76) is given by eq.(65) for the $S+E$ and by
eq.(72) for the $S'+E'$ structure. So for the cases (a) and (b),
the two model-structures are formally {\it fully isomorphic}. The
{\it physical} distinctions between the two
structures--correlations in the initial state for the new
structure, and non-stationary state of the new environment--are
known {\it not} to change the physical picture (Lutz 2003, Romero
and Paz 1996, Bellomo et al 2005, Anglin et al 1997). Hence  we
can  conclude  that (Dugi\' c and Jekni\' c-Dugi\' c 2012):

\smallskip

\centerline{\emph{The "new" open system, $S'$, is a Brownian
particle itself}.}

\smallskip

\noindent The case of the initially non-equal temperatures of the
open system $S$ and the environment $E$ is not  easy to
handle\footnote{\small See Section 6.3.4.}. In such case, as
emphasized above, due to QCR, the two models are not fully
isomorphic\footnote{\small It is worth noticing: this situation
refers physically to the situation, in which the open system is
suddenly brought in touch with the thermal bath.}. Along with the
existing literature, here we do not tempt to offer a general
description of this situation. Nevertheless, we offer a scenario
that is a {\it speculative}  physical picture, which bears some
generality compared to the non-realistic ansatz eq.(67) for the
$S+E$ structure.

\smallskip

\noindent Having in mind Section 6.2 and eq.(70), we collect the
constraints for deriving  master equation for the $S'$ system.
First, from eq.(70), we can see that interaction $H_{S'E'}$ is of
the strength of the order of $H_{SE}$, i.e. it's weak. From
Section 6.2 we learn that the initial state for the $S'+E'$
structure is in general a mixture $\rho_{S'+E'}(t=0) = \sum_i
\lambda_i \rho_{S'i} \otimes \rho_{E'i}$. Thus one cannot adopt
Markovian approximation for the $S'$ system (Rivas and Huelga
2011). Having in mind that the $S$ and $S'$ are one-particle
systems, the two environments $E$ and $E'$ are also of equal
dimensionality (of equal number of oscillators).

\smallskip

\noindent Then we start from eq.(3.113) of (Breuer and Petruccione
2002) [in the interaction picture]: %%
\begin{equation}
{d\rho_{S'}(t) \over dt} = - \int_0^t ds tr_{E'}
[H_{S'E'}^I(t),[H_{S'E'}^I(s), \rho_{S'+E'}(s)]].
\end{equation}

\noindent Without further ado, we resort to the following {\it
ansatz}: %%
\begin{equation}
\rho(t) = (1 - \epsilon_{\circ}(t)) \sum_i
\sigma_{S'i}^{t>t_{\circ}} \sigma_{E'\circ} + \sum_i
\epsilon_i(t)\sigma_{S'i}^{t>t_{\circ}}
\delta_{E'i}^{t>t_{\circ}}, \quad tr_{E'}
\delta_{E'i}^{t>t_{\circ}} = 1, \forall{i,t},
\end{equation}

\noindent where $\epsilon(t) = \max \{\vert\epsilon_i(t)\vert\}
\ll 1, \forall{t}>t_{\circ}$.

\smallskip

\noindent Physically, eq.(78) means: the environment $E'$
undergoes thermal relaxation much faster than the open system
$S'$. From eq.(78): $\rho_{E'} \approx \sigma_{E'\circ}$ after
some instant $t > t_{\circ}$ [the initial instant is assumed $ t =
0$]. The open system's state  $\rho_{S'}(t) \approx \sum_i
\sigma_{S'i}(t), t>t_{\circ}$.

\smallskip

\noindent Further, instead of the total environment $E'$, we
consider a small part of the $E'$ system that should appear in
eq.(78). Namely, we divide $E' = E'_1 + E'_2$, where the part
$E'_2$ monitors $E'_1$ and does not interact with the $S'$
system\footnote{\small One can think of this tripartition in
analogy with the DISD method of (Dugi\' c 2000).}. Then eq.(78)
reduces to the standard {\it ansatz} eq.(67): if the limit
$\epsilon \to 0$ is available for the time intervals of
$\gamma^{-1}$, then eq.(78) reduces to $\rho_{S'}(t)
\sigma_{E'_1\circ}$; $\gamma$
 is the $E'_1$-system's relaxation rate.

 \smallskip

\noindent Substituting eq.(78) into eq.(77): %%
\begin{equation}
{d\rho_{S'}(t) \over dt} \approx - \int_0^t ds tr_{E'_1}
[H_{S'E'_1}^I(t),[H_{S'E'_1}^I(s), \rho_{S'}(s)\otimes
\sigma_{E'_1\circ}]].
\end{equation}

\noindent If we assume thermal state for the $E'_1$ environment,
$\sigma_{E'_1\circ} = \rho_{th}$, then eq.(79) strongly suggests
applicability of the Markov approximation\footnote{\small Remind:
the environment state $\sigma_{E'_1\circ}$ is thermal--no state
change in eq.(79) for the environment (except in very short time
intervals). The exact form of eq.(79) contains non-Markovian
corrections that are, according to the ansatz eq.(78), assumed to
be small.} and hence the master equation (see (Breuer and
Petruccione 2002) for details): %%
\begin{equation}
{d\rho_{S'}(t) \over dt} \approx - \int_0^{\infty} ds tr_{E'_1}
[H_{S'E'_1}^I(t),[H_{S'E'_1}^I(s), \rho_{S'}(t)\otimes
\rho_{th}]],
\end{equation}

\noindent which is formally the starting point for deriving the
master equation eq.(68)--but this time for the $S'$ system. So,
again, we obtain the above-brought conclusion, which is the title
of this section.

\smallskip

\noindent Eq.(80) is approximate ($\epsilon(t) \ll 1$) and valid
for the time instants $t > t_{\circ}$. Nevertheless, it's
applicable for arbitrary initial state of the composite system. Of
course, whether or not this plausible derivation may be used in
the more general context remains unanswered as Markovian
approximation for the rhs in eq.(79) is another plausible ansatz,
not  a rigorous physical condition yet.

\bigskip

\noindent {\bf {\emph {6.3.4 A limitation of the Nakajima-Zwanzig
projection method}}}

\bigskip

\noindent Derivation of eq.(80) assumes Markov approximation,
which, as we already know (Rivas and Huelga 2011), cannot be valid
for the $S'+E'$ structure. Nevertheless, plausibility of this
simplification stems from the form of the rhs of eq.(79) as well
as from the fact that the neglected term is non-Markovian. So we
hope for {\it approximate} Markovianity of the $S'$ system's
dynamics (assuming that $\epsilon \ll 1$, and only after some time
interval $t_{\circ}$).

\smallskip

\noindent At first sight, we could have used the methods adapted
to description of non-Markovian dynamics, notably the so-called
time-convolutionless method (Breuer and Petruccione 2002).
However, as we show below, this method is {\it not adapted} to the
structural considerations.

\smallskip

\noindent The Nakajima-Zwanzig projection method is the central
method of modern open quantum systems theory (Breuar and
Petruccione 2002, Rivas and Huelga 2011). It's the very basis of
modern open systems theory and application that include the
time-convolutionless method.

\smallskip

\noindent The key idea behind the Nakajima-Zwanzig projection
method consists of the introduction of a certain projection
operator, $\mathcal{P}$, which acts on the operators of the state
space of the total system "system+environment" ($S+E$). If $\rho$
is the density matrix of the total system, the projection
$\mathcal{P}\rho$ (the "relevant part" of the total density
matrix) serves to represent a simplified effective description
through a reduced state of the total system. The complementary
part (the "irrelevant part" of the total density matrix),
$\mathcal{Q} \rho = (I - \mathcal{P})\rho$. For the "relevant
part", $\mathcal{P}\rho(t)$, one derives closed equations of
motion in the form of integro-differential equation. The open
system's density matrix $\rho_S(t) = tr_E \mathcal{P}\rho(t)$
contains {\it all} necessary information about the open system
$S$. Here, we refer to the mostly-used kind of projection: %%
\begin{equation}
\mathcal{P} \rho(t) = tr_E \rho(t) \otimes \rho_E \equiv \rho_S(t)
\otimes \rho_E
\end{equation}

\noindent where $\rho_E \neq tr_S \rho$ for any instant in time.
The open system's state %%
\begin{equation}
\rho_S(t) = tr_E \rho(t) = tr_E \mathcal{P}\rho(t) \Leftrightarrow
tr_E \mathcal{Q} \rho(t) = 0, \forall{t};
\end{equation}

\noindent the $\mathcal{Q}$ projector satisfies $\mathcal{P} +
\mathcal{Q} = I$.

\smallskip

\noindent The Nakajima-Zwanzig projection method assumes a
concrete, in advance chosen and fixed, system-environment split (a
"structure"), $S+E$, which is uniquely defined by the associated
tensor product structure  of the total system's Hilbert space,
$\mathcal{H} = \mathcal{H}_S \otimes \mathcal{H}_E$. The division
of the composite system into "system" and "environment" is
practically motivated. In principle, the projection method can
equally describe arbitrary system-environment split i.e. arbitrary
factorization of the total system's Hilbert state.

\smallskip

\noindent However, our task points out a limitation of the
Nakajima-Zwanzig method. In the more general terms, the task reads
as: for a pair of open systems, $S$ and $S'$, pertaining to the
different system-environment splits of a composite system, can the
Nakajima-Zwanzig and/or the related projection methods provide
{\it simultaneous} dynamical description of the open systems, $S$
and $S'$?

\smallskip

\noindent The answer is provided by the following theorem:

\smallskip

\noindent {\bf Theorem 6.1}. {\it Quantum correlations relativity
precludes simultaneous pro\-jec\-tion-method-based description of
a pair of system-environment splits.}

\smallskip

\noindent {\it Proof}: There are only  two options for writing the
{\it simultaneous} master equations for the $S$ and $S'$ systems.
First, if the projection adapted to the $S+E$ structure can be
used for deriving the master equation for the $S'$ system. Then it
is required, that $tr_E\mathcal{P}\rho(t) = \rho_S(t)$ {\it and}
$tr_{E'} \mathcal{P}\rho(t) = \rho_{S'}(t)$,  i.e. $tr_E
\mathcal{Q}\rho(t) = 0$ {\it and} $tr_{E'} \mathcal{Q}\rho(t) = 0$
for {\it every} instant in time. Second, if we perform in
parallel, i.e. if we use the different projection operators,
$\mathcal{P}$ and $\mathcal{P}'$, for the two structures
independently of each other. Then it is required,
$tr_E\mathcal{P}\rho(t) = \rho_S(t)$ {\it and}
$tr_{E'}\mathcal{P}'\rho(t) = \rho_{S'}(t)$, i.e.
$tr_E\mathcal{Q}\rho(t) = 0 $ {\it and}
$tr_{E'}\mathcal{Q}'\rho(t) = 0$ for {\it every} instant in time.
We use the following lemmas.

\smallskip

\noindent {\bf Lemma 6.1}. {\it Quantum correlations relativity in
dynamical terms, for the mixed states, reads as: $\rho_S (t)
\otimes \rho_E = \sum_i \lambda_i \rho_{S'}(t) \otimes
\rho_{E'}(t) \nonumber$. The possible time dependence of the
weights $\lambda$ is irrelevant.}

\smallskip

\noindent {\bf Lemma 6.2}. {\it For the most part of the composite
system's dynamics, $tr_E \mathcal{Q}\rho(t)$ $= 0$ implies
$tr_{E'} \mathcal{Q}\rho(t) \neq 0$, and {\it vice versa}.}

\smallskip

\noindent {\bf Lemma 6.3}. {\it The two structure-adapted
projectors $\mathcal{P}$ and $\mathcal{P}'$  do not mutually
commute and cannot be simultaneously performed.}

\smallskip

\noindent Lemma 6.1 establishes  time-dependence of states of {\it
both} subsystems, $S'$ and $E'$. Bearing eqs.(28)-(29) in mind, we
realize that it may happen that there are only classical
correlations for the $S'+E'$ structure.

\smallskip

\noindent Lemma 6.2 establishes: for the most part of the
composite system's dynamics, projection $\mathcal{Q}\rho$ (or
$\mathcal{Q}'\rho$) brings some information about the open system
$S'$ (or $S$)--in {\it contradiction} with the basic {\it idea} of
the Nakajima-Zwanzig projection method.

\smallskip

\noindent On the other hand, Lemma 6.3 establishes: for any pair
of structures, $S+E$ and $S'+E'$, one {\it cannot}
choose/construct a pair of compatible projectors defined by
eq.(81). So Lemma 6.3 precludes simultaneous (for the same time
interval) derivation of master equations for the two open systems,
$S$ and $S'$.

\smallskip

\noindent From Lemma 6.2 and 6.3, it directly follows the claim of
the theorem. Q.E.D.

\smallskip

\noindent Thus the Nakajima-Zwanzig projection method faces a
limitation. While it can be separately performed for any structure
(either $\mathcal{P}$ or $\mathcal{P}'$), it cannot be {\it
simultaneously} used for a pair of structures. Once performed,
projection does not in general allow for drawing complete
information about an alternative structure of the composite
system--projecting is non-invertible ("irreversible").

\smallskip

\noindent Our finding refers to {\it all projection-based
methods}--including the above mentioned time-convolutionless
method. In formal terms: Lemma 2 implies that, in an instant of
time, $d\mathcal{P}\rho(t)/dt$ allows tracing out over only one
structure of the composite system. If that structure is $S+E$,
then $tr_{E'} d\mathcal{P} \rho(t)/dt \neq d\rho_{S'}(t)/dt$ [as
long as $\rho_{S'}(t) = tr_{E'}\rho(t)$]. On the other hand, Lemma
3 excludes simultaneous projecting, i.e. simultaneous master
equations for the two structures. E.g., $d\mathcal{P}\rho(t)/dt =
d\rho_S(t)/dt \otimes \rho_E$ is in conflict with
$d\mathcal{P}'\rho(t)/dt = d\rho_{S'}(t)/dt \otimes \rho_{E'}$:
due to QCR, Section 3.2, only one of them can be correct for
arbitrary instant in time.

\smallskip

\noindent Despite the fact that  quantum correlations relativity
can have exceptions for certain states, our findings presented by
Theorem 6.1 do not. Even if QCR does not apply to an instant in
time (i.e. to a special state of the total system), it is most
likely to apply already for the next instant of time in the
unitary (continuous in time) dynamics of the total system $C$.
This general argument makes the above lemmas universal, i.e.
applicable for every Hilbert state space and every model and
structure (the choice of the open systems $S$ and $S'$) of the
total system. Hence our findings and conclusions refer to the
finite- and infinite-dimensional systems and to all kinds of the
transformations of variables.

\smallskip

\noindent These findings do not present any inconsistency with the
open systems theory or with the foundations of the
Nakajima-Zwanzig method. Rather, our findings point out that the
Nakajima-Zwanzig projection method has a {\it limitation}, i.e. is
{\it not suitable} for the above-posed task.

\smallskip

\noindent Everything told in this section equally refers to the
open systems $S$ and $S'$ that are macroscopically "almost
equal"--e.g. in number of their respective constituent particles.
Even if the $S'$ system follows from the (local) transformation of
joining a single particle of the $E$ system with the $S$ system,
thus obtaining the new $S'$ and $E'$ systems, the projection
method cannot be straightforwardly used to derive master equation
for the $S'$ system. Hence we conclude: there is not a priori a
hope that the small changes in the system-environment split would
effect in small changes in dynamics. In other words, we can
conclude: the "shortcuts" for describing the
alternative-open-systems dynamics may be non-reliable and
delicate.

\smallskip

\noindent {\it Proof of Lemma 6.1}. Without any loss of
generality, and in order to eliminate the weights $\lambda$ from
consideration, consider $\rho_S(t)\otimes \rho_E =
\rho_{S'}(t)\otimes \rho_{E'}$. Then calculate $tr x_{E'}$, where
$x_{E'} = \alpha x_S + \beta x_E$. Then $tr x_{E'} = \alpha tr_S
x_S \rho_S(t) + \beta tr_E x_E\rho_E$, which is time dependent. On
the other hand, $tr x_{E'} = tr_{E'} x_{E'} \rho_{E'}$, which is
time-independent. In order to reconcile the two, we conclude that
also $\rho_{E'}$ must be time dependent. \hfill Q.E.D.

\smallskip

\noindent We borrow the proofs of the lemmas from the original
paper (Arsenijevi\' c et al 2013b).

\smallskip

\noindent {\it Proof of Lemma 6.2}. Given  $tr_E
\mathcal{Q}\rho(t) = 0, \forall{t}$, we investigate the conditions
that should be fulfilled in order for $tr_{E'} \mathcal{Q}\rho(t)
= 0, \forall{t}$. The $\mathcal{Q}$ projector refers to the $S+E$,
not to the $S'+E'$ structure. Therefore, in order to calculate
$tr_{E'}\mathcal{Q}\rho(t)$, we use ER. We refer to the projection
eq.(81) in an instant of time: %%
\begin{equation}
\label{eq.10}
 \mathcal{P} \rho = (tr_E \rho) \otimes \rho_E.
\end{equation}

\noindent A) Pure state $\rho = \vert \Psi \rangle\langle \Psi
\vert$, while $tr_E \mathcal{Q} \vert \Psi \rangle\langle \Psi
\vert = 0$.

\smallskip

\noindent We consider the pure state presented in its (not
necessarily unique) Schmidt form %%
\begin{equation}
\label{eq.11}
 \vert \Psi \rangle = \sum_i c_i \vert i\rangle_S
\vert i\rangle_E,
\end{equation}

\noindent where $\rho_S = tr_E \vert \Psi\rangle\langle \Psi\vert
= \sum_i p_i \vert i\rangle_S\langle i\vert$, $p_i = \vert
c_i\vert^2$ and for arbitrary $\rho_E \neq tr_S \vert
\Psi\rangle\langle\Psi\rangle$. Given $\rho_E = \sum_{\alpha}
\pi_{\alpha} \vert \alpha \rangle_E\langle \alpha \vert $, we
decompose $\vert \Psi\rangle$ as: %%
\begin{equation}
\label{eq.12} \vert \Psi\rangle = \sum_{i, \alpha} c_i C_{i\alpha}
\vert i\rangle_S \vert \alpha\rangle_E,
\end{equation}

\noindent with the constraints: %%
\begin{equation}
\label{eq.13}
 \sum_i \vert c_i\vert^2 = 1 = \sum_{\alpha}
\pi_{\alpha}, \sum_{\alpha} \vert C_{i\alpha} \vert^2 = 1,
\forall{i},
\end{equation}

\noindent Then %%
\begin{equation}
\label{eq.14}
 \mathcal{Q} \vert \Psi\rangle\langle \Psi \vert =
\vert \Psi \rangle\langle \Psi \vert - \sum_{i, \alpha} p_i
\pi_{\alpha} \vert i\rangle_S\langle i\vert \otimes \vert \alpha
\rangle_E\langle \alpha \vert.
\end{equation}

\noindent We use ER: %%
\begin{equation}
\label{eq.15}
 \vert i\rangle_S \vert \alpha\rangle_E = \sum_{m,n}
D^{i\alpha}_{mn} \vert m\rangle_{S'} \vert n \rangle_{E'}
\end{equation}

\noindent with the constraints: %%
\begin{equation}
\label{eq.16}
 \sum_{m,n}  D^{i\alpha}_{mn} D^{i'\alpha'\ast}_{mn}
= \delta_{ii'} \delta_{\alpha\alpha'}.
\end{equation}

\noindent With the use of eqs.(\ref{eq.12}) and (\ref{eq.15}),
eq.(\ref{eq.14}) reads as: %%
\begin{equation}
\label{eq.17}
 \sum_{m,m'n,n'}[\sum_{i,i', \alpha, \alpha'} c_i
C_{i\alpha} c_{i'}^{\ast}C_{i'\alpha'}^{\ast} D^{i\alpha}_{mn}
D^{i'\alpha' \ast}_{m'n'} -  \sum_{i,\alpha} p_i \pi_{\alpha}
D^{i\alpha}_{mn} D^{i\alpha\ \ast}_{m'n'}] \vert
m\rangle_{S'}\langle m' \vert \otimes \vert n\rangle_{E'}\langle
n'\vert.
\end{equation}

\noindent After tracing out, $tr_{E'}$: %%
\begin{equation}
\label{eq.18}
 \sum_{m,m'} \left.\{  \sum_{i,\alpha,n}
\sum_{i',\alpha'} c_i C_{i\alpha}
c_{i'}^{\ast}C_{i'\alpha'}^{\ast} D^{i\alpha}_{mn}
D^{i'\alpha'\ast}_{m'n}\right. -  \left. p_i \pi_{\alpha}
D^{i\alpha}_{mn} D^{i\alpha\ast}_{m'n} \right.\} \vert
m\rangle_{S'}\langle m'\vert
\end{equation}

\noindent Hence %%
\begin{equation}
\label{eq.19}
 tr_{E'} \mathcal{Q} \vert \Psi\rangle\langle
\Psi\vert = 0 \Leftrightarrow \sum_{i,\alpha,n} [\sum_{i',\alpha'}
c_i C_{i\alpha} c_{i'}^{\ast} C_{i'\alpha'}^{\ast}
D^{i\alpha}_{mn} D^{i'\alpha'\ast}_{m'n} -  p_i \pi_{\alpha}
D^{i\alpha}_{mn} D^{i\alpha\ast}_{m'n}] = 0, \forall{m,m'}
\end{equation}

\noindent Introducing notation, $\Lambda^m_n \equiv
\sum_{i,\alpha} c_i C_{i\alpha} D^{i\alpha}_{mn}$, one obtains: %%
\begin{equation}
\label{eq.20}
 tr_{E'} \mathcal{Q} \vert \Psi\rangle\langle
\Psi\vert = 0 \Leftrightarrow \\  A_{mm'} \equiv \sum_{n}
[\Lambda^m_n \Lambda^{m'\ast}_n - \sum_{i,\alpha}p_i \pi_{\alpha}
D^{i\alpha}_{mn} D^{i\alpha\ast}_{m'n}] = 0, \forall{m,m'}.
\end{equation}

Notice: %%
\begin{equation}
\label{eq.21} \sum_m A_{mm} = 0.
\end{equation}

\noindent which is equivalent to $tr \mathcal{Q} \vert \Psi\rangle
\langle \Psi \vert = 0$, see eq.(\ref{eq.14}).

\noindent B) Mixed (e.g. non-entangled) state. %%
\begin{equation}
\label{eq.22}
 \rho = \sum_i \lambda_i \rho_{Si}\rho_{Ei}, \quad
\rho_{Si} = \sum_m p_{im} \vert \chi_{im}\rangle_S
\langle\chi_{im} \vert,   \\ \rho_{Ei} = \sum_n \pi_{in}\vert
\phi_{in}\rangle_E\langle \phi_{in}\vert,
\end{equation}

\noindent In eq.(\ref{eq.22}), having in mind eq.(\ref{eq.10}),
$tr_E \mathcal{Q} \rho = 0$, while $tr_E\rho = \sum_p \kappa_p
\vert \varphi_{p}\rangle_S\langle \varphi_{p}\vert$, and $\rho_E =
\sum_q \omega_q \vert \psi_{q}\rangle_E\langle \psi_{q}\vert \neq
tr_S \rho$.

\noindent Constraints: %%
\begin{equation}
\label{eq.23}
 \sum_i \lambda_i = 1 = \sum_p \kappa_p = \sum_q
\omega_q, \quad \sum_{m} p_{im} = 1 = \sum_n \pi_{in}, \forall{i}.
\end{equation}

\noindent Now we make use of ER and, for comparison, we use the
same basis $\{\vert a\rangle_{S'} \vert b \rangle_{E'}\}$ %%
\begin{equation}
\label{eq.24}
 \vert \chi_{im}\rangle_S \vert \phi_{in} \rangle_E =
\sum_{a,b} C^{imn}_{ab} \vert a \rangle_{S'} \vert b \rangle_{E'},
 \\ \vert\varphi_p\rangle_S \vert \psi_q\rangle_E = \sum_{a,b}
D^{pq}_{ab} \vert a \rangle_{S'} \vert b \rangle_{E}.
\end{equation}

\noindent Constraints: %%
\begin{equation}
\label{eq.25}
 \sum_{a,b} C^{imn}_{ab} C^{im'n'\ast}_{ab} =
\delta_{mm'} \delta_{nn'}, \quad \sum_{a,b} D^{pq}_{ab}
D^{p'q'\ast}_{ab} = \delta_{pp'} \delta_{qq'}.
\end{equation}

\noindent So %%
\begin{eqnarray}
&\nonumber&  \label{eq.26} \mathcal{Q}\rho = \rho - (tr_E\rho)
\otimes \rho_E = \sum_{a,a',b,b'} \{\sum_{i,m,n} \lambda_i p_{im}
\pi_{in} C^{imn}_{ab} C^{imn\ast}_{a'b'}
\\&&
- \sum_{p,q} \kappa_p \omega_q D^{pq}_{ab} D^{pq\ast}_{a'b'}\}
\quad \vert a \rangle_{S'}\langle a'\vert \otimes \vert
b\rangle_{E'} \langle b'\vert.
\end{eqnarray}

\noindent Hence %%
\begin{equation}
\label{eq.27}
 tr_{E'} \mathcal{Q} \rho = 0 \Leftrightarrow
\Lambda_{aa'} \equiv \sum_{i,m,n,b} \lambda_i p_{im} \pi_{in}
C^{imn}_{ab} C^{imn\ast}_{a'b}-  \sum_{p,q,b} \kappa_p \omega_q
D^{pq}_{ab} D^{pq\ast}_{a'b} = 0, \forall{a,a'}.
\end{equation}

\noindent Again, for $a=a'$: %%
\begin{equation}
\label{eq.28}
 \sum_a \Lambda_{aa} = 0,
\end{equation}

\noindent as being equivalent with $tr \mathcal{Q} \rho = 0$, see
eq.(\ref{eq.26}).

\smallskip

\noindent Both eq.(\ref{eq.20}) and eq.(\ref{eq.27}) represent the
sets of the simultaneously satisfied equations. We do not claim
non-existence of the particular solutions to eq.(\ref{eq.20})
and/or to eq.(\ref{eq.27}), e.g. for the finite-dimensional
systems. We just emphasize, that the number of states they might
refer to, is apparently negligible compared to the number of
states for which this is not the case. For instance, already for
the fixed $a$ and $a'$, a small change e.g. in $\kappa$s (while
bearing eq.(\ref{eq.23}) in mind) undermines equality in
eq.(\ref{eq.27}).

\smallskip

\noindent Quantum dynamics is continuous in time. Provided $tr_E
\mathcal{Q}\rho(t) = 0$ is fulfilled, validity of $tr_{E'}
\mathcal{Q}\rho(t) = 0$ might refer {\it only} to a special set of
the time instants. So we conclude: for the most part of the open
$S'$-system's dynamics, $tr_{E'} \mathcal{Q}\rho(t) = 0$ is not
fulfilled. By exchanging the roles of the $S$ and the $S'$ systems
in our analysis, we obtain the reverse conclusion, which completes
the proof. \hfill Q.E.D.

\smallskip

\noindent {\it Proof of Lemma 6.3}. The commutation condition,
$[\mathcal{P}, \mathcal{P}']\rho(t) = 0, \forall{t}$. With the
notation $\rho_P(t) \equiv \mathcal{P}\rho(t)$ and $\rho_{P'}(t)
\equiv \mathcal{P}'\rho(t)$, the commutativity reads as:
$\mathcal{P}\rho_{P'}(t) = \mathcal{P}'\rho_P(t), \forall{t}$.
Then,  $\mathcal{P}\rho_{P'}(t) = tr_E\rho_{P'}(t) \otimes \rho_E
= \rho_S(t) \otimes \rho_E$, while, according to Lemma 6.1,
$\mathcal{P}'\rho_{P}(t) = tr_{E'}\rho_P(t) = \sigma_{S'}(t)
\otimes \sigma_{E'}(t)$. So, the commutativity requires the
equality $\sigma_{S'}(t)\otimes \sigma_{E'}(t) =
\rho_{S}(t)\otimes \rho_{E}, \forall{t}$. However, quantum
dynamics is continuous in time. Like in Proof of Lemma 2, quantum
correlations relativity guarantees, that, for the most of the time
instants, the equality will not be fulfilled. \hfill Q.E.D.

\bigskip

\noindent {\bf 6.4 The LCTs preserve linearity of a
composite-system's model}

\bigskip

\noindent Physically, the structure $\mathcal{S'} = \{x_{CM},
Q_{Ri} \}$--Brownian particle is the total system's center of
mass, while the environment is composed of the normal modes for
the relative positions for the original structure $S+E$.
Nevertheless, the  $\mathcal{S'}$ structure is not very special.
The procedure presented in Sections 6,3,2 and 6.3.3 is applicable
formally for arbitrary LCTs eq.(6). In other words: linear
canonical transformations preserve linearity\footnote{\small
"Linearity" means that the Hamiltonian is quadratic, with the
bilinear interaction, eq.(66), and uncoupled environment
oscillators.} of the original structure $S+E$. The whole
structural transformation can be presented as: %%
\begin{equation}
\mathcal{S}_{\circ} = \{x_S, x_{Ei}\} \to \mathcal{S} = \{x_{S'},
x_{E'i}\} \to \mathcal{S'} = \{x_{S'}, Q_{E'i}\},
\end{equation}

\noindent for every LCT eq.(6).

\smallskip

\noindent However, there are certain constraints for the LCTs, in
order to make the alternate structure physically sensible. For the
case presented in Section 6.3, the constraints are:
$M\Omega_{S'}^2
> 0$ and $\mu_i \nu_{E'i}^2 > 0, \forall{i}$. The analogous constraints
appear for all the alternate structures. Thereby the physical
relevance of the alternate structure $\mathcal{S'}$ is not
unconditional.

\smallskip

\noindent The "linear model" refers also to some other physically
relevant models. e.g. In (Bellomo et al 2005),  a
[non-relativistic] charged particle is embedded in the
electromagnetic-field modes at zero temperature. Interestingly
enough,  initial state of the total system is correlated, and is
shown that decoherence is related to the time dependent "dressing"
of the particle. A similar analysis (of a linear model) is
provided by (Stokes et al 2012) with the explicit LCTs performed
on the total system "atom+EM-field". An emphasis is placed on the
range of validity of the quantum optical master equations for the
composite system; for details see Section 7.3.

\smallskip

\noindent Here we do not elaborate on this any further. The reason
is probably apparent: depending on the choice of  LCTs, i.e. of
the new variables and the related parameters range, the results
may vary--i.e. are {\it case dependent}. Subtlety and complexity
of investigating the occurrence of decoherence is fairly expressed
by (Anglin et al 1997) [our emphasis]:

\smallskip

\noindent "{\it In this paper we will effectively argue that many
perceived universalities in the phenomenology of decoherence are
artifacts of studying toy models, and that the single neat border
checkpoint should be replaced as an image for decoherence by} the
picture of a wide and ambiguous No Man's Land, filled with pits
and mines, {\it which may be crossed on a great variety of more or
less tortuous routes. Once one has indeed crossed this region, and
travelled some distance away from it, the going becomes easier: we
are not casting doubt on the ability of the very strong
decoherence acting on macroscopic objects to enforce effective
classicality. ... By presenting a number of theoretically
tractable examples in which various elements of phenomenological
lore can be seen to fail explicitly, we make the point that each
experimental scenario will have to be examined theoretically on
its own merits, and from first principles.}".

\smallskip

\noindent Nevertheless, as emphasized above, there is the
following, generally valid, observation: linearity of the total
system's model is preserved by
 linear canonical transformations. Encouraged by Section 6.3, we dare to state the
following

\smallskip

\noindent {\bf Conjecture 1}. {\it For the linear models one can
expect in principle the occurrence of decoherence also for some
alternate degrees of freedom, which are  provided by the linear
canonical transformations.}

\bigskip

\noindent {\bf 6.5 More than one "classical world"}

\bigskip

\noindent Section 6.3 teaches us: if the open system $S$ is a
Brownian particle, then also the open system $S'$ is necessarily a
Brownian particle. This seemingly naive observation is physically
remarkable.

\smallskip

\noindent Every Hamiltonian generates the {\it simultaneously}
unfolding dynamics for  different structures of a  composite
system. Of all possible structures, for the standard QBM model, we
distinguish and consider only those emphasized in Section 2.2.2:
mutually global, non-trivial and irreducible structures. In
Section 6.3, we consider a pair of such structures and find the
{\it parallel occurrence of decoherence} for the structures.

\smallskip

\noindent The standard "decoherence program" (Giulini et al 1996,
Schlosshauer 2004) bases itself on the following assumption:
quantum decoherence is in the root of the appearance of "classical
world" in quantum theory. Now, as we elaborate below, our result
on the parallel occurrence of decoherence suggests: for a
composite (closed) system $C$, Section 6.3, there are at least
 two, mutually autonomous and irreducible "classical world{\it
s}"--{\it one classical world for one structure, i.e. for one
Brownian particle}\footnote{\small Needless to say, not every
structure bears classicality. Therefore "classicality" of a
physical system is {\it relative}--it's a matter of the system's
structure.}.

\smallskip

\noindent In a set of mutually global and irreducible structures,
every [physically reasonable]  structure\footnote{\small $C = S+E
= S'+E' = ...$} has the following characteristics:

\smallskip

\noindent (a) It is completely describable by the universally
valid quantum mechanics;

\smallskip

\noindent (b) It has its own set of  "elementary particles" and
the interactions between them, and is (cf. Section 2.2)
irreducible and information-theoretically separated from any
alternative, global and irreducible structure;

\smallskip

\noindent (c) It dynamically evolves in time, simultaneously with
but totally independently (autonomously) of any other structure;

\smallskip

\noindent (d) It has its own Brownian particle;

\smallskip

\noindent (e) Is locally indistinguishable from the others: an
observer belonging to one (and to {\it only one}) structure cannot
say which structure
 he belongs to;

 \smallskip

\noindent (f) Physically is {\it not}, {\it a priori}, less
realistic than any other.

\smallskip

\noindent Thereby, physically, the model-universe $C$, Section
6.3, {\it hosts} more than one {\it dynamical classical world}. As
the worlds are mutually global and irreducible, there is more than
one "classical world"  for one and the unique (a single)
"universe" $C$. Thereby, if the standard decoherence program
provides the "appearance of {\it a} Classical World" (Giulini et
al, 1996), our results suggest the "appearance of {\it the}
Classical World{\it s}".

\smallskip

\noindent This observation challenges  foundations of the standard
decoherence program and requires additional interpretational
analysis, which will  be presented in Chapter 8.

\bigskip

\noindent {\bf 6.6 A few general notions}

\bigskip

\noindent The composite system $C$ is closed--a model-universe
subject to the Schr\" odinger law. For the closed systems, which
are  not observable from the outside, there does not seem to exist
a privileged fundamental decomposition into subsystems
(structure). Regarding the open composite systems, see the next
chapter.

\smallskip

\noindent Our considerations are explicit only for the linear
models. So, the parallel occurrence of decoherence is in its
infancy yet. The natural question whether or not our
considerations can be applied to the more realistic models of the
many-particle open systems here remains unanswered. To this end,
see the quote from (Anglin et al 1997) in Section 6.4.

\smallskip

\noindent Of course, instead of decoherence, one can use some
other criteria for classicality, i.e. for the "appearance of the
classical world". e.g. One can use the absence of non-classical
correlations as such a criterion. An example in this regard for an
open system is given in Section 7.2.  For some results concerning
the information theoretic description of the decoherence process,
see (Coles 2012).

\newpage

\noindent {\bf Chapter 7}

\noindent\textbf{\large Decoherence-Induced Preferred Structure}

\vspace*{55mm}

\noindent "Observing" is {\it local}. Of a composite system, only
a small fraction of degrees of freedom is accessible to
observation. That is, only a small amount of information about a
composite system is acquired in realistic experimental situations.

\smallskip

\noindent There is no observer outside the Universe. Observer is a
part of the structure he observes. From Section 6.5 we learn:
there are certain structures of the Universe that cannot be
observed by an observer belonging to another structure. More on
this in Chapter 8.  Here we are interested in structures of  local
systems, which are  {\it open}--i.e. in unavoidable interaction
with their environments. This means, as distinct from Chapter 6,
the transformations of variables leave  environmental degrees of
freedom {\it intact}.

\smallskip

\noindent Which degrees of freedom are accessible to an observer
(Def.5.1), i.e. what constitutes  "system" in a given physical
situation? Are there some general rules and/or limitations? What
is origin of the classical prejudice, which is described in
Section 2.1? Those are the main questions of interest in the
remainder of this chapter. For a couple of models, we obtain that
the environment selects a "preferred" structure of the open
system.

\bigskip

\noindent {\bf 7.1 Decoherence-based classicality}

\bigskip

\noindent Decoherence Program (Giulini et al 1996, Schlosshauer
2004) offers a clue regarding the above posed questions:
environment decoheres only a fraction of the open system's degrees
of freedom. The decoherence-preferred degrees of freedom are
considered to be accessible (directly measurable in the sense of
Def.5.1) and therefore "objective" for an observer.

\smallskip

\noindent For instance, quantum vacuum monitors  atomic $R$
system, not atomic $CM$ system (Breuer and Petruccione 2002, Rivas
and Huelga 2011). Atomic de-excitation, i.e. the state decay,
refers to the atomic $R$ system: detection of a photon  reveals
the atomic internal-energy decay, which, typically, does not
affect the atomic $CM$ system. Bearing in mind Section 5.1, it is
now clear: quantum vacuum only partially monitors  atomic
electron(s) and proton(s)\footnote{\small For some details see
(Jekni\' c-Dugi\' c et al 2011).}.

\noindent Interaction between the open system $S$ and its
environment $E$: %%
\begin{equation}
H_{SE} = A_S \otimes B_E
\end{equation}

\noindent models a measurement of the system's observable, $A_S$,
that is performed by the environment $E$. If this interaction
dominates the composite system's ($S+E$'s) dynamics, then the
eigenstates of $A_S$ appear as the preferred (e.g. the approximate
"pointer basis") states for the open system $S$. In general,
spectral form of the interaction Hamiltonian gives only a
hint--not necessarily a definition--of the pointer basis states
(Dugi\' c 1996, 1997).

\smallskip

\noindent So, the environment-induced decoherence naturally offers
the following {\it basis} for answering the above posed questions:

\smallskip

\noindent {\bf Clue.} {\it Decoherence-selected preferred states
(and the related preferred observables) determine the preferred
structure of the open system.}

\smallskip

\noindent For instance, it is easy to design the
phenomenologically inspired effective interaction that promotes
the $CM$ system as a preferred
 subsystem:
\begin{equation}
H_{SE} = X_{CM} \otimes B_E; \quad S = CM + R.
\end{equation}

\noindent For  Brownian particle,  neglecting  the particle's $R$
system, the model eq.(104) is presented by eq.(66), i.e. $B_E =
\sum_i \kappa_i x_{Ei}$.

\smallskip

\noindent Similarly, Stipulation 1 of Section 4.2.2 assumes, at
least approximate, commutation $[H_{K_NE}, K_N] = 0$, that can be
modelled e.g. as\footnote{\small See eq.(3.164) in (Giulini et al
1996).}: %%
\begin{equation}
H_{K_NE} = K_N \otimes B_E.
\end{equation}

\noindent However,  eqs. (104)-(105) are {\it constructed}, i.e.
{\it designed} (or stipulated)  in order to fit with
phenomenology. However, in order to answer the above posed
questions, it is desirable to have {\it derived} (not merely
stipulated) the preferred structure of an open system.

\smallskip

\noindent In the next sections we will justify the Clue. We
consider specific models without posing any stipulation. Thereby
we come to the conclusion: {\it decoherence may provide} a unique
preferred structure {\it of the open system}.

\smallskip

\noindent In this context it is natural to reject physical reality
for the degrees of freedom representing linear combinations of the
decohered degrees of freedom. E.g., cf. Section 2.1 [and Footnote
27], the Earth's and the Venus' $CM$ systems are decohered, but
the $CM$ system for the Earth's and the Venus' $CM$-systems is
not--and is therefore an empty point in space, not an object--in
full agreement with the classical intuition described in Section
2.1; for more details see Section 8.2.

\smallskip

\noindent Of course, classicality of certain [decohered] degrees
of freedom does not imply non-observability of the alternative
(non-decohered) degrees of freedom.  As it is emphasized in
Sections 2.4 and 5.1, "local action" is a relative concept, which
is usually well defined in a concrete physical situation.

\smallskip

\noindent So, we are concerned with the following task:

\smallskip

\noindent $\mathcal{T}$. {\it Are there some realistic models that
do not require "construction" or stipulation of the preferred
structure of an open composite system}?

\smallskip

\noindent In the next sections we give just a few such models
referring to the few-degrees-of-freedom open systems. The models
employ different criteria for classicality--a definition of the
open system's pointer basis (and correlation in the composite
system), and the validity range of certain kind of master
equations, respectively. We are not aware of any other technically
elaborated considerations.

\bigskip

\noindent {\bf 7.2 Asymptotic dynamics of a two-mode system}

\bigskip

\noindent For a pair of modes, we investigate  asymptotic ($t \to
\infty$) behavior\footnote{\small For Markovian bipartite open
systems, which is our case, cf. eq.(106), (Ferraro et al 2010)
pointed out non-occurrence of discord sudden death, i.e. the
smooth disappearance of non-classical correlations. This is the
reason we, in search for classicality, stick to the asymptotic
solutions.} of the environment-induced preferred states. Following
generally accepted decoherence procedures, we find that there is
only one structure of the composite system which allows for the
preferred states to be regarded to bear classicality.

\smallskip

\noindent We consider a pair of uncoupled modes in the "phase
space" representation (as a pair of non-interacting linear
harmonic oscillators) that are independently subjected to the
quantum amplitude damping channels. A pair of noninteracting
linear oscillators, 1 and 2, with the respective frequencies and
masses, $\omega_1, \omega_2$. and $m_1, m_2$. The "phase space"
position variables, $x_1$ and $x_2$, and the conjugate momentums,
$p_1$ and $p_2$, respectively. The total Hilbert state space
factorizes $\mathcal{H} = \mathcal{H}_1 \otimes \mathcal{H}_2$ and
the total Hamiltonian $H = H_1 + H_2$, $H_i = p^2_i/2m_i + m_i
\omega_i^2 x^2_i/2, i = 1, 2$.

\smallskip

\noindent We analytically (exactly) solve the Heisenberg equations
of motion in the Kraus representation (Fan and Hu 2009, Jiang et
al 2011, Zhou et al 2011, Ferraro et al 2005, Kraus 1983, Breuer
and Petruccione 2002, Rivas and Huelga 2011) and analyze the
obtained results  for the original, as well as for some
alternative, degrees of freedom of the open system. The considered
structures are {\it local} in the sense that the environmental
degrees of freedom remain intact. We find that the environment
non-equally "sees" the different structures. It appears, that
there is {\it only one} structure that is distinguished by
classicality and locality of the environment influence.

\smallskip

\noindent For an oscillator (mode) subjected to a lossy channel
(or cavity at zero temperature), the master equation reads (Jiang
et al 2011): %%
\begin{equation}
{d \rho \over dt} = - \kappa \left[ 2a \rho a^{\dag} -
 \{a^{\dag} a, \rho\} \right]
\end{equation}

\noindent with the bosonic "annihilation" operator $a$ and the
damping parameter $\kappa$.

\smallskip

\noindent The master equation eq.(106) is known to be
representable in the Kraus form (Ferraro et al 2005, Fan and Hu
2009, Jiang et al 2011, Zhou et al 2011): %%
\begin{equation}
\rho(t) = \sum_{n=0}^{\infty} K_n(t) \rho K_n^{\dag}(t)
\end{equation}

\noindent with the completeness relation $\sum_{n=1}^{\infty}
K^{\dag}_n(t) K_n(t) = I, \forall{t}$. For the amplitude damping
process, i.e. for the master equations eq.(106),  Kraus operators
(Ferraro et al 2005, Fan and Hu 2009, Jiang et al 2011, Zhou et al
2011): %%
\begin{equation}
K_n(t) = \sqrt{{(1 - e^{-2kt})^n\over n!}}  e^{-kNt} a^n, \quad N
= a^{\dag}a.
\end{equation}

\noindent In the Heisenberg picture, the state $\rho$ does not
evolve in time. Then, in the Kraus representation, dynamics of an
oscillator's observable A reads: %%
\begin{equation}
A(t) = \sum_{n=0}^{\infty} K_n^{\dag}(t) A(t=0) K_n(t) =
\sum_{n=0}^{\infty} {(1-e^{-2kt})^n \over n!} a^{\dag n} e^{-ktN}
A(t=0) e^{-ktN} a^n.
\end{equation}

\noindent The infinite sum in Eq.(109) is often approximated by a
few first terms, e.g. in (Liu et al 2004). However, below we give
exact solutions to Eq.(109) without calling for or imposing any
approximation.

\bigskip

\noindent {\bf {\emph {7.2.1 Original degrees of freedom}}}

\bigskip

\noindent The structure we are interested in %%
\begin{equation}
(1 + E_1) + (2 + E_2)
\end{equation}

\noindent can be described by the following form of interaction:
$V = \alpha \sum_{k_{=1}}^2 (A_{S_1k} \otimes B_{E_1k} + A_{S_2k}
\otimes B_{E_2k})$ (Rivas and Huelga 2011). If the open system's
dynamics is Markovian [of the Lindblad form], one can totally
separate dynamics of the two subsystems, $1$ and $2$, for the
initial tensor-product\footnote{\small For correlated state
$\rho_{12}$, the $S_i$ system is in initial correlation with the
effective environment, $S_j+E, j \neq i = 1,2$, and thus its
dynamics is not Markovian (Rivas and Huelga 2011); see Section
6.2.} state $\rho_{12}$.

\smallskip

\noindent For the above interaction $V$, the master equation for
$C=1+2$ is of the form (Rivas et al 2010b, Rivas and Huelga 2011):
\begin{eqnarray}
&\nonumber&  {d\rho_{12} \over dt} = -\imath \sum_i [H_i +
\alpha^2 H^{(i)}_{LS}, \rho_{12}]
\\&&
+ \alpha^2 \sum_{\omega, i, k, l} \gamma_{kl}^{(i)}(\omega)
\left[A^{(i)}_k(\omega) \rho_{12} A^{(i)\dag}_l(\omega) - {1 \over
2} \{A^{(i)\dag}_l(\omega) A^{(i)}_k(\omega), \rho_{12} \}\right
].
\end{eqnarray}

\noindent By tracing out eq.(111), $\rho_i = tr_j \rho_{12}, i
\neq j = 1,2$, with the use of $tr_i [B_j, \rho_{12}] = 0, i=j$
and $tr_i [B_j, \rho_{12}] = [B_j, \rho_j]$ for $i \neq j$,
$i,j=1,2$, one easily obtains the following master equation: %%
\begin{eqnarray}
&\nonumber&  {d\rho_i \over dt} = -\imath [H_i + \alpha^2
H^{(i)}_{LS}, \rho_i]
\\&&
+ \alpha^2 \sum_{\omega, k, l} \gamma_{kl}^{(i)}(\omega)
\left[A^{(i)}_k(\omega) \rho_i A^{(i)\dag}_l(\omega) - {1 \over 2}
\{A^{(i)\dag}_l(\omega) A^{(i)}_k(\omega), \rho_i \}\right ]
\end{eqnarray}

\noindent for {\it both}, $i=1,2$.

\smallskip

\noindent We are interested in the independent,
environment-induced amplitude-dam\-ping processes for the two
oscillators, $1$ and $2$. For the "amplitude damping channel" for
one oscillator, there is only one Lindblad operator, $a$--the
"annihilation" boson operator. So, eq.(112) now obtains the form
of eq.(106) for both oscillators (modes) with the respective
damping parameters $\kappa_i$.

\smallskip

\noindent To facilitate our considerations, we switch to the Kraus
representation of the master equation eq.(106).  We do that in the
Heisenberg picture.

\smallskip

\noindent Independent amplitude damping channels for the two modes
are presented by mutually non-correlated, local, Kraus operators,
$K^{{1}}_m \otimes I_2$ and $I_1 \otimes K^{(2)}_n$, i.e. by the
separable total operation $K^{(1)}_m \otimes K^{(2)}_n$. This
operation gives for a one-mode operator, $A_1$: %%
\begin{eqnarray}
&\nonumber&  \sum_{m,n=0}^{\infty} K^{(1)\dag}_m \otimes
K^{(2)\dag}_n A_1(t=0) \otimes I_2 K^{(1)}_m \otimes K^{(2)}_n =
\\&&
\sum_{m=0}^{\infty} K^{(1)\dag}_m  A_1(t=0)  K^{(1)}_m \otimes
\sum_{n=0}^{\infty} K^{(2)\dag}_n K^{(2)}_n = A_1(t) \otimes I_2
\end{eqnarray}

\noindent Similarly for a $A_1B_1$ operator: %%
\begin{eqnarray}
&\nonumber&  \sum_{m,n=0}^{\infty} K^{(1)\dag}_m \otimes
K^{(2)\dag}_n A_1(t=0) B_1(t=0) \otimes I_2 K^{(1)}_m \otimes
K^{(2)}_n =
\\&& \nonumber
\sum_{m=0}^{\infty} K^{(1)\dag}_m  A_1(t=0) B_1(t=0)  K^{(1)}_m
\otimes \sum_{n=0}^{\infty} K^{(2)\dag}_n K^{(2)}_n \equiv
\\&&
(A_1B_1)(t) \otimes I_2.
\end{eqnarray}

\noindent Above we used the completeness relation for the $2$
system's Kraus operators. Of course, the completeness relation for
the two-mode Kraus operators is fulfilled:  $\sum_{m,n=0}^{\infty}
K^{(1)\dag}_m \otimes K^{(2)\dag}_n K^{(1)}_m \otimes K^{(2)}_n =
I_1 \otimes I_2 \equiv I_{12}$. For the two-mode operators: %%
\begin{eqnarray}
&\nonumber&  \sum_{m,n=0}^{\infty} K^{(1)\dag}_m \otimes
K^{(2)\dag}_n A_1(t=0) \otimes A_2(t=0) K^{(1)}_m \otimes
K^{(2)}_n =
\\&& \nonumber
\sum_{m=0}^{\infty} K^{(1)\dag}_m  A_1(t=0)  K^{(1)}_m \otimes
\sum_{n=0}^{\infty} K^{(2)\dag}_n A_2(t=0)K^{(2)}_n \equiv
\\&&A_1(t) \otimes A_2(t),
\end{eqnarray}

\noindent which exhibits independence of the actions of the two
environments, $E_1$ and $E_2$.

\smallskip

\noindent We use the following  generalization of the
Baker-Hausdorff lemma (Menda\v s and Popovi\' c 2010): %%
\begin{equation}
e^{-sA} B e^{-sA} = B - s \{A,B\} + {s^2 \over 2!} \{A, \{A,B\}\}
- {s^3 \over 3!}\{A, \{A, \{A,B\}\}  \}+ ...
\end{equation}

\noindent where the curly brackets denote anti-commutator,
$\{A,B\} = AB + BA$. So %%
\begin{equation}
e^{-ktN} a e^{-ktN} = e^{kt} a e^{-2ktN}, \quad e^{-ktN} a^{\dag}
e^{-ktN} = e^{-kt} a^{\dag} e^{-2ktN}
\end{equation}

\noindent Substituting eq.(1117) into eq.(109) one directly
obtains: %%
\begin{equation}
a^{\dag}(t) = e^{-kt} a^{\dag} \sum_{n=0}^{\infty}
{(1-e^{-2kt})^n\over n!} a^{\dag n} e^{-2ktN} a^n = e^{-kt}
a^{\dag}.
\end{equation}

\noindent Similarly: %%
\begin{equation}
a(t) = - e^{kt} \sum_{n=0}^{\infty} {(1-e^{-2kt})^n\over (n-1)!}
a^{\dag n-1} e^{-2ktN} a^n + e^{kt} a.
\end{equation}

\noindent With the use of $\sum_{n=0}^{\infty}
{(1-e^{-2kt})^n\over (n-1)!} a^{\dag n-1} e^{-2ktN} a^n = (1-
e^{-2kt}) a$, we obtain: %%
\begin{equation}
a(t) = e^{-kt} a.
\end{equation}

\noindent In the completely analogous way one obtains: %%
\begin{eqnarray}
&\nonumber&  (a^2)(t) = e^{-2kt} a^2
\\&&
(a^{\dag 2})(t) = e^{-2kt} a^{\dag 2} \nonumber\\&&
 (a^{\dag} a)(t) = e^{-2kt} a^{\dag} a.
\end{eqnarray}

\noindent Now with the aid of %%
\begin{equation}
x = \left({\hbar \over 2m \omega}\right)^{1/2} (a + a^{\dag}),
\quad p = \imath \left( {m\hbar \omega \over 2} \right)^{1/2}
(a^{\dag} - a),
\end{equation}

\noindent we obtain the following solutions to the Heisenberg
equations for the position and momentum observables: %%
\begin{eqnarray}
&\nonumber&  x(t) = e^{-kt} x, \quad p(t) = e^{-kt} p
\\&&
x^2(t) = e^{-2kt} x^2 +  {\hbar  \over 2 m \omega} (1 - e^{-2kt})
\nonumber\\&& p^2(t) = e^{-2kt} p^2 + {m \hbar \omega \over 2} (1
- e^{-2kt}).
\end{eqnarray}

\noindent From eq.(123), one directly obtains  asymptotic
solutions: %%
\begin{equation}
\lim_{t \to \infty} x(t) = 0 =\lim_{t \to \infty} p(t), \quad
\lim_{t\to \infty} x^2(t) = {\hbar \over 2m\omega}, \lim_{t\to
\infty} p^2(t) = {m\hbar \omega \over 2}.
\end{equation}

\noindent From eq.(124) directly follows product of the standard
deviations in the asymptotic limit for {\it both} oscillators: %%
\begin{equation}
\lim_{t\to\infty} \Delta x(t) \Delta p(t) = {\hbar \over 2}.
\end{equation}

\noindent Physically, eq.(125) is clear:  asymptotic states, for
both oscillators, are the minimum uncertainty  states. In the
position-representation, those states are the minimum uncertainty
Gaussian states--the well-known Sudarshan-Glauber coherent states.

\smallskip

\noindent For the Markovian bipartite open systems (which is our
case for the structure eq.(110) and for both oscillators, $1$ and
$2$) it is well known, that non-classical correlations smoothly
disappear in the asymptotic limit--there is no  discord sudden
death (Ferraro et al 2010). On the other hand, for  Gaussian
states (Adesso and Datta 2010), the only bipartite-system states
that have zero discord are the tensor product states--no
correlations at all. Hence, we directly conclude about the {\it
preferred asymptotic states}, (the approximate pointer basis) for
the pair of oscillators, that satisfy eq.(125): %%
\begin{equation}
\vert \alpha \rangle_1 \vert \beta \rangle_2,
\end{equation}

\noindent where $\vert \alpha \rangle_1$ and $\vert \beta
\rangle_2$ are the Sudarshan-Glauber coherent states, i.e. the
minimum uncertainty Gaussian states for the two oscillators, $1$
and $2$.

\bigskip

\noindent {\bf {\emph {7.2.2 Alternative degrees of freedom}}}

\bigskip

\noindent We introduce formally a pair of the degrees of freedom,
$X_A$ and $\xi_B$, with the conjugate momentums, $P_A$ and
$\pi_B$, $[X_A, P_A] = \imath \hbar$ and $[\xi_B, \pi_B] = \imath
\hbar$; of course, $[X_A, \pi_B] = 0 = [\xi_B, P_A]$. Without loss
of generality, let us consider the following linear canonical
transformations\footnote{\small With the constraints: $\alpha_i
\gamma_i = 1 = \beta_i \delta_i$, and $\alpha_i \delta_i = 0
=\beta_i \gamma_i $.}: %%
\begin{eqnarray}
&\nonumber&  X_A = \sum_i \alpha_i x_i, \quad P_A = \sum_j
\gamma_j p_j, \quad i,j=1,2
\\&&
\xi_B = \sum_m \beta_m x_m, \quad \pi_B = \sum_n \delta_n p_n,
\quad m,n=1,2.
\end{eqnarray}

\noindent for the pair of oscillators considered in Section 7.2.1.

\smallskip

\noindent Then the total system's Hilbert state space factorizes,
$\mathcal{H} = \mathcal{H}_A \otimes \mathcal{H}_B$, while the
Hamiltonian obtains the general form $H = H_A + H_B + H_{AB}$.

\smallskip

\noindent According to the above task, $\mathcal{T}$, we are
interested in
 solutions to the Heisenberg equations for the alternative
degrees of freedom.

\smallskip

\noindent However, we cannot directly use the master equation
eq.(106)   in order to derive the Kraus operators  for the new
subsystems, $A$ and $B$.

\smallskip

\noindent The transformations eq.(127) are local, i.e., they leave
the environmental degrees of freedom intact. On the other hand, as
the oscillators are out of any external (classical) field, bearing
in mind experience with the model in Section 6.3, we directly
conclude that there is not any interaction between the new
subsystems, $A$ and $B$, i.e. that $H_{AB} = 0$. However,  the
LCTs eq.(127) change the character of interaction with the
environment. This is easily seen from the forms of the Kraus
operators for the original oscillators, eq.(108). Placing the
inverse to eq.(127) into eq.(108)  directly provides the following
conclusion: for the new subsystems, $A$ and $B$, the environment
$E=E_1+E_2$ acts as a {\it common} environment, non-locally. This
is, one can easily show: %%
\begin{equation}
K^{(1)}_m \otimes K^{(2)}_n \neq K^{(A)}_m \otimes K^{(B)}_n.
\end{equation}

 \noindent Physically, it means that {\it non-local} action is exerted by the {\it
 total} environment $E$ on the pair $A+B$.

 \smallskip

 \noindent This conclusion also [directly] follows from eq.(25) in the context of
 Entanglement Relativity: according to ER, the preferred states
 eq.(126) typically obtain entangled form for the new structure, $A+B$. So,
 while the environment $E$ independently acts on the $1$ and $2$
 systems, its action on the $A$ and $B$ systems is [typically] non-local.
 Finally, one can  deal  with eq.(106) by expressing the old
 Lindblad operators via the new subsystem's operators. On this basis one expects
 nonseparation of master equations for the
 new subsystems $A$ and $B$, i.e. nonvalidity of the master equation eq.(112) for the new subsystems $A$ and $B$.

 \smallskip

 \noindent In the structure terms, the model of the total system (as distinct from eq.(110)) reads:
 \begin{equation}
(A + B) + E.
 \end{equation}

 \noindent Fortunately enough, the Kraus operators formalism deals with the
 infinite {\it sums}, not with the individual Kraus
 operators. So, we can circumvent all the technical problems by
 dealing with the infinite sums of the Kraus operators for the
 original structure, eq.(110), while expressing the new variables through the old ones.

 \smallskip

\noindent With the use of eqs.(123), (127), we can directly write
for the new position and momentum observables: %%

\begin{eqnarray}
&\nonumber&  X_A(t) = \sum_i \alpha_i x_i(t), P_A(t) = \sum_i
\gamma_i p_i(t)
\\&&
\xi_B(t) = \sum_i \beta_i x_i(t), \pi_B(t) = \sum_i \delta_i
p_i(t).
\end{eqnarray}

\noindent Similarly, e.g. %%
\begin{equation}
X_A^2(t) = \sum_{i,j} \alpha_i \alpha_j (x_ix_j)(t), P_A^2(t) =
\sum_{i,j} \gamma_i \gamma_j (p_i p_j)(t);
\end{equation}

\noindent of course [cf. eqs.(114), (115)], $(ab)(t) \equiv
\sum_{m,n=0}^{\infty} K^{(1)\dag}_m(t) K^{(2)\dag}_n(t) \quad ab $
$K^{(1)}_m(t)$ $K^{(2)}_n(t)$. Analogous expressions can be
directly written for the $B$ system.

\smallskip

\noindent For the tensor product initial state $\rho_{12}(0) =
\rho_1(0) \rho_2(0)$, with the aid of eq.(123), we obtain: %%
\begin{eqnarray}
&\nonumber&  \left(\Delta X_A(t)\right)^2 = tr_{12} \left[
\sum_{i,j} \alpha_i \alpha_j (x_i x_j)(t) \rho_{12}(0)\right] -
\left[ tr_{12} \sum_i \alpha_i x_i(t) \rho_{12}(0)\right]^2 =
\\&&
\sum_i \alpha_i^2 (\Delta x_i(t))^2 + \sum_{i, j \neq i} \alpha_i
\alpha_j \left [ \langle (x_ix_j)(t)\rangle - \langle
x_i(t)\rangle \langle x_j(t)\rangle\right] = \nonumber\\&& \sum_i
\alpha_i^2 (\Delta x_i(t))^2.
\end{eqnarray}

\noindent In complete analogy, one can calculate all the other
standard deviations finally to obtain in the asymptotic limit: %%
\begin{eqnarray}
&\nonumber&  \Delta X_A(\infty) \Delta P_A(\infty) = \hbar \sqrt{
\left({ \alpha_1^2 \over 2m_1\omega_1} + { \alpha_2^2 \over
2m_2\omega_2}\right) \left({\gamma_1^2 m_1  \omega_1\over 2} +
{\gamma_2^2 m_2 \omega_2\over 2}\right) }
\\&&
\Delta \xi_B(t) \Delta \pi_B(t) = \hbar\sqrt{ \left({ \beta_1^2
\over 2m_1\omega_1} + { \beta_2^2 \over 2m_2\omega_2}\right)
\left({\delta_1^2 m_1
 \omega_1\over 2} + {\delta_2^2 m_2  \omega_2\over 2}\right) }.
\end{eqnarray}

\noindent In general, the rhs of both expressions in eq.(133) are
larger than $\hbar /2$.

\smallskip

\noindent On the other hand, the common environment ($E=E_1+E_2$)
for the subsystems $A$ and $B$ is expected to induce correlations
for the $A$ and $B$ systems, even if the initial state is tensor
product. This can be easily justified by the use of the
"covariance function", e.g. $C(t) = \langle X_A(t) \xi_B(t)\rangle
- \langle X_A(t)\rangle \langle \xi_B(t) \rangle$. From eqs.(123)
and (127) we obtain: %%
\begin{eqnarray}
&\nonumber&  C(\infty) = \lim_{t \to \infty} \sum_{i,j} \left[
(x_i x_j)(t) - x_i(t) x_j(t) \right] =
\\&&
\sum_i \alpha_i \beta_i \left( \Delta x_i(\infty)\right)^2 =
\alpha_1 \beta_1 {\hbar \over 2 m_1 \omega_1} + \alpha_2 \beta_2
{\hbar \over 2 m_2 \omega_2},
\end{eqnarray}

\noindent which, typically, is non-zero; in eq.(134), likewise for
eq.(123), we used notation $(x_i x_j)(t)$  $= x_i(t) x_j(t), i
\neq j$, for the tensor product initial state.

\smallskip

\noindent Non-zero covariance function reveals  presence of
correlations\footnote{\small  However, zero covariance function
does not guarantee the absence of correlations. The correlations
can be classical or quantum--to distinguish between them, one
should use discord, Section 3.2.} in the $A+B$ structure of the
composite system, even in the asymptotic limit.

\bigskip

\noindent {\bf {\emph {7.2.3 Preferred structure}}}

\bigskip

\noindent For the  $A$ representing the center of mass and the $B$
representing the "relative particle" for the pair of equal-mass
($m_1=m_2$) and resonant ($\omega_1 = \omega_2$)
oscillators\footnote{\small These are the common assumptions--that
simplify calculation, see e.g. (Paz and Roncaglia 2008)--that are
absent from our considerations.}, i.e. for $\alpha_1 = \ 1/2 =
\alpha_2, \beta_1 = 1 = - \beta_2, \gamma_1 = 1 = \gamma_2$ and
$\delta_1 = 1/2 = - \delta_2$, one obtains equalities on the rhs
of eq.(133), $\Delta X_A(\infty) \Delta P_A(\infty) = \hbar /2$
and
 $\Delta \xi_B(\infty) \Delta \pi_B(\infty) = \hbar/2$, as well as
the absence of correlations, $C(\infty) = 0$.

\smallskip

\noindent However, typically, the preferred states for the $A+B$
structure are not the minimum uncertainty states and are
correlated.

\smallskip

\noindent The states eq.(126) are   arguably the most classical
bipartite system states of all. They are free of any kind of
correlations (classical or quantum) and are of the minimum quantum
uncertainty.
 Hence,
[in the asymptotic limit], one can imagine the pair 1+2 as a pair
of "individual", mutually distinguishable and non-correlated
systems. In a sense, this is a definition of "classical systems"
(Giulini et al 1996, Zurek 2003, Schlosshauer 2004). On the other
hand, none of these is in general valid for the alternate
structure $A+B$: even for the initial tensor product state for
$A+B$, non-local action of the common environment induces
correlations, even in the asymptotic limit. So, one can say that
the environment composed of two noninteracting parts, which induce
independent (local) amplitude damping processes, makes the 1 + 2
structure special ("preferred").

\bigskip

\noindent {\bf 7.3 Atom in electromagnetic field}

\bigskip

\noindent The composite system $C$ consists of an atom's internal
degrees of freedom ($A$) in electromagnetic field ($EM$);
$C=A+EM$.\footnote{\small The $A$ represents the atomic internal
($R$) system for the standard $CM+R$ atomic structure, Chapter 5.
Only the subsystem $R$ is in interaction with the electromagnetic
field. For this reason, the atomic $CM$ system is omitted from
considerations--compare to Section 5.3.} The $C$ system is
monitored by the environment, which is {\it supposed} to be photon
absorbing and quickly to thermalize (Stokes et al 2012). A similar
analysis can be found in (Stokes 2012).

\smallskip

\noindent Structural changes in $C$ are performed by certain
unitary operations that give rise to the different forms of the
$C$'s Hamiltonian, $H$.

\smallskip

\noindent Different forms of the Hamiltonian are expected to give
rise to different master equations for different structures of the
open system. The authors introduce the following {\it criterion of
classicality}: the preferred structure is the one that provides
predictions in accordance with the presence of the
photon-absorbing environment. Some other characteristics of the
environment, i.e. of physical situation, could lead to different
conclusions about the preferred structure of the open system $C$.

\smallskip

\noindent The following forms of the composite system's
Hamiltonian are of interest. The original structure, $A+EM$, is
defined by the so-called minimal-coupling Hamiltonian: %%
\begin{eqnarray}
&\nonumber&  H = H^{(min)} = {1 \over 2m_A} [\vec p_A+ e\vec
A_{EM}(\vec 0)]^2 + V(\vec r_A) + \\&&{1 \over 2}   \int d^3\vec x
\left[ {1 \over \mu_{\circ}} \vert\vec B_{EM}(\vec x)\vert^2 +
{1\over \epsilon_{\circ}} \vert\vec \Pi_{EM}(\vec
x)\vert^2\right].
\end{eqnarray}

\noindent In eq.(135), one can easily recognize the $EM$
self-energy (the last term) and the minimal coupling of the atom
with the electromagnetic field. The conjugate observables are
$\vec r_A, \vec p_A$, for the atom, and $\vec A_{EM}, \vec
\Pi_{Em}$ for the EM field; $\vec B = \vec \nabla \times \vec A$.

\smallskip

\noindent With the use of the specific unitary transformations,
the following forms of the Hamiltonian are obtained: %%
\begin{eqnarray}
&\nonumber& H = H^{(mult)} = {\vec p_1^2 \over 2m_A} + V(\vec r_1)
+ \\&& \nonumber {1 \over 2} \int d^3\vec x \left[ {1 \over
\mu_{\circ}} \vert\vec B_{2}(\vec x)\vert^2 + {1\over
\epsilon_{\circ}} \vert\vec \Pi_{2}(\vec x)\vert^2\right] \quad
[\rm multipolar],
\\&& \nonumber
H = H^{(rw)} = \sum_{\vec k, \lambda} \hbar g_{\vec k \lambda}
\sigma_{ex}^+ a_{F\vec k \lambda} + H.c. + \hbar \omega_{\circ}
\sigma_{ex 3} +  \\&& \sum_{\vec k \lambda} \hbar \omega_k
a_{F\vec k\lambda}^{\dag} a_{F\vec k \lambda}, \quad [{\rm
rotating-wave}].
\end{eqnarray}

\noindent The transformations behind these forms of the total
system's Hamiltonian are specific in that they do not change the
degrees of freedom: $\vec r_1 = \vec r_A$ and $\vec A_2 = \vec
A_{EM}$, hence $\vec B_2 = \vec B_{EM}$.  However, their conjugate
momentums change so as $[\vec p_A, \vec p_1] \neq 0$. The rotating
wave form of the total system's Hamiltonian is obtained after
approximating the atomic internal system by the exciton two-level
model (compare to Section 4.1)--hence the Pauli operators $\sigma$
and the annihilation and creation operators on the bosonic Fock
space in eq.(136). The "rotating wave" structure is thus a variant
of the famous spin-boson model.

\smallskip

\noindent The structure transformations can be described as: %%
\begin{equation}
\mathcal{S}_{min} = \{\vec r_A, \vec A_{EM}, \vec p_A, \vec
\Pi_{EM}\} \to \mathcal{S}_{mult} = \{\vec r_A, \vec A_{EM}, \vec
p_1, \vec \Pi_2\} \to \mathcal{S}_{rw} = \{\vec{\sigma}_{ex},
N_{F}\}
\end{equation}

\noindent $N_F \equiv a_F^{\dag} a_F$. Definitions of the momentum
observables as well as  physical interpretation of the structures
can be found in the original paper (Stokes et al 2012). These
subtle details are not substantial for our considerations.

\smallskip

\noindent By applying the second order perturbation approximation,
the authors derive  master equations\footnote{\small  Not yet
emphasizing the complications originating from  QCR.} for the
different structures. Expectably, these master equations are both
formally and physically different.

\smallskip

\noindent As it is emphasized above, the criterion for
classicality relies on the experimental evidence with the
photon-absorbing environment. Such environment (that quickly
thermalizes after the photon absorption) does not support
spontaneous emission for the atom in the ground state--there are
not photons in the field that could re-excite the atom. For this
specific, yet realistic, physical situation, the authors were able
to show that only the "rotating wave" Hamiltonian provides the
proper master equation [with notation adapted to eq.(136)] (see
their eq.(67)): %%
\begin{equation}
{d\rho_{ex} \over dt} = - \imath \omega_{\circ} [\sigma_{ex 3},
\rho_{ex}] + {1 \over 2} A_- (2 \sigma_{ex}^- \rho_{ex}
\sigma_{ex}^+ - \{\sigma_{ex}^+ \sigma_{ex}^-, \rho_{ex}\}).
\end{equation}

\noindent Formally, this is the master equation eq.(80). From
eq.(138) is calculated the stationary state photon emission rate,
$I_{ss} = 0$, which is  {\it in accordance} with the experimental
evidence. The other master equations, that correspond to  other
structures, give physically unreasonably large photon emission
[without any external driving]. Thereby, for the considered
physical {\it situation}, the {\it preferred structure} is
$\mathcal{S}_{rw}$.

\smallskip

\noindent The authors emphasize that a change of the criterion for
classicality, which would correspond to another physical
situation, would, in general, distinguish some other structure of
the composite system as "preferred structure".

\bigskip

\noindent {\bf 7.4 Outlook}

\bigskip

\noindent Interaction determines correlation of "system" and its
environment (Dugi\' c 1996, 1997); see Supplement for some
technical details. The kind and the type of  correlation
determines the preferred structure.

\smallskip

\noindent If the total system ("system+environment") is in
entangled state: %%
\begin{equation}
\vert \Psi\rangle_{SE} = \sum_i c_i \vert i\rangle_S \vert
i\rangle_E,
\end{equation}

\noindent

\noindent both the open system's and the environment's state are
unique for every structure.  Regarding the model of Section 7.2,
eq.(110), in accordance with the solutions to eqs.(111)-(112), the
total system's state in the asymptotic limit is tensor product [in
simplified form] $\vert \phi\rangle_{1+E_2} \vert \chi
\rangle_{2+E_2} = \sum_{\alpha} c_{\alpha} \vert \alpha\rangle_1
\vert \epsilon(\alpha)\rangle_{E_1}$ $\sum_{\beta} d_{\beta} \vert
\beta\rangle_2 \vert \epsilon(\beta) \rangle_{E_2}$. However, for
the structure, $S + E = (1+2) + (E_1 + E_2)$, the state takes the
form of eq.(139): %%
\begin{equation}
\vert \Psi \rangle_{SE} \equiv \vert \Psi\rangle_{12E} =
\sum_{\alpha,\beta} C_{\alpha\beta} \vert \alpha \rangle_1 \vert
\epsilon(\alpha)\rangle_{E_1} \vert \beta \rangle_2 \vert
\epsilon(\beta)\rangle_{E_2} \equiv \sum_k d_k \vert k\rangle_{12}
\vert k \rangle_E,
\end{equation}

\noindent where $C_{\alpha\beta} \equiv c_{\alpha} d_{\beta}$ and
$ k \equiv (\alpha, \beta)$, for both $S$ and $E$.

\smallskip

\noindent However, for the alternate  $A+B$ structure, eq.(140)
acquires the form: %%
\begin{equation}
\vert \Psi\rangle_{SE} \equiv \vert \Psi\rangle_{ABE} = \sum_k d_k
\vert k\rangle_{AB} \vert k \rangle_E,
\end{equation}

\noindent where, of course, the preferred states for the total
system $S$\footnote{\small The nonorthogonal Gaussian states
constitute an approximate pointer basis--compare to eq.(148).},
$\vert k\rangle_S = \vert k\rangle_{12} = \vert k\rangle_{AB},
\forall{k}$, but, in general, $\vert k \rangle_{AB}$ is not of the
tensor-product form relative to the $A+B$ structure. The
environment $E$ acts as a whole, simultaneously on both subsystems
$A$ and $B$--the subsystems $A$ and $B$ have  common environment
$E$. From Section 3.1 we know that, for at least some states
$\vert k \rangle_{12}$ in eq.(140), $\vert k \rangle_{12} =
\sum_{m} c_{km} \vert m\rangle_A \vert m\rangle_B$. So collecting
eq.(140) and eq.(141) we can write ($1+2=S=A+B$): %%
\begin{equation}
\sum_k d_k \vert k\rangle_{12} \vert k\rangle_E = \sum_k d_k
\left(\sum_{m} c_{km} \vert m\rangle_A \vert m\rangle_B\right)
\vert k\rangle_E,
\end{equation}

\noindent which clearly exhibits the {\it preferred states},
$\vert k \rangle_S$, {\it and} the {\it preferred structure},
$1+2$, of the open system $S$--$\vert k\rangle_S = \vert
k\rangle_{12} = \vert \alpha\rangle_1 \vert \beta\rangle_2$,
eq.(126)--which justifies the Clue from Section 7.1.

\smallskip

\noindent The model considered in Section 7.3 does not offer such
a clear picture on the choice of the preferred structure. For this
to be provided, solutions to {\it all}  master equations referring
to the different structures are needed. Needless to say, this is a
complicated task. Nevertheless, bearing in mind eq.(142), one can
expect analogous conclusions.

\smallskip

\noindent So we find the results of Sections 7.2 and 7.3 mutually
qualitatively consistent and also consistent with the told in
Section 7.1. This consistency encourages us to state

\smallskip

\noindent {\bf Conjecture 2}. {\it The environment (i.e. its
interaction with the open system) is responsible for existence of
the "preferred" structure of the open system.}

\smallskip

\noindent In this context, it becomes clear: there is not any
reason to claim or suppose existence of "preferred structure" (or
of the preferred states and/or observables) for a closed physical
system (i.e. for the Universe)--cf. Chapter 6 and Chapter 8 for
 details.

\smallskip

\noindent The above conjecture is in intimate relation with the
recent suggestion in (Harshman 2012a): "{\it the
physically-meaningful observable subalgebras are the ones that
minimize entanglement in typical states.}" Rigorously speaking,
this is {\it another criterion of classicality}, which is not
considered in this book: {\it minimum correlations} in open
system\footnote{\small The opposite, i.e. the requirement for the
maximum correlations, is essential for quantum information
processing. To this end  see (Fel'dman and Zenchuk, 2012).}. This
condition is already fulfilled for the model considered in Section
7.2: the state eq.(126) is without any correlations {\it relative}
to the preferred structure $1+2$.  Similarly, regarding the {\it
more general} considerations of the Markov open systems
(Arsenijevi\' c et al 2012), the minimum correlations as a
criterion of classicality distinguishes the model eq.(110).

\smallskip

\noindent In summary,  we can conclude: typically, the environment
[i.e. its {\it interaction} with the open system's degrees of
freedom--compare e.g. eq.(110) with eq.(129)] distinguishes
preferred structure of the open composite system. We conjecture
that this is a universal rule of the open systems theory.
Regarding the closed systems, of course, this is not the case, as
we elaborate in the next chapter.

\newpage

\noindent {\bf Chapter 8}

\noindent\textbf{\large Some Interpretation-Related Issues}

\

\vspace*{55mm}

\noindent Classical "phase space" of a physical system is {\it
unique}. All  degrees of freedom (and their conjugate momenta) of
the system, that are mutually related by  linear transformations,
belong to the same phase space.  This analogously  applies to the
quantum mechanical counterpart. As we have emphasized in Section
2.3, state space of a quantum system is unique Hilbert space.
Quantum state of a system is unique in every instant of time.

\bigskip

\noindent {\bf 8.1 Global irreducible structures with decoherence}

\bigskip

\noindent Let us consider the Universe as a closed quantum system.
By definition, there is nothing outside the Universe, including
"observer". To this end, there is no room for any operational
definition of the Universe preferred structure. Without any
additional condition/criterion, all structures of the Universe are
equally physically valid. Every structure, denoted $\sigma_i$, is
defined by a  set of observables, whose locality is adapted to the
 tensor factorization of the Hilbert state space.
All the structures share the same physical space and time, and
their dynamics are uniquely determined--there is unique (pure)
quantum state of the Universe in every instant in time.

\smallskip

\noindent Every structure $\sigma_i$ is composed of some
"elementary particles" and by structu\-re-specific fundamental
interactions (i.e. symmetries) between them. Every instantaneous
universal state can be expressed (cf. Section 3.1) in a unique way
for every  structure separately.

\smallskip

\noindent In every $\sigma_i$ structure, additional LCTs are
allowed {\it locally} to re-define the structure\footnote{\small
E.g., in our structure, instead of $e+p$, the hydrogen atom can be
described as $CM+R$--see Chapter 5.}. If a local transformation is
indexed by $\alpha$, then $\sigma_{i\alpha}$ represents the
$\alpha$th local variation of the $\sigma_i$ structure. This
subtle topic will be considered in Section 8.2.

\smallskip

\noindent Of all  possible structures, $\sigma_i$, we are
interested in the  structures, $\mathcal{S'}_n$, that are,
including our structure  (denoted $\mathcal{S}_{\circ}$), mutually
global and irreducible; $\{\mathcal{S'}_n\} \subset \{\sigma_i\}$.
For such structures, the respective sets of "elementary particles"
are mutually irreducible. Dynamics of such structures, although
unique on the level of the Universe, are mutually independent,
autonomous. Physical interactions, as well as the related symmetry
conditions, may be totally different (Anderson 1993, 1994,
Harshman 2012b, Manzano et al 2013). Of course, provided the full
details for one structure are known, one can mathematically
describe all  other structures, in full detail. Such structures
are (Section 2.2) mutually information-theoretically independent,
separated. Symmetry fixed for one structure uniquely determines
(induces) the symmetry rules for every other structure.

\smallskip

\noindent One can still ask if a local measurement in one
structure can represent a measurement of certain variables
characteristic for an alternative structure. This subtle question
regards both  local structures, which will be considered in
Section 8.2, as well as the issue of the "quantum reference
frame", which will be discussed in Section 8.3.2. One should still
keep in mind: "observation" [cf. Chapter 7] is local. So, {\it an
observer can directly "see" only the systems, which belong to the
structure he lives in}. Acquiring information about the subsystems
of the alternate structures is inevitably indirect and limited,
i.e. partial, Lemma 2.1.

\smallskip

\noindent Having in mind equal physical status of the considered
structures $\mathcal{S'}_n$, it is apparent: {\it observer cannot
operationally conclude which structure he is a part of}. Due to
invertibility of the LCTs, our structure is alternative relative
to the other structures. Which structure is then
primary\footnote{\small  Our structure may look like the $CM+R$
structure for some alternate structure of the Universe.}?

\smallskip

\noindent We place a special emphasis on the $\mathcal{S}_i$
structures which are subject to decohe\-rence-induced classicality
(for an example see Section 6.3).  This additionally shrinks the
set of the structures of interest for us: $\{\mathcal{S}_i\}
\subset \{\mathcal{S'}_n\}$. So we are interested in the
structures that are, {\it relative to each other as well as to our
structure}, global and irreducible, and of all of them, we are
interested only in the structures that {\it support decoherence}
for certain local degrees of freedom.

\bigskip

\noindent {\bf 8.2 Local structures and classicality}

\bigskip

\noindent In Section 8.1 we  distinguished the set of mutually
global and irreducible structures, $\{\mathcal{S}_i\}$, which
carry decoherence for some of their respective degrees of freedom.
Let us denote by $\mathcal{S}_{\circ}$ the one we belong to, and
the other by $\mathcal{S}_j, j=1,2,...$.

\smallskip

\noindent Every such structure is defined by a set of the
fundamental degrees of freedom, which can be subject to local
transformations of variables. An illustration is given by Example
2 in Section 2.2. Here we use notation of Section 2.2 to emphasize
 local transformations:
\begin{equation} \mathcal{S}_1 = \{1e,2e,1p,2p\} \to \mathcal{S}_2
= \{1H, 2e, 2p\} \to \mathcal{S}_4 = \{1CM,1R, 2CM, 2R\},
\end{equation}

\noindent for every pair of "neighbor" structures. Analogous
transformations can be performed for every structure
$\mathcal{S}_j$.

\smallskip

\noindent Grouping the particles and imposing certain boundary
conditions can lead to formation of a local alternative structure
$\mathcal{S}'_{\circ}$ for our structure. Physically, the new
subsystems can be  some composite particles, like mesons, atoms,
molecules, compounds, large bodies and so on. This  local
re-structuring can be performed for every  structure
$\mathcal{S}_j$.

\smallskip

\noindent How many local structures, $\mathcal{S}_{i\alpha}$, of
the $\mathcal{S}_i$ structure (here: $i=0,1,2,3,...$), can carry
decoherence? According to Sections 7.2 and 7.3, there is only one
such a  local structure for every $\mathcal{S}_i$. However, this
is only a plausible conjecture--cf. Conjecture 2 in Section 7.3.
Justifying/unjustifying this conjecture, i.e. defining local
"systems" for a structure $\mathcal{S}_i$, is an open issue yet
(Arsenijevi\' c et al 2012, Harshman 2012a, Zanardi
2001)\footnote{\small  e.g. The atoms presented by the structure
$\mathcal{S}_3 = \{1H, 2H\}$, where "$H$" denotes a hydrogen atom,
which, as a subsystem of the Universe, can be differently
decomposed. See also the  molecule structures $\mathcal{S}_1$ and
$\mathcal{S}_5$ in Section 4.1.}.

\smallskip

\noindent This subtlety of "what is 'system'?" (Dugi\' c and
Jekni\' c 2006, Dugi\' c and Jekni\' c-Dugi\' c 2008) clearly
exhibits limitations of the pure theoretical considerations.
Rather, {\it some phenomenological facts are needed}. It is not
surprising, as quantum mechanics offers much more than our
experience can support\footnote{\small Recall the efforts to
introduce the one-particle models, {\it in order to avoid quantum
entanglement}, for the composite systems in nuclear physics and
condensed matter physics.}.

\smallskip

\noindent As emphasized in Section 7.1:  local structural
variations {\it are meaningless for the already decohered
structure}. Decohered degrees of freedom (subsystems of a
composite system) are quasi-classical. For the quasi-classical
degrees of freedom, the transformations of variables become
nonphysical, a mathematical artifact. As it is emphasized
throughout this book (cf. e.g. Chapter 5): the center-of-mass
positions of the macroscopic bodies are quasi-classical and hence
regarded as (locally) physically {\it realistic}. Due to the very
meaning of "realistic"\footnote{\small "Realistic" serves
primarily to define what is "not realistic".},  linear
combinations of the centers of mass of the macroscopic bodies
cannot be realistic. This way, {\it we justify the classical
prejudice} distinguished in Section 2.1: our structure [{\it
phenomenologically}] supports decoherence, and decohered degrees
of freedom have classical reality. Of course, this reality is {\it
local}, i.e. of relevance and interest {\it only} for {\it our}
structure, which we are a part of--what's realistic in one
structure does not determine what is realistic in any other
structure.

\smallskip

\noindent Local structures, including the "classical" ones
provided by decoherence, are {\it reducible}. Observing  classical
structures gives rise to the classical prejudice (Section 2.1) and
directly results in classical intuition on finite decomposability
of the physical matter. Apotheosis of this position brings about
the naive reasoning as described in Section 2.1--classical
reasoning precludes the idea of the alternative quantum structures
$\mathcal{S}_j$.

\smallskip

\noindent Composition of our decoherence-defined structure of the
Universe is a consequence of a {\it dynamical change} of local
structures: the systems are constantly exchanging particles, some
systems are in formation while some other are splitting (or
decaying). This dynamical particles-exchange is an instance of the
trivial LCTs (Section 2.2.1). It provides dynamical local changes
in our structure and stresses the fact, that the time-independent
models, typical for the decoherence theory (cf. Chapters 6 and 7),
as well as for the standard open systems theory (Breuer and
Petruccion 2002, Rivas and Huelga 2011), are somewhat artificial.
In such dynamical system, some {\it effective} and approximate
structures are expected, and the physical description can be
complicated. Furthermore, there may appear some "emergent"
properties of the macroscopic bodies that are poorly understood in
physical sciences, but vastly referred to, in biology, economy,
social sciences, psychology etc. (Auyang 1998).

\smallskip

\noindent It is worth repeating: everything told for our structure
$\mathcal{S}_{\circ}$ and local observations can be in principle
applied to every other structure $\mathcal{S}_i$. While we do not
claim existence of "observer" in any of the alternate structures,
we use this standard vocabulary to highlight  physical equivalence
of the Universe structures we are interested in.

\bigskip

\noindent {\bf 8.3 A unifying physical picture}

\bigskip

\noindent The group-theoretic character of the LCTs can formally
link the structures, which are endowed by decoherence. However,
this mathematical possibility is physically irrelevant. Actually,
according to Section 8.2: performing LCTs on the local
(decoherence endowed) structures is physically pointless. So one
can say:

\smallskip

\noindent {\it The group-theoretic character of LCTs does not
imply} physical {\it reducibility of all structures onto one and
only one structure of the Universe.}

\smallskip

\noindent Nevertheless, there is a subtlety, which might not be
obvious. In Section 8.3.1 we connect the microscopic and
macroscopic domains for our structure of the Universe. In Section
8.3.2, we discuss the recently elevated issue of "quantum
reference frames".

\bigskip

\noindent {\bf {\emph {8.3.1 Microscopic vs macroscopic domain}}}

\bigskip

\noindent The microscopic physical domain is phenomenologically
described in Chapter 5. In theoretical analysis, all  kinds of
LCTs are in principle allowed--that constitutes the core of the
"what is 'system'?" issue (Dugi\' c and Jekni\' c 2006, Dugi\' c
and Jekni\' c-Dugi\' c 2008).

\smallskip

\noindent The "macroscopic domain" is the classical physics
domain. Its quantum mechanical origin seems naturally to appear
within the universally valid quantum mechanics (Giulini et al
1996, Zurek 2003, Schlosshauer 2004). For the Universe as a whole
there may be different structures, whose local variations can be
endowed by decoherence for certain (local) degrees of freedom. We
can think of these structures in analogy with our own, which, in
turn, is a dynamical system with the local particles exchange,
i.e. with weakly defined border-line between the systems, and
between the systems and their environments. Of course,
classicality of the alternative structures is a matter of the
"parallel occurrence of decoherence", which is as yet rigorously
established only for the quantum Brownian motion model, Section
6.3.

\smallskip

\noindent In our structure, which is paradigmatic for our
considerations, there is continuous exchange of particles between
the systems. The Universe split into subsystems is subtle and, as
yet, not well known a topic. This may be a sign for a need for a
new methodology, which could encompass all the subtleties
regarding the many-particle systems structures.

\smallskip

\noindent In the absence of such methodology, below we collect the
findings from the previous Chapters as kind of "algorithm" for
defining the possible structures of the Universe:

\smallskip

\noindent (a) For our structure, we phenomenologically learn about
the set of  elementary particles and their local compositions such
as the atoms, molecules etc. Its dynamics allows local structural
variations on the both micro- and macro-scopic level (domain).
Regarding the macroscopic domain, we recognize decoherence as the
{\it fundamental} and also {\it universal} quantum mechanical
process. This process provides the (quasi)classical dynamics of
certain degrees of freedom and defines what's "local" and
"realistic" in a given instant in time. This bases our classical
intuition, and, as yet, the dominant view of the quantum Universe.

\smallskip

\noindent (b) Perform the LCTs of the  fundamental (microscopic)
degrees of freedom of our structure. Of all thus obtained
structures, consider those that are global and irreducible
relative to each other as well as to our structure. Of all such
structures, choose only those supporting decoherence for some of
their local degrees of freedom\footnote{\small In principle,
existence of such structures is provided in Section 6.3.}.
Everything told in the (a) above, should be analogously expressed
for  such alternate structures of the Universe. The structures of
the kind are mutually information-theoretically separated, and,
due to decoherence in all of them, application of the LCTs on
their (quasi)classical structures loses its physical meaning and
relevance.

\smallskip

\noindent (c) An observer belongs to one and only one such
structure. The only degrees of freedom he is able directly to
observe are the parts of the structure he belongs to. While
existence of  intelligent observer is not required for all the
structures (Dugi\' c et al 2002), we use this terminology to
highlight physically equal status of all the structures  as well
as to emphasize the possibility to choose the classical reference
frame in every of them. Some subtleties regarding the recently
raised issue of quantum reference frames will be considered in the
next section.

\pagebreak

\noindent {\bf {\emph {8.3.2 The quantum reference frame issue}}}

\bigskip

\noindent Quantum decoherence provides quasiclassical behavior of
certain open-sys\-tem's degrees of freedom. "Observer" is usually
assumed to be a classical system that can collect (classical)
information about the open system.

\smallskip

\noindent However, "observation" assumes existence of "reference
frame" where "observer" is spatially placed. Hence the familiar
classical reference frames. To this end, it is important to
stress: if $x_S$ is a system's position observable, its
measurement from the {\it O} reference frame is defined by the
$x_S - X_O$ variable, where  $X_O$ is the reference-frame
position, which is a {\it classical} variable--a $c$-number. In
order to highlight this definition, we introduce the standard
"hat" mark for the quantum mechanical observables: $\hat x_S -
X_O\hat I$, where the $\hat I$ is the identity operator. So it is
clear: the variable $\hat x_S - X_O\hat I$ is {\it not} obtained
via canonical transformation. If there are more than one reference
frame, then $\hat x_S - X_i\hat I$ denotes the system's position
seen from the $i$th (classical) reference frame. Transitions from
a classical reference frame to another one are the standard
symmetry transformations--e.g. the spatial translation--that do
not change  structure of the observed system.

\smallskip

\noindent However, the {\it quantum} reference frames refer to the
microscopic (quantum) systems, which are not subjected to
decoherence. This raises non-trivial  questions, e.g., as to how
the electron in the hydrogen atom can see the atomic proton, and
{\it vice versa} (for a similar considerations see, e.g., (Angelo
et al 2011)). Then, by definition, the proton's position measured
by the electron is defined as $\hat{\vec{r}}_p - \hat{\vec{r}}_e$.
Notice that the reference-system's position is not a classical
$c$-number but a {\it dynamical} quantum observable,
$\hat{\vec{r}}_e$. For a many-particle system, if the particle $1$
is the reference system, then the position observables of all the
other particles are defined in the $1$ reference system: %%
\begin{equation}
\hat{\vec{\rho}}_{i}^1 = \hat{\vec{r}}_i - \hat{\vec{r}}_1.
\end{equation}

\noindent As distinct from the classical reference frames, the
quantum reference frames (QRF) raise a number of interesting
observations and open questions. Below, we consider one out of
plenty of structures $\mathcal{S}_i$ introduced in Section
8.1--e.g. our structure $\mathcal{S}_{\circ}$. We consider
exclusively the QRFs belonging to the same Universe structure
$\mathcal{S}_{\circ}$\footnote{\small Everything equally applies
to the case that there existed a preferred structure of the
Universe.} of the Universe.

\smallskip

\noindent First, it is obvious that the QRF-defined variables,
eq.(144), are obtained via the global LCTs, which introduce a
specific kind of the "relative positions" observables for the
considered structure.

\smallskip

\noindent Second, the different QRFs perceive the different
structures of the $R$ system. To see this, we emphasize that, in
principle, the structure's $CM$ system
 cannot be observed
from a local QRF (Angelo et al 2011)\footnote{\small See also
Chapter 6.}. So there remains the "relative positions"--i.e. the
$R$ system is the {\it only} one that (of course, locally) can be
observed.

\smallskip

\noindent For a quantum observer in the $1$ QRF system, the $R$'s
subsystems are described by the relative positions eq.(144). As a
consequence, every distance-dependent interaction\footnote{\small
We drop the "hat"-mark for the observables.} $V(\vert \vec{r}_i -
\vec {r}_1\vert)$ is an external ("classical") field, $V(\vert
\vec{\rho}^1_i\vert)$, for the $i$th particle. The interactions
$V(\vert \vec{r}_i - \vec {r}_j\vert), i,j\neq 1$ remain
interactions. However, for an observer in the $2$ QRF system, the
physical picture is different. Then $V(\vert \vec{r}_i - \vec
{r}_2\vert)$ become the external fields, while the $V(\vert
\vec{r}_i - \vec {r}_j\vert), i,j\neq 2$ remain  interactions in
the total system Hamiltonian.

\smallskip

\noindent Of course,  different interactions produce  different
correlations between the constituents of the structure's $R$
system. In effect, two quantum observers perceive  two {\it
different structures} of the  $R$ system. These structures are
global to each other, but are local to the $\mathcal{S}_{\circ}$
structure, which consists also of the $CM$ system. So, according
to the quantum correlations relativity, Section 3.2, we conclude
about the different correlations at the
 quantum observers disposals.

 \smallskip

\noindent Now, quantum correlations relativity points out the
following striking observation: as "quantum observer" can perceive
only the
 $R$ system, the correlations in the $R$ system he can
perceive need not be present in the considered structure
$\mathcal{S}_{\circ}$, eq.(145) below.

\smallskip

\noindent Finally, the requirement that "{\it all} [QRF] {\it
perspectives must agree on the probability of detector clicks}"
(Angelo et al 2011) does not seem to resolve the following, now
naturally appearing foundational question.

\smallskip

\noindent Assume that the structure $\mathcal{S}_{\circ}$  is
defined by the set $\{x_m, p_m\}$ of the fundamental degrees of
freedom. The fact that no QRF can see the structure's $CM$ system,
but only the "relative positions", which are different for the
different reference systems, raises the following question:

\smallskip

\noindent {\it is there a fundamental lack of information about
the Universe structure for  QRFs?}

\smallskip

\noindent Having in mind Entanglement Relativity, i.e. the
possible entanglement between the structure's $CM$ and the
QRF-defined "internal" degrees of freedom, we re-phrase the above
question:

\smallskip

\noindent {\it may it be the case, that the [inevitably local]
QRFs cannot provide the fundamental description of any of the
Universe structures?}

\smallskip

\noindent Formally, introduction of the QRFs gives rise to the
following structure transformation: %%
\begin{equation}
\mathcal{S}_{\circ} = \{x_m, p_m, m=1,2,...,N\} \to
\mathcal{S}_{\circ}^{(i)} = \{CM_{\circ}, R^{(i)}_{\circ}\}
\end{equation}

\noindent for the $i$th QRF. Regarding the universal state for the
considered structure of the Universe, entanglement may be expected
for the $\mathcal{S}_{\circ}^{(i)}$ structure: %%
\begin{equation}
\sum_p c_{pi} \vert p \rangle_{CM}\ \vert p\rangle_{R^{(i)}},
\quad \forall{i}.
\end{equation}

\noindent Then for the $i$th "quantum observer", quantum state of
the degrees of freedom in his disposal is {\it mixed}: %%
\begin{equation}
\rho_{R^{(i)}} = \sum_p \vert c_{pi}\vert^2 \vert
p\rangle_{R^{(i)}}\langle p \vert.
\end{equation}

 \noindent That is,

 \smallskip

\noindent  {\it Even the universal validity of the Schr\" odinger
law may be
  at stake for a quantum observer.}

  \smallskip

\noindent Leaving these questions out of further consideration, we
stick to the standard understanding of the classical reference
frames and of "classical observer" as presented above and used
throughout this book.

\bigskip

\noindent {\bf {\emph {8.3.3 The unifying picture}}}

\bigskip

\noindent Bearing in mind that all  basic concepts--of the
composite system's degrees of freedom, locality, correlations,
classical reality--are {\it relative}, i.e. {\it structure
dependent}, we face the following physical picture of the
Universe:

\smallskip

\noindent {\it The Universe {\it hosts} a number of dynamical
structures. All these structures are equally described by quantum
mechanical formalism. While the local\footnote{\small The only
"global" physical law valid for all  structures is the Schr\"
odinger law.} laws and symmetries may be different, physical
reality of the structures cannot be a priori rejected. Of all
structures,  we consider only those that are global and
irreducible relative to each other as well as to our structure,
and that support decoherence for some of their local degrees of
freedom. An observer, in principle, cannot say which structure he
belongs to. An observer is a part of one and only one structure
and cannot directly observe subsystems of the alternative
structures of the Universe}.

\bigskip

\noindent {\bf 8.4 Some interpretational issues}

\bigskip

\noindent {\it Prima facie}, Section 8.3.3 may seem to re-phrase
the Ithaca interpretation of quantum mechanics (Mermin 1998).
However, as distinct from the Ithaca interpretation, we consider a
limited set, $\{\mathcal{S}_i\}$ (Section 8.1), of the Universe
structures. In other words, we go beyond  the Ithaca
interpretation: not arbitrary structures of the Universe are
relevant for our study. Only the structures that are mutually
nontrivial, global and irreducible, and endowed by decoherence,
are of interest. Thus Lemma 2.1 makes our conclusions irreducible
to the Ithaca interpretation--our conclusions are basically in
agreement with phenomenology.

\smallskip

\noindent Bohmian theory (Durr et al 2012) contradicts our
considerations. In Bohmian theory   existence of the fundamental,
ontological structure of the Universe is postulated. The
transformations of variables are mathematical artifacts. This
contradiction tackles the issue of completeness of the standard
quantum theory. For our position see Section 8.6.

\smallskip

\noindent As the Complementarity principle [but properly
understood, cf. (Dugi\' c 2012)] remains intact by our
considerations, we believe that the standard Copenhagen
interpretation, as well as the collapse-based interpretations, are
in no conflict with quantum correlations relativity.

\smallskip

\noindent However, this does not apply to the Everett Many Worlds
Interpretation (MWI) as we are going to reproduce from (Jekni\'
c-Dugi\' c et al 2011).

\bigskip

\noindent {\bf {\emph {8.4.1 Non-branching of the Everett worlds
}}}

\bigskip

\noindent It is a universal requirement in the context of
interpretation of quantum mechanics: every physically reasonable,
even {\it gedanken}, situation must be consistent with the
interpretation foundations. This, however, is not the case with
the Everett MWI {\it in the context} of the quantum Brownian
motion model, Section 6.3.

\smallskip

\noindent To see this, we first show, that  "branching" of one
structure excludes the alternative-structure "branching". Consider
a decoherence-induced "history" for the $S+E$ structure of the QBM
setup for the different instants of time, $t_{\circ} < t_1 < t_2$,
as required by Everett interpretation: %%
\begin{eqnarray}
&\nonumber&  \vert x_S(t_{\circ}), p_S(t_{\circ})\rangle_S \vert
\epsilon (x_S(t_{\circ}), p_S(t_{\circ})) \rangle_E \to \vert
x_S(t_1), p_S(t_1)\rangle_S \vert \epsilon (x_S(t_1), p_S(t_1)
\rangle_E
\\&&
\to \vert x_S(t_2), p_S(t_2)\rangle_S \vert \epsilon (x_S(t_2),
p_S(t_2)) \rangle_E.
\end{eqnarray}

\noindent In eq.(148), we introduce the tensor-product states for
the subsystems $S$ and $E$ as a consequence of the
decoherence-induced "branching"\footnote{\small This is essential
for the Everett MWI in order to be able [at least approximately]
to mimic the "state collapse".}. The "history" eq.(148) describes
dynamics of one, out of  plenty, of Everett worlds. The states of
the $S$ system are Gaussian states [not necessarily of the minimal
uncertainty]\footnote{\small See, e.g. Wallace, chapter 1 in
(Saunders et al 2010).}. These states represent the approximate
pointer basis, i.e. the preferred set of (non-orthogonal) states
for the open system $S$. The environment states, that appear in
eq.(148), bring information about the open system's states.

\smallskip

\noindent According to Section 6.3, there exists another
structure, $S'+E'$, for which the open system $S'$ undergoes
quantum Brownian motion. That is, both  open systems ($S$ and
$S'$) are  quantum Brownian particles.

\smallskip

\noindent Now, due to Entanglement Relativity, Section 3.1, it
becomes clear and {\it unavoidable}: at least some of the
instantaneous states in eq.(148) will be endowed by
entanglement--i.e. are {\it non-branched}--for the $S'+E'$
structure. So, Everett branching for the $S + E$ structure
excludes Everett branching for the $S'+E'$ structure. More
precisely: for a time interval for which the $S+E$ structure is
branched, the $S'+E'$ structure cannot be branched. Due to the
assumption that branching is fast, i.e. of the order of the
decoherence time (Schlosshauer 2004, Saunders et al 2010),
non-branching for the $S'+E'$ structure refers to the most of the
composite system's dynamics.

\smallskip

\noindent Physical equivalence of the two structures (see Section
8.2) directly provides the following observation:  Everett
branching for the $S + E$ structure excludes Everett branching for
the $S'+E'$ structure, and {\it vice versa}. As the only
consistent statement now appears:

\smallskip

\noindent {\it World branching is not allowed for the QBM
structures $S+E$ and $S'+E'$.}

\smallskip

\noindent So we conclude (Jekni\' c-Dugi\' c et al
2012)\footnote{\small Note that neither ER nor POD separately are
sufficient for the conclusion. Even ER+POD is not sufficient. The
point is that ER+POD, {\it when} applied to a pair of mutually
global and irreducible structures ($S+E$ and $S'+E'$), makes the
case: the Everett interpretation is not applicable to the QBM
model.}:

\smallskip

\noindent {\it There is at least one physically relevant model of
a composite system in decoherence theory which cannot be described
by the Everett interpretation.}

\bigskip

\noindent {\bf {\emph {8.4.2 Emergent structures and decoherence
}}}

\bigskip

\noindent Decoherence is typically studied starting from a fairly
unprincipled choice of system-environment split. In this sense,
decoherence is by its nature an approximate process and so the
process of branching is likewise approximate. In other words
(Wallace, chapter 1 in (Saunders et al 2010)) [our emphasis]:
"{\it ...decoherence is an emergent process occurring within an
already stated microphysics: unitary quantum mechanics. It is} not
{\it a mechanism to define a part of that microphysics}".

\smallskip

\noindent Within this new wisdom, one may suppose that there
should be an emergent structure for the QBM model of Section 6.3,
i.e. that world-branching refers to some "emergent" Brownian
particle, $B$, {\it not} directly to the "microscopic", $S$ and
$S'$, Brownian particles. In the absence of a general {\it
physical} definition of "emergent properties" (i.e. of the "higher
level ontology") of complex systems (Auyang 1998), we are forced
to speculate about the possible ways to obtain a
branching-eligible structure for the QBM model. To this end, we
are able to detect only two possibilities. We find both of them
inappropriate for defining an emergent QBM structure.

\smallskip

\noindent We distinguish the following bases for emergentism.
First, it is
 dynamical exchange of particles between the "system" and the
"environment", which  encompasses the standard choice of the
"dividing line" in the von Neumann sense (the von Neumann "chain",
(von Neumann 1955)). Second, one may suppose, that there is an
alternate, third structure providing an emergent Brownian
particle, $B$, for the pair of Brownian particles, $S$ and $S'$.
To see that the first doesn't work for the QBM model is
straightforward. Actually, both Brownian particles are
one-dimensional and there is not, by definition, any possibility
of exchanging particles of the $S$ system with the environment $E$
(or of the $S'$ system with the environment $E'$); of course, due
to irreducibility of the two structures, exchange of the particles
between the $S$ system with the environment $E'$ (i.e. of the $S'$
system with $E$) is not even defined. The variant that an
environmental oscillator takes the role of Brownian particle is
also not allowed. For both structures, the environmental particles
do not mutually interact and therefore there is not a properly
defined environment for the variant--not even to mention that this
{\it a priori} excludes the possibility (Section 6.3) that the $S$
system is a "free particle" (not an oscillator).

\smallskip

\noindent The second option is a bit more subtle yet. To this end
we justify the statements of Sections 2.2.2 and 6.5: (1) obtaining
information about one Brownian particle (e.g. $S$) provides no
information about the other one (e.g. the particle $S'$); (2)
there does not exist any observable, $X_B$ (of the subsystem $B$
of the composite system $C$), which could approximate a pair of
observables of the two Brownian particles, $S$ and $S'$. In
effect, there does not exist any structure $B+E_B$ that
 could be
emergent for the structures $S+E$ and  $S'+E'$.

\smallskip

\noindent Regarding the point (1), we first remind (cf. Section
3.1): the $S'$ system is the original-structure's ($S+E$'s)
center-of-mass. So the position-observable of the $S'$ system is
subject to Lemma 2.1. Thereby one can say: Brownian particles, $S$
and $S'$, cannot approximate each other, neither there is any
information flow between them.

\smallskip

\noindent The arguments for the point (1) apply to the point (2).
As the only probability density that can provide probability
density for arbitrary subsystem  is the universal state, $\vert
\Psi \rangle$, there is not any subsystem's ($B$'s) probability
density, $\rho(X_B, X'_{B})$, that could provide probability
density for both the $S$ and the $S'$ systems. e.g. The definition
$X_B = f(x_S, x_{S'})$  gives rise to the probability density
$\rho(X_B, X'_B) = \rho (x_S,x'_S, X_{S'}, X'_{S'})$, which, as
emphasized in the proof of Lemma 2.1, cannot provide the
probability densities $\rho (x_S, x'_S)$ or $\rho (x_{S'},
x'_{S'})$ by integrating over $X_{S'}$ and $x_{S}$, respectively.
So, there is not any observable of the $B$ system whose
measurement might approximate simultaneous measurement of any pair
of observables for the two Brownian particles, $S$ and $S'$.
Physically, this means that we cannot imagine a third system $B$,
which undergoes Brownian-motion-like dynamics {\it and} can
approximately describe both "microscopic" Brownian particles, $S$
and $S'$. As we cannot recognize any other basis for emergentism,
we are forced to conclude that the above-distinguished
inconsistency between the QBM model and the modern Everett
interpretation remains intact.

\smallskip

\noindent Finally, we emphasize: the standard QBM model, Section
6.3, is a (paradigmatic theoretical) decoherence model pertaining
to the realistic macroscopic situation of "Brownian motion". There
are not any structural phenomenological facts about Brownian
motion known to us that go beyond the standard QBM model--there is
no need for any "emergent" Brownian particle.

\smallskip

\noindent Bearing this in mind, the possibility that the
structures considered in Section 6.3 are not susceptible to the
Everett interpretation directly raises the following foundational
question: Whether or not decoherence is sufficient for the Everett
branching? If it is, then the told above is unavoidable. If not,
then some additional requirement for branching, i.e. for
completeness of the Everett interpretation, is needed. e.g. One
may require some amount of "complexity" of the composite system to
be subject to the modern Everett interpretation (Saunders et al
2010). Certainly, then the range of applicability of the modern
Everett interpretation shrinks, as distinct from the competitive
interpretations. As the "additional requirement" is not a part of
the present state of the art in the field, we will not elaborate
on this any further, and we finally return to the conclusion of
Section 8.4.1.

\bigskip

\noindent {\bf 8.5 There are no "particles"}

\bigskip

\noindent Physical picture presented in Section 8.1 strongly
suggests:

\smallskip

\noindent {\it There are no "particles".}

\smallskip

\noindent "Particles" pertains to  some special states of the
Universe. The kind and  behavior of the elementary particles is
structure dependent, and is otherwise determined by the Universe
global symmetry as well as by the local laws (interactions) that
are characteristic for the structure. Decoherence independently
occurs in the different structures, and the {\it contents} of
physical reality is relative, i.e. structure dependent.

\smallskip

\noindent Section 8.2, and the parallel occurrence of decoherence,
Section 6.3, naturally support the position, that "there are no
particles" even for the local structures, i.e. on the
lower-ontological level (Zeh 1993, Primas 1994).

\smallskip

\noindent This position naturally describes certain experiments
without raising any further puzzles. Some points presented below
are already raised in Chapter 5.

\bigskip

\noindent{\bf {\emph {8.5.1 Delayed choice experiments }}}

\bigskip

\noindent In order to exhibit weirdness of the quantum world,
Wheeler (Wheeler 1978) emphasized, that classical reasoning can
lead to inconsistency with quantum mechanical conclusions. A
recent elaboration due to Peres (Peres 2000) abandons the
classical prejudice on individuality of "quantum particles".
Instead, Peres distinguishes operational reality of quantum
entanglement and implicitly points out entanglement relativity.

\smallskip

\noindent In this picture, that is experimentally tested (Ma et al
2012), entanglement relativity, Section 3.1, naturally appears.
Everything can be expressed in terms of correlations for different
partitions  of the composite system (entanglement swapping),
without even mentioning the constituent "particles" (i.e. the
qubits).\footnote{\small In a sense, physical picture is easily
described in terms of "correlations without correlata" (Mermin
1998).}

\smallskip

\noindent Of course, the use of the concept of particles may be
physically correct. Our point is, that it is {\it neither
necessary nor the simplest description} of the
entanglement-swapping-based delayed-choice experiment.

\bigskip

\noindent{\bf {\emph {8.5.2 Interaction-free quantum measurements
}}}

\bigskip

\noindent Recently, based on the "interaction-free measurement",
see e.g. (Elitzur and Vaidman 1993), a theoretical proposal for
"direct counterfactual quantum communication"--that claims that
there may be communication without any particles exchange between
the parties--has been formulated (Salih et al 2013). While quantum
communication without the particle exchange is interesting, this
does not provide any spectacular result in "quantum mechanics
without particles". Rather, as we briefly point out below, it
provides another argument "against particles".

\smallskip

\noindent All the phrases and spectacular statements simply
disappear if we abandon the classical prejudice, which underlies
the phrase "interaction-free" in the original theoretical proposal
(Elitzur and Vaidman 1993). In the interference situations, there
is not "particle trajectory". Finding a particle in an arm of the
interferometer (provided by the click of a detector) does {\it
not} necessarily mean that the particle was there before
detection.

\smallskip

\noindent The following picture removes the puzzles and the
spectacular statements: {\it Every} detector {\it is in
interaction} with the  system of interest (a photon). This gives
rise to entanglement of {\it all} the detectors with the system,
{\it without assumption on the definite spatial position of the
system--before, during, or after the measurement}--even if the
system's position is measured. By applying e.g. the von Neumann's
projection postulate (von Neumann 1955), one easily obtains the
standard final state for every detector and for the system after
the measurement. A detector's click is a local effect that neither
precludes nor implies existence of a "particle" in any arm of the
interferometer before [or even after] detection. The puzzling
click of one and only one detector is a particular instance of the
long-standing problem of quantum measurement--the apparent state
collapse--but not more or less than this.

\pagebreak

\noindent{\bf {\emph {8.5.3 Relativistic quantum processes }}}

\bigskip

\noindent Although it is not subject of our considerations,
certain relativistic quantum effects provide striking
confirmations of relativity of structure as well as of the "there
are no particles" position.

\smallskip

\noindent Quantum-particles annihilation/creation is really
striking. e.g. A pair "electron + positron" transforms into "pair
of photons". In such process, one cannot, even in principle, say
that the pair "electron + positron" can be decomposed or  imagined
to consist of a pair of photons, and {\it vice versa}. The place
of quantum vacuum, which is responsible for the effect, is yet to
be properly described in the canonical formalism of Chapter 2,
(Stokes 2012).

\bigskip

\noindent {\bf 8.6 The universally valid and complete quantum
theory}

\bigskip

\noindent The phrase "there are no 'particles' ", Section 8.5,
naturally fits with the hypothesis of the universally valid {\it
and} complete quantum mechanics. In this section we adopt this
hypothesis and extend the picture obtained in Section 8.3.3.

\bigskip

\noindent{\bf {\emph {8.6.1 Why universally valid and complete
quantum theory? }}}

\bigskip

\noindent Throughout  this book, we respect the hypothesis of the
universally valid quantum mechanics by employing universal
validity of the Schr\" odinger law for the closed, isolated,
quantum systems. Of course, our findings are susceptible to
different interpretations. Then, one can ask the question from the
title: which arguments may justify the choice of universally valid
and complete quantum mechanics?

\smallskip

\noindent The arguments are as follows.

\smallskip

\noindent First of all, our approach is {\it minimalistic}: we do
not introduce or add any additional assumption or hypothesis. In
this context, it is easiest to get rid of the problematic concepts
of "particles", "individuality" and "classical intuition" and to
try to  {\it derive} them as the approximate and relative
concepts, whose contents are different for the different
decompositions (the structures) of the Universe.

\smallskip

\noindent Second, modern open systems theory (Breuer and
Petruccione 2002, Rivas and Huelga 2011) provides, that
practically every physically reasonable dynamics of a system can
be described by the unitary (Schr\" odinger) dynamics on the
extended system "system+environment". The inverse, however, as
yet, is not the case--unitary dynamics is not derived from the
open system's dynamics. Therefore, the unitary quantum mechanics
is methodologically more primary than the open system's theory.
Furthermore, the unitary quantum theory encompasses the collapse
models (Markovian\footnote{\small See e.g. (Bassi et al 2013).},
or non-Markovian), and still can describe the models not
presenting the state collapse. The inverse, however, is not the
case. Therefore we choose the universally valid quantum mechanics.

\smallskip

\noindent Third, modern quantum information theory provides the
following {\it conjecture}: quantum state saturates the
information contents of a quantum system (Brukner and Zeilinger
1999, Pusey et al 2012). That is, it is conjectured that every
possible information about a system can be drawn from the quantum
state--there is no room for "hidden variables" of any kind,
including those of the modern Bohm's theory (Durr et al
2012)\footnote{\small This means that the Bohmian theories are not
"deeper" than the standard quantum mechanical theory.}. Therefore
we choose complete quantum theory.

\smallskip

\noindent So, modern open systems and information theories
strongly support the hypothesis of the universally valid and {\it
complete} quantum mechanics. Nevertheless, as we show in Section
8.4, the Everett interpretation is at stake. Hence a new view of
the quantum world is needed. For a hint see (Dugi\' c et al 2012).

\pagebreak

\noindent{\bf {\emph {8.6.2 Completing the picture}}}

\bigskip

\noindent Now, the unifying picture of Section 8.3.3 may be
extended by the phrase:

\smallskip

\noindent "{\it There are  particles neither on the most
fundamental, i.e. on the ontological, physical  level nor on the
level of the Universe decompositions (structures). All that we can
assume is a fundamental quantum field, whose states can be
[non-relativistically] described by  different decompositions of
the Universe and their local, decoherence-defined structures.}"

\smallskip

\noindent While this picture may seem pessimistic, it is not
necessarily so. Actually, we do not think that ontological
existence of the Universe, seen as a fundamental physical quantum
field, should be considered to be non-realistic. Such an option
(Vedral 2010) is essentially an {\it additional condition}, which
is absent from the universally valid quantum theory.

\newpage

\noindent {\bf Chapter 9}

\noindent\textbf{\large Outlook and Prospects}

\vspace*{55mm}

\noindent In Chapter 1 we posed the questions that are worth
repeating: Is there unique fundamental structure of a composite
quantum system? How do the classical structures (and intuition)
appear from the quantum substrate? Can the structural variations
 be of any practical use that is not known to
the classical physics wisdom?

\smallskip

\noindent Issued answers are partial--some answers imply another
questions, which nevertheless sharpen our view of the quantum
world. The only assumption of our considerations is the universal
validity of quantum mechanics--for an isolated (closed) quantum
system we assume validity of the Schr\" odinger law. The recently
provided technical tools we introduce and use are as follows: (a)
Quantum correlations relativity (Section 3.2); (b) Parallel
occurrence of decoherence for the quantum Brownian motion model
(Section 6.3); (c) Preferred local structures for bipartite
decompositions (Section 7.2 and 7.3). Those are the corollaries of
the universally valid quantum mechanics.

\smallskip

\noindent In Chapters 6 and 8, we found the classical intuition
(emphasized in Section 2.1) "mechanistic"--structure is a
fundamental and  ontological notion that precludes a deeper
physical analysis of certain structural variations. However, in
the quantum context it seems that there is not any reason to claim
existence of the ontologically unique structure of the Universe.
Even more, there may be more than one structure that bears the
decoherence-induced classicality. In other words: there may be
more than one classical world (a structure) hosted by one and only
one, unique quantum Universe. For every such quasiclassical world,
classical intuition (Section 2.1) is justified as a local rule,
which does not preclude reality of the alternative worlds and
their local (internal), quasiclassical structures and physical
laws. Parallel occurrence of decoherence is conjectured
(Conjecture 1, Section 6.4) for all linear models. Whether this
conjecture can be justified, and probably extended to the more
general models of the many-particle systems, remains an open
question of our considerations.

\smallskip

\noindent An observer belongs to one and only one such world and
can only partly observe the alternative quantum worlds. From a set
of the possible local structures of a composite system, the
environment chooses {\it the} preferred structure, Conjecture 2
(Section 7.4). This structure can be directly observable
(accessible, Def.5.1) for an observer. In effect, the preferred
structure of an open composite system can be considered to be
"objective" and "realistic" for the observer. The choice of the
preferred local structure of a composite system cannot be provided
on the purely theoretical basis--some phenomenological facts are
needed. Occurrence of decoherence should be equipped with another
theoretical tools, which, as yet we can only speculate about. To
this end, we introduce the assumption, that the minimum quantum
correlations should be required for "classicality" (Arsenijevi\' c
et al 2012, Harshman 2012a), which also opens the following
speculation. May it be the case that "classicality" is mainly a
matter of structure--i.e. that structural studies may provide a
basic clue for answering the long-standing problem of the
transition from quantum to classical; compare to (Ragy and Adesso
2012)? Needless to say, this new perspective offers a basis for a
brand new approach to some old foundational questions in
non-relativistic quantum theory.

\smallskip

\noindent Quantum interpretation studies provide a unique lecture:
"classicality", as we usually see or feel it, may be idealized.
 Once we better understand
"classicality", we might be in a better position to perceive and
eventually to solve the long standing problem of quantum
measurement. To this end, the quantum structures studies may
nontrivially help, as we already know that "system", locality and
correlations--are {\it relative}, i.e. the structure-dependent
concepts.

\smallskip

\noindent Unfortunately, [as emphasized above], quantum formalism
seems to be much  richer than we might ever need. So there does
not seem to be any other way but to refer to phenomenology (e.g.
to the decoherence and quantum information phenomenology) as a
precursor as well as to validate our theories. Experimental
evidence, even more, the use, of entanglement relativity, we
believe, is  a precursor for the new and exciting applications
that will emerge from the quantum structure studies.

\newpage

{\flushleft {\bf References}

\bigskip

Adesso G. and Datta A., 2010, Quantum versus classical
correlations in Gaussian states. {\it Phys. Rev. Lett.} {\bf 105},
030501

Amann A., 1991, Chirality: A Superselection rule generated by the
molecular environment?. {\it J. Math. Chem. } {\bf  6}, 1

Anderson A, 1993, Quantum Canonical Transformations and
Integrability: Beyond Unitary Transformations. {\it Phys. Lett.} B
{\bf 319}, 147

Anderson A, 1994,  Canonical Transformations in Quantum Mechanics.
{\it Annals Phys.} {\bf 232}, 292

Angelo R. M., Brunner N., Popescu S., Short A. J. and Skrzypczyk
P., 2011, Physics within a quantum reference frame. {\it J. Phys.
A: Math. Theor.} {\bf 44}, 145304

Anglin J. R., Paz J. P., Zurek W. H., 1997, Deconstructing
Decoherence. {\it Phys. Rev. A} {\bf 55}, 4041

Arsenijevi\' c M., Jekni\' c-Dugi\' c J. and Dugi\' c M., 2012,
Zero Discord for Markovian Bipartite Systems. arXiv:1204.2789
[quant-ph]

Arsenijevi\' c M., Jekni\' c-Dugi\' c J. and Dugi\' c M., 2013a,
Asymptotic dynamics of the alternate degrees of freedom for a
two-mode system: An analytically solvable model. {\it Chin. Phys.
B} {\bf 22}, 020302 (2013)

Arsenijevi\' c M., Jekni\' c-Dugi\' c J. and Dugi\' c M., 2013b, A
Limitation of the Nakajima-Zwanzig projection method.
arXiv:1301.1005 [quant-ph]

Atkins P. and Friedman R, 2005, Molecular Quantum Mechanics
(Oxford: Oxford University Press)

Auyang S.,  1998, Foundations of Compex-System Theories. Cambridge
University Press, Cambridge

Bahrami M., Shafiee A., Bassi A., 2012, Decoherence Effects on
Superpositions of Chiral States in a Chiral Molecule. {\it Phys.
Chem. Chem. Phys.} {\bf 14}, 9214

Bassi A., Durr D., Hinrichs G., 2013. Uniqueness of the equation
for state-vector collapse.  arXiv:1303.4284 [quant-ph]

Bellomo B.,Campogno G. and Petruccione F., 2005, Initial
correlations effects on decoherence at zero temperature. {\it J.
Phys. A: Math. Gen.} {\bf 38}, 10203

Bellomo B., Compagno G., Lo Franco R., Ridolfo A., Savasta S.,
2011, Dynamics and extraction of quantum discord in a multipartite
open system, International. {\it J. Qu. Inform.} {\bf 9}, 1665

Bennett C. H., Brassard G., Crepeau C., Jozsa R., Peres A., and
Wootters W. K., 1993, Teleporting an unknown quantum state via
dual classical and Einstein-Podolsky-Rosen channels. {\it Phys.
Rev. Lett.} {\bf 70}, 1895

Breuer H.-P., Petruccione F., 2002, The Theory of Open Quantum
Systems, Oxford University Press, Oxford

Breuer H.P., Gemmer J. and Michel M., 2006, Non-Markovian quantum
dynamics: Correlated projection superoperators and Hilbert space
averaging. {\it Phys. Rev. A} {\bf 73}, 016139.

Breuer H.-P., Laine E.-M. and Piilo J., 2009,  Measure for the
Degree of Non-Markovian Behavior of Quantum Processes in Open
Systems. {\it Phys. Rev. Lett.} {\bf 103}, 210401

Brodutch A., Datta A., Modi K., Rivas A. and Rodriguez- Rosario C.
A., 2013, Vanishing quantum discord is not necessary for
completely-positive maps. |\it Phys. Rev. A} {\bf 87}, 042301

Brukner \v C., Zeilinger A., 1999, Operationally Invariant
Information in Quantum Measurements. {\it Phys. Rev. Lett.} {\bf
83} 3354

Brus D., 2002, Characterizing entanglement. {\it J. Math. Phys.}
{\bf 43}, 4237

Caban P., Podlaski K., Rembielinski J., Smolinski K. A., Walczak
Z., 2005, Entanglement and Tensor Product Decomposition for Two
Fermions. {\it J.Phys. A} {\bf 38}, L79

Caldeira A. O., Leggett A. J., 1983, Path integral approach  to
quantum Brownian motion. {\it Physica} {\bf 121A}, 587

Caspi S., Ben-Jacob E., 2000, Conformation changes and folding of
proteins mediated by Davydov's soliton. {\it Physics Letters A}
{\bf 272}, 124

Ciancio E., Giorda P., Zanardi P., 2006, Mode transformations and
entanglement relativity in bipartite Gaussian states. {\it Phys.
Lett.} A {\bf 354}, 274

Cohen-Tanoudji C.,  Dalbart J., 2006, Manipulating Atoms with
Photons, in {\it The New Physics for the Twenty-first Century},
Ed. G. Fraser, Cambridge University Press, Cambridge UK, pp.
145-171

Coles P. J., 2012,  Unification of different views of decoherence
and discord. {\it Phys. Rev. Lett.} {\bf 85}, 042103

De la Torre A. C., Goyeneche D., Leitao L., 2010, Entanglement for
all quantum states. {\it Europ. J. Phys.} {\bf 31}, 325

Dieks D., 1998, Preferred Factorizations and Consistent Property
Attribution, in {\it Quantum Measurement: Beyond Paradox}, Richard
E. Healey and Geoffry Hellman, eds., Univer. Minnesota Press,
pp.144-151

Dill K. A., Chan H. S., 1997, From Levinthal to pathways to
funnels. {\it Nature Struct. Biol.} {\bf 4}, 10

Dugi\' c M. 1996,  On the occurrence of decoherence in
nonrelativistic quantum mechanics. {\it Phys. Scr.} {\bf 53}, 9

Dugi\' c  1997,   On diagonalization of a composite-system
observable. Separability. {\it Phys. Scr.} {\bf 56}, 560

Dugi\' c M., 2000, Decoherence-Induced Suppression of Decoherence
in Quantum Computation. {\it Quantum Computers and Computing} {\bf
1}, 102

Dugi\' c M., \' Cirkovi\' c M. M., Rakovi\' c D., 2002, On a
Possible Physical Metatheory of Consciousness. {\it Open Syst.
Inf. Dyn.} {\bf 9}, 153

Dugi\' c M., 2012, Delayed choice without choice. arXiv:1211.1574
[quant-ph]

Dugi\' c M. Arsenijevi\' c M., Jekni\' c-Dugi\' c J., 2013.
Quantum Correlations Relativity for Continuous Variable Systems.
{\it Sci. China PMA} {\bf 56}, 732

Dugi\' c M., Jekni\' c J., 2006, What is "System": Some
Decoherence-Theory Arguments. {\it Int. J. Theor. Phys.} {\bf 45},
2249

Dugi\' c M., Jekni\' c-Dugi\' c J., 2008, What Is "System": The
Information- Theoretic Arguments. {\it Int. J. Theor. Phys.} {\bf
47}, 805

Dugi\' c M.,  Jekni\' c-Dugi\' c J., 2009, On a consistent quantum
adiabatic theory of molecules.  {\it The Old and New Concepts of
Physics} Vol. VI, 477

Dugi\' c M., Jekni\' c-Dugi\' c J., 2012, Parallel decoherence in
composite quantum systems. {\it Pramana - J. Phys.} {\bf 79}, 199

Dugi\' c M., Rakovi\' c D., Jekni\' c-Dugi\' c J., Arsenijevi\' c
M., 2012, The Ghostly Quantum Worlds. {\it NeuroQuantology}, {\bf
4}, 619

Dunningham J. A., Rau A. V., Burnett K., 2004, Transitivity of the
relative localization of particles. {\it J. Mod. Opt.}, {\bf 51},
2323

Durr D., Goldstein S., Zanghi N., 2012, Quantum Physics Without
Quantum Philosophy, Springer, Berlin

Elitzur A.C., Vaidman L., 1993, Quantum mechanical
interaction-free measurements. {\it Found. Phys.} {\bf 23}, 987

Facchi P., Florio G., Pascazio S., 2006,
Probability-density-function characterization of multipartite
entanglement.  {\it Phys. Rev. A} {\bf 74}, 042331

Fan H. Y., Hu L. Y., 2009, Infinite-dimensional Kraus operators
for describing amplitude-damping channel and laser process. {\it
Opt. Commun.} {\bf 282}, 932

Fel'dman E., Zenchuk A. I., 2012, Quantum correlations in
different density-matrix representations of spin-1/2 open chain.
{\it Phys. Rev.} A {\bf 86}, 012303

Fel'dman E. B., Zenchuk A. I., 2014a, Systems with stationary
distribution of quantum correlations: open spin-1/2 chains with XY
interaction, {\it Qu. Inf. Proc.} {\bf 13}, 201

E. B. Fel'dman, A. I. Zenchuk, 2014b,
 Robust stationary distributed discord in Jourdan-Wigner fermion system under perturbations of initial
 state, arXiv:1405.0811 [quant-ph]

Ferraro A., Olivares S., Paris M. G. A., 2005, Gaussian States in
Quantum Information. Bibliopolis, Napoli

Ferraro A,, Aolita L,, Cavalcanti D,, Cucchietti F, M., Acin A.,
2010,  Almost all quantum states have non-classical correlations.
{\it Phys. Rev. A} {\bf 81}, 052318

Gelin M. F., Egorova D., Domcke W., 2011, Exact quantum master
equation for a molecular aggregate coupled to a harmonic bath.
{\it Phys. Rev. A} {\bf 84}, 041139

Gharibian S., 2010, Strong NP-Hardness of the Quantum Separability
Problem. {\it Qu. Inform. and Comput.} {\bf 10}, 343

Giulini, D., Joos, E., Kiefer, C., Kupsch, J., Stamatescu, I.-O.,
Zeh, H. D., 1996, Decoherence and the Appearance of a Classical
World in Quantum Theory. Springer, Berlin

Gribov L. A., Mushtakova S. P., 1999, Quantum Chemistry.
Gardariki, Moscow (in Russian)

Gribov L. A., Magarshak Y. B., 2008, To the problem of formulation
of basic principles in the theory of molecular structure and
dynamics. {\it The Old and New Concepts of Physics} Vol. V, 191

Gr\" oblacher S., Trubarov A., Prigge N., Aspelmeyer M., Eisert
J., 2013, Observation of non-Markovian micro-mechanical Brownian
motion.\\* arXiv:1305.6942v1

Hackermuller L., Uttenthaler S., Hornberger K., Reiger E., Brezger
B., Zeilinger A., Arndt M., 2003, The wave nature of biomolecules
and fluorofullerenes. {\it Phys. Rev. Lett} {\bf 91}, 90408

Hackermuller L., Hornberger K., Brezger B., Zeilinger A., Arndt
M., 2004, Decoherence of matter waves by thermal emission of
radiation. {\it Nature}, {\bf 427}, 711

Haikka P., Cresser J. D., Maniscalco S., 2011,  Comparing
different non-Markovianity measures in a driven qubit system. {\it
Phys. Rev. A} {\bf 83}, 012112

Harshman N.L., 2012a, Observables and Entanglement in the Two-Body
System. {\it AIP Conf. Proc.} {\bf 1508}, 386

Harshman N.L., 2012b, Symmetries of Three Harmonically-Trapped
Particles in One Dimension. {\it Phys. Rev. A} {\bf 86}, 052122

Harshman1 N. L., Ranade K. S., 2011. Observables can be tailored
to change the entanglement of any pure state. {\it Phys. Rev. A}
{\bf 84}, 012303

Harshman N. L., Wickramasekara S, 2007, Galilean and Dynamical
Invariance of Entanglement in Particle Scattering. {\it Phys. Rev.
Lett.} {\bf 98}, 080406

Henderson L, Vedral V., 2001, Classical, quantum and total
correlations. {\it J. Phys. A: Math. Gen.} {\bf 34}, 6899

Huang Y., 2014, Computing quantum discord is NP-complete. {\it New
J. Phys.} {\bf 16}, 033027

Hund F., 1927, Zur Dentung der Molekelspektren.III. {\it Z. Phys.}
{\bf 43}, 805

Jekni\' c-Dugi\' c J., 2009a, The
environment-induced-superselection model of the large molecules
conformational stability and transitions. {\it Eur. Phys. J.} D
{\bf 51}, 193

Jekni\' c-Dugi\' c J., 2009b, Protein folding: the optically
induced electronic excitations model. {\it Phys. Scr.} T{\bf 135},
014031

Jekni\' c-Dugi\' c J., Dugi\' c M., 2008, Multiple
System-Decomposition Method for Avoiding Quantum Decoherence. {\it
Chin. Phys. Lett.} {\bf 25}, 371

Jekni\' c-Dugi\' c J., Dugi\' c M., Francom A., 2014,  Quantum
Structures of a Model-Universe: Questioning the Everett
Interpretation of Quantum Mechanics, {\it Int. J. Theor. Phys.}
{\bf 53}, 169

Jekni\' c-Dugi\' c J., Dugi\' c M., Francom A., Arsenijevi\' c M.,
2012, Quantum Structures of the Hydrogen Atom, accepted {\it
OALib} (2014); arXiv:1204.3172 [quant-ph]

Jiang N. Q., Fan H. Y., Xi L. S., Tang L. Y., Yuan X. Z., 2011,
Evolution of a two-mode squeezed vacuum in the amplitude
dissipative channel. {\it Chin. Phys. B}  {\bf 20}, 120302

Jona-Lasinio G., Claverie P, 1986, Symmetry Breaking and Classical
Behaviour. {\it Prog. Theor. Phys., Suppl.}, {\bf 86}, 54

Kraus K., 1983, States Effects and Operations Fundamental Notions
of Quantum Theory, Lecture Notes in Physics Vol. 190. Springer-
Verlag, Berlin

Laine E.M., Piilo J., Breuer H.-P., 2010, Measure for the
non-Markovianity of quantum processes. {\it Phys. Rev. A} {\bf
81}, 062115

Lendlein A, Jiang H, Junger O, Langer R., 2005, Light-induced
shape-memory polymers. {\it Nature} {\bf 14}, 879

Levinthal C, 1968,  Are there pathways for protein folding? {\it
J. Chim. Phys. et de Phys.-Chim. Biol.} {\bf 65}, 44

Lim J., Tame M., Yee K. H., Lee J.-S., Lee J., 2014, Comment on
"Energy transfer, entanglement and decoherence in a molecular
dimer interacting with a phonon bath", {\it New J. Phys.} {\bf
16}, 018001

Liu Y. X., Ozdemir S. K., Miranowicz A., Imoto N., 2004, Kraus
representation of a damped harmonic oscillator and its
application. {\it Phys. Rev. A} {\bf 70} 042308

Luo L, Lu J., 2011,  Temperature Dependence of Protein Folding
Deduced from Quantum Transition. arXiv:1102.3748 [q-bio.BM]

Lutz E., 2003, Effect of initial correlations on short-time
decoherence. {\it Phys. Rev. A} {\bf 67},  022109

Lychkovskiy O., 2013, Dependence of decoherence-assisted
classicality on the ways a system is partitioned into subsystems.
{\it Phys. Rev. A} {\bf 87}, 022112

Ma X-S, Zotter S., Kofler J., Ursin R., Jennewein T., Brukner \v
C., Zeilinger A., 2012, Experimental delayed-choice entanglement
swapping. {\it Nature Phys.} {\bf 8}, 480

Maeda H., Norum D. V. L., Gallagher T. F., 2005, Microwave
Manipulation of an Atomic Electron in a Classical Orbit. {\it
Science} {\bf 307}, 1757

Manzano G., Galve F., Zambrini R., 2013, Avoiding dissipation in a
system of three quantum harmonic oscillators. {\it Phys. Rev. A}
{\bf 87}, 032114

Mc Weeney R., 1978, Methods of Molecular Quantum Mechanics.
Academic press, New York

Menda\v s  I. P., Popovi\' c D. B., 2010,  A generalization of the
Baker-Hausdorff lemma. {\it Phys. Scr.} {\bf 82}, 045007

Mermin N. D., 1998, The Ithaca Interpretation of Quantum
Mechanics. {\it Pramana - J. Phys.} {\bf 51}, 549

Modi K., Brodutch A., Cable H., Paterek T., Vedral V., 2012,
 The classical-quantum boundary for correlations: discord and related measures.
{\it Rev. Mod. Phys.} {\bf 84}, 1655

Nakajima S., 1958, On quantum  theory of transport phenomena. {\it
Prog. Theor. Phys.} {\bf 20}, 948

Nielsen M. A., Chuang I. A., 2000,  Quantum Computation and
Quantum Information. Cambridge University Press, Cambridge

Ollivier H., Zurek W. H., 2001, Quantum Discord: A Measure of the
Quantumness of Correlations. {\it Phys. Rev. Lett.} {\bf 88},
017901

Paz J. P., Roncaglia A. J., 2008, Dynamics of the Entanglement
between Two Oscillators in the Same Environment. {\it Phys. Rev.
Lett.} {\bf 100}, 220401

Peres A. 2000, Delayed choice for entanglement swapping. {\it J.
Mod. Opt.} {\bf 47}, 139

Primas H., 1994, in Logic, Methodology and Philosophy of Science
IX, D. Prawitz et al (Eds.). Elsevier Science, Singapore

Pusey M. F., Barrett J., Rudolph T., 2012. On the reality of the
quantum state, {\it Nature Phys.} {\bf 8}, 476

Ragy S., Adesso G. 2012, Nature of light correlations in ghost
imaging, {\it Sci. Rep.} {\bf 2},  651

Rakovi\' c D. et al, 2014, ON MACROSCOPIC QUANTUM PHENOMENA IN
BI\-O\-MO\-LECULES AND CELLS: FROM LEVINTHAL TO HOPFIELD, {\it
BioMed Res. Int.} (in press)

Rau A. V., Dunningham  J. A., Burnett K., 2003,
Measurement-induced relative-position localization through
entanglement. {\it Science} {\bf 301}, 1081

Rivas A., Huelga S. F., Plenio M. B., 2010a, Entanglement and
Non-Markovianity of Quantum Evolutions. {\it Phys. Rev. Lett.}
{\bf 105}, 050403

Rivas A., Douglas A., Plato K., Huelga S. F., Plenio M. B., 2010b,
Markovian master equations: a critical study. {\it New J. Phys.}
{\bf 12}, 113032

Rivas A., Huelga S. F., 2011, Open Systems Theory. An
Introduction. SpringerBriefs, Springer, Berlin

Rodriguez-Rosario C. A., Sudarshan E. C. G., 2011, Non-Markovian
Open Quantum Ssstems: System-environment Correlations in Dynamical
Maps. {\it Int. J. Quantum Inf.} {\bf 9}, 1617

Saunders S., Barret J., Kent A., Wallace D., (Eds.), 2010, Many
Worlds? Everett, Quantum Theory, and Reality. Oxford University
Press, Oxford

Salih H., Li Z.-H., Al-Amri, Suhail Zubairy M. S., 2013, Protocol
for direct counterfactual quantum communication. {\it
Phys.Rev.Lett.} {\bf 110}, 170502

Schlosshauer M., 2004, Decoherence, the measurement problem, and
interpretations of quantum mechanics. {\it Rev. Mod. Phys.} {\bf
76}, 1267

Stokes A., 2012, Noncovariant gauge fixing in the quantum Dirac
field theory of atoms and molecules. {\it Phys. Rev. A} {\bf 86},
012511

Stokes A., Kurcz A., Spiller T. P., Beige A.,  2012,  Extending
the validity range of quantum optical master equations. {\it Phys.
Rev. A} {\bf 85}, 053805

Svr\v cek M., 2012, Centre-of-Mass Separation in Quantum
Mechanics: Implications for the Many-Body Treatment in Quantum
Chemistry and Solid State Physics, in {\it Advances in the Theory
of Quantum Systems in Chemistry and Physics}, Volume 22 , Part 7,
pp. 511-552, Eds., P. E. Hoggan, E. J. Brandas, J. Maruani, P.
Piecuch  and G. Delgado-Barrio

Terra Cunha M. O., Dunningham J. A., Vedral V., 2007, Entanglement
in single-particle systems. {\it Proc. R. Soc.} A {\bf 463}, 2277

Tommasini P., Timmermans E., Piza A. F. R. D., 1998, The hydrogen
atom as an entangled electronproton system. {\it Am. J. Phys.}
{\bf 66}, 881

Trost J., Hornberger K., 2009, Hund's paradox and the collisional
stabilization of chiral molecules. {\it Phys. Rev. Lett} {\bf
103}, 023202

Vedral V., 2003, Entanglement in The Second Quantization
Formalism. {\it Central Eur.J.Phys.} {\bf 1}, 289

Vedral V., 2010, Decoding Reality: The universe as a Quantum
Computer. Oxford Unviersity Press, Oxford

von Neumann J., 1955, Mathematical Foundations of Quantum
Mechanics. Princeton University Press, Princeton

Wheeler J. A., 1978, in Mathematical Foundations of Quantum
Theory. Academic, New York, pp. 9-48

Wichterich H. C., 2011, Entanglement Between Noncomplementary
Parts of Many-Body Systems. Springer Theses, Springer

Xu J-S, Li C-F, 2013, Quantum discord under system-environment
coupling: the two-qubit case. {\it Int. J. Mod. Phys. B} {\bf 27},
1345054

Zanardi P., 2001, Virtual Quantum Subsystems. {\it Phys. Rev.
Lett.} {\bf 87}, 077901

Zanardi P., Lidar D. A., Lloyd S., 2004, Quantum Tensor Product
Structures are Observable Induced. {\it Phys. Rev. Lett.} {\bf
92}, 060402

Zeh, H. D., 1993, There are no Quantum Jumps, nor are there
Particles. {\it Physics Letters A} {\bf 172}, 189

Zeh H-D, 2005, Roots and Fruits of Decoherence. {\it Seminaire
Poincare} {\bf 2}, 1; available also as arXiv:quant-ph/0512078v2

Zeppenfeld M., Englert B. G. U., Glockner R., Prehn A., Mielenz
M., Sommer C., van Buuren L. D., MotschM. , Rempe G., 2012,
Sisyphus Cooling of Electrically Trapped Polyatomic Molecules.
{\it Nature} {\bf 491}, 570

Zhou N. R., Hu L. Y. and Fan H. Y., 2011, Dissipation of a
two-mode squeezed vacuum state in the single-mode amplitude
damping channel. {\it Chin. Phys. B} {\bf 20}, 120301

Zurek W. H., 1998, Decoherence, einselection and the existential
interpretation (the rough guide). {\it Philos. Trans. R. Soc.
London Ser.} A {\bf 356}, 1793

Zurek W. H., 2003, Decoherence, einselection, and the quantum
origins of the classical.  {\it  Rev. Mod. Phys.} {\bf 75}, 715

Zwanzig R., 1960, Ensemble method in the theory of
irreversibility. {\it J. Chem. Phys.} {\bf 33}, 1341

\cleardoublepage

{\flushleft {\bf Index}
\bigskip

Accessible/accessibility [direct measurability]  2, 4, 16, 27, 36,
47, 48, 49, 52, 57, 84, 125

Bohm's theory/interpretation 114, 122

Center of mass 1, 7, 11, 45, 47, 48, 50, 51, 57, 62, 80, 96, 106,
117

Chemical model of molecule 28, 29, 56

Classicality 59, 81, 83, 84, 86, 87,  97, 99, 101, 105, 108, 124

Coarse graining 1, 10, 14, 30, 54, 55, 56

Collapse [of state] models 58, 114, 115, 120, 122

Conjectures

\hskip10mm Conjecture 1 81, 124

\hskip10mm Conjecture 2 101, 106, 125

Consistent histories 114, 115

Delayed choice 54, 119

Emergent structures 116

Entanglement relativity (ER) 20, 21, 24, 54, 55, 57, 94, 112, 115,
119, 126

Entanglement swapping 11, 19, 53, 54, 119

Everett branching 114, 115, 116,  118

Everett theory/interpretation 114, 115, 116,  118, 122

Fine graining 1, 10, 14, 44, 54

Gaussian states 6, 26, 55,  65, 92, 101, 115

Interaction-free quantum measurement 120

Levinthal paradox 4, 29, 34,  35, 40, 42, 43

Linear canonical transformations (LCTs) and Structures

\hskip10mm Global/Local LCTs (Structures) 10-14, 16, 17,  34, 44,
47, 49, 50, 54, 55, 59, 60, 68, 82, 85, 88, 94, 103-114

\hskip10mm Reducible/Irreducible LCTs (Structures) 4, 10-13, 33,
60, 68, 82, 103-107, 109, 113

\hskip10mm Trivial/Nontrivial LCTs (Structures) 10-14, 18, 26, 34,
45, 54, 56, 59, 68, 82, 107, 114

Many Worlds Interpretation  [see Everett interpretation]

Master equations 43, 61, 65, 67-76, 80, 87-89, 93, 94, 97, 99, 101

Nakajima-Zwanzig [projection] method 27, 61, 72

Parallel occurrence of decoherence 4, 27, 58, 82, 83, 108, 119,
124

Protein (un)folding 4, 27, 29, 34-36, 54

Quantum Brownian motion 4, 43, 52, 62, 65, 67, 70, 80, 83,  85,
108, 114-116-118

Quantum chemistry model of molecule 12, 28, 30, 35

Quantum correlations relativity 17, 25, 45, 54-57, 61, 73, 75, 79,
101, 111, 114, 124

Quantum discord relativity (QDR) 4, 22, 25

Quantum reference frame 4,  38, 104, 108, 110

Relative positions 1, 7, 12, 47, 52, 53, 62, 63, 80, 111

Relativity of

\hskip10mm Discord (see Discord relativity)

\hskip10mm Entanglement (see Entanglement relativity)

\hskip10mm Locality 16, 47, 50, 57-59

\hskip10mm System 16, 47, 58, 59

Solid-state molecule model 28, 29

Symmetries 8, 10, 13, 103, 105, 110, 113, 119

The Caldeira-Leggett model 64

Thermal

\hskip10mm Bath 43, 61, 64, 70

\hskip10mm Equilibrium 41, 43, 51, 62, 67

\hskip10mm Relaxation 41, 42, 71}

\pagebreak

\noindent {\bf Supplement}

\bigskip

\noindent This Supplement serves to complete and partially extend
the contents of the body text.

\bigskip

\noindent {\bf S.1 Decoherence preferred states and observables}

\bigskip

\noindent The "orthodox" approach to decoherence deals directly
with the unitary operator of time evolution for the composite
system "(open)system+environment" ($S+E$), %%
\begin{equation}
U = e^{-\imath t H/\hbar},
\end{equation}

\noindent where the total system's Hamiltonian: %%
\begin{equation}
H = H_S + H_E + H_{SE}.
\end{equation}

\noindent If the self-Hamiltonians, $H_S, H_E$, can be neglected
(e.g. in the collisional decoherence), then the total-system's
(pure) state evolves, approximately, as: %%
\begin{equation}
e^{-\imath t H_{SE}/\hbar} \vert \psi\rangle_S \vert \chi
\rangle_E = \sum_i c_i \vert \phi_i\rangle_S \vert
\chi_i(t)\rangle_E.
\end{equation}

\noindent In eq.(151): $\vert \phi_i\rangle_S$ is a basis
diagonalizing the interaction term, $H_{SE}$, and the precise form
of the $\vert \chi_i(t\rangle_E)$ depends of the form of $H_{SE}$.

\smallskip

\noindent It can be shown that eq.(151) cannot be valid, in
principle, if the interaction is not of the so-called separable
kind (Dugi\' c 1996, 1997), which, in turn, can be always
diagonalized by an orthonormalized basis in the system's Hilbert
state space, $\mathcal{H}_S$. However, whenever such basis exists
[and need not be unique], the states $\vert \phi_i\rangle_S$
diagonalizing $H_{SE}$ represent the "pointer basis" states for
the $S$ system--the carriers of the quasi-classical behavior of
the open system $S$.

\smallskip

\noindent If the interaction $H_{SE}$ is only approximately of the
separable-kind, then there exists a not-necessarily-exactly
orthonormalized basis, which approximately diagonalizes $H_{SE}$.
Such a basis represents the approximate pointer basis states--the
"preferred" set of states carrying the quasiclassical behavior of
$S$. For the continuous-variable systems, the minimal-uncertainty
gaussian states (the Sudarshan-Glauber coherent states) are
typical "preferred" states [see Chapters 6 and 7].

\smallskip

\noindent Whenever the self-Hamiltonians cannot be neglected,
non-commutativity $[H_S,$ $H_{SE}] \neq 0$ does not allow a choice
of the exact pointer basis states. Then reading out the spectral
form of the [separable kind] interaction $H_{SE}$ can point out
the candidates for the approximate pointer basis. There are the
cases not allowing existence of any "preferred" states for the
open system (Dugi\' c 1996, 1997).

\smallskip

\noindent Formally, the spectral form of the interaction provides
information about the pointer basis: %%
\begin{eqnarray}
&\nonumber& H_{SE} = \sum_{i,j} h_{ij} P_{Si} \otimes \Pi_{Ej}
\quad [{\rm exact \quad pointer \quad basis}]
\\&&
H_{SE} = \sum_{i,j} h_{ij} P_{Si} \otimes \Pi_{Ej} + H', \Vert H'
\Vert \ll \Vert H_{SE} \Vert \quad [{\rm approx-pointer-basis}]
\nonumber \\&& [H_S, H_{SE}] \neq 0 \quad [{\rm approximate \quad
or \quad no \quad pointer \quad basis}].
\end{eqnarray}

\bigskip

\noindent {\bf S.2 The continuous-variable-system transformations}

\bigskip

\noindent {\it The center-of-mass and the relative positions}

\smallskip

\noindent For a pair of one-dimensional particles see eq.(4) in
the body text. Generalization to realistic three-dimensional
particles is straightforward. Here we consider a many-particles
system\footnote{\small For further details and some proofs see
e.g. McWeeney 1978.}.

\smallskip

\noindent There are formally similar,  not yet equivalent,
variants of  the relative positions ($R$) variables. Here we adopt
the following definitions as a direct generalization of eq.(4) in
the body text: %%
\begin{equation}
\vec R_{CM} = {\sum_i m_i\vec r_i\over \sum_i m_i}, \vec \rho_{Rl}
= \vec r_i - \vec r_j, \quad (i,j) \equiv l =1, 2, 3,...,N-1.
\end{equation}

\noindent Eq.(153) gives rise to the following transformations of
the kinetic terms: %%
\begin{equation}
\sum_i {\vec p_i^2 \over 2 m_i} \to {\vec P_{CM}^2 \over 2 M} +
\sum_i {\vec p_{Ri}^2 \over 2 \mu_i} + \sum_{i, j} {m_{i+1}
m_{j+1} \over m_i m_jM} \vec p_{Ri} \cdot \vec p_{Rj}
\end{equation}

\noindent In eq.(154): $M = \sum_i m_i$,  the reduced masses,
$\mu_i = m_{i+1} (M - m_{i+1})/M$, while $[X_{CMi}, P_{CMj}] =
\imath \hbar \delta_{ij}$, and analogously for the $R$ system's
variables. Of course, then the total system's Hilbert state space
factorizes as $\mathcal{H}_{CM} \otimes \mathcal{H}_R$;
$\mathcal{H}_R = \otimes_{i=1}^{N-1} \mathcal{H}_{Ri}$. The third
term in eq.(154) is the so-called "mass polarization" term which,
for $i \neq j$, becomes internal interaction for the $R$ system.
For the $i = j$, the mass polarization term gives rise to the
following contribution to the kinetic energy of the $i$th
"relative particle": $\sum_i m_{i+1}^2 p_{Ri}^2 /m_i^2M$, which
will further be neglected.

\smallskip

\noindent The inverse to eq.(153) gives rise to: %%
\begin{equation}
\vec r_i = \vec R_{CM} + \sum_j \omega_{ij} \vec \rho_{Rj}
\end{equation}

\noindent with the real parameters $\omega$. Then the
distant-dependent interactions: %%
\begin{equation}
V(\vert \vec r_i - \vec r_j\vert ) = V(\vert \vec \rho_{Rl}\vert )
\end{equation}

\noindent become the external fields for the $R$ system. However,
interestingly enough, the external one-particle field: %%
\begin{equation}
V(\vec r_i) = V(\vec R_{CM} + \sum_j \omega_{ij} \vec \rho_{Rj})
\end{equation}

\noindent becomes the interaction between the $CM$ and $R$
systems.

\smallskip

\noindent Regarding the {\it molecules model}, eq.(48), the
atomic-nuclei $R_N$ subsystem is divided into two subsystems, the
$Rot_N$ (rotation of the atomic nuclei system as a whole described
by the Euler angles) and $K_N$ (the conformation system). The
kinetic term for rotation $T_{Rot_N} = \vec L^2_N/2I_N$, where
$\vec L_N$ is the molecule angular momentum and $I_N$ is the
moment of inertia [in the simplest form]. The kinetic term
$T_{K_N}$ is of the standard form eq.(154) for the remaining
"relative positions" variables.

\smallskip

\noindent Regarding the {\it QBM model}, eq.(66), the LCTs give
rise to the terms proportional to $X_{CM}^2$ and to $\rho_{Rl}
\rho_{Rl'}$, while there is the bi-linear coupling $X_{CM} \sum_i
\kappa_i$ $\sum_l \omega_{il} \rho_{Rl}$. So, even for the
original free particle model, the new structure provides the
harmonic-oscillator Brownian particle $S'$ ($S' \equiv CM$). The
above terms $\rho_{Rl} \rho_{Rl'}$ give for $l=l'$ the additional
terms $\rho_{Rl}^2$ thus providing a harmonic term for the new
environment. On the other hand, the original harmonic term (for
the Brownian particle and/or for the environment), $\vec r^2_i =
(\vec R_{CM} + \sum_j \omega_{ij} \vec \rho_{Rj})^2$, thus
providing the harmonic terms for both the new Brownian particle as
well as for the new environment, and the bilinear coupling,
$X_{CMi} \rho_{Rj}$, which directly gives rise to the coupling in
eq.(70).

\smallskip

\noindent Regarding the {\it quantum reference system}, the
expression eq.(156) is {\it not }  applicable. The reason is
rather simple. For the $1$ system as a QRF system: the relative
positions $\vec \rho_{Rl} = \vec r_1 - \vec r_l$. But then
$V(\vert \vec r_i - \vec r_j\vert) = V(\vert \sum_l \omega_{il}
\vec \rho_{Rl} - \sum_{l'} \omega_{jl'} \vec \rho_{Rl'}\vert)$ for
$i,j \neq 1$, in contrast to eq.(156).

\smallskip

\noindent {\it Building the boson Fock space}

\smallskip

\noindent For a single harmonic oscillator described by the
canonical position and momentum, $x$ and $p$, with the mass $m$
and frequency $\omega$, one can introduce the "annihilation" and
the "creation" operators: %%
\begin{equation}
a = \sqrt{m\omega\over 2\hbar} (x + {\imath \over m\omega} p),
a^{\dag} = \sqrt{m\omega\over 2\hbar} (x - {\imath \over m\omega}
p),
\end{equation}

\noindent with the commutation relation: %%
\begin{equation}
[a, a^{\dag}] = I.
\end{equation}

\noindent The inverse to eq.(158): %%
\begin{equation}
x = \sqrt{\hbar \over 2m\omega}(a+ a^{\dag}), p = \imath
\sqrt{m\omega\hbar \over 2} (a^{\dag} - a).
\end{equation}

\noindent The Hermitian operator $N \equiv a^{\dag}a$ defines the
eigenstates: %%
\begin{equation}
N \vert n\rangle = n \vert n \rangle, n = 0,1,2,3,...
\end{equation}

\noindent The states $\{\vert n\rangle\}$ constitute an
orthonormalized basis in the so-called Fock space,
$\mathcal{H}_F$. The equalities hold: %%
\begin{equation}
a \vert n\rangle = \sqrt{n} \vert n-1\rangle, a^{\dag} \vert
n\rangle = \sqrt{n+1} \vert n+1\rangle,
\end{equation}

\noindent from which: %%
\begin{equation}
\vert n \rangle = {a^{\dag n} \over \sqrt{n!}} \vert 0 \rangle,
\quad a\vert 0\rangle = 0; \quad \langle 0 \vert 0 \rangle = 1,
\end{equation}

\noindent In wide use (especially in quantum optics) are the
so-called quadratures (Hermitian) operators defined as $X =
\sqrt{m\omega /\hbar} x$ and $P = \sqrt{1/m\hbar \omega} p$.

\smallskip

\noindent Generalization to the many-particle (multimode) system
is straightforward: $(x_i, p_i) \to (a_i, a^{\dag}_i)$, so that
$[a_i, a^{\dag}_j] = \delta_{ij}$. Then the total system's Fock
space $\mathcal{H}_F = \otimes_i \mathcal{H}_{Fi}$, where
$\mathcal{H}_{Fi}$ is the Fock space for individual oscillators
(modes). The eigenbasis of $N = \sum_i N_i$, $\{\vert n \rangle =
\otimes_i \vert n_i \rangle\}$, $N_i \vert n_i\rangle  = n_i \vert
n_i \rangle$, while $n_i = 0,1,2,...$, $\forall{i}$. For the
vacuum state, $\vert 0\rangle = \otimes_i \vert 0\rangle_i$, $a_i
\vert 0 \rangle = 0, \forall{i}$.

\smallskip

\noindent {\it The Bogoliubov-like transformations}

\smallskip

\noindent For one mode (or one harmonic oscillator) defined on
appropriate Fock space $\mathcal{H}_F$, a pair of the
"annihilation" and "creation" operators are defined by the
commutator relation eq.(159).

\smallskip

\noindent The one-mode {\it boson-translation transformation} is
defined by: %%
\begin{equation}
a \to a(\theta) = a + \theta, \quad \theta \in \mathcal{C}.
\end{equation}

\noindent This is a canonical transformation as %%
\begin{equation}
[a(\theta), a^{\dag}(\theta)] = I, \forall{\theta}.
\end{equation}

\noindent It can be shown this is an unitary transformation.
However, this transformation does not preserve the original Fock
space as: %%
\begin{equation}
a(\theta) \vert 0 \rangle = \theta \vert 0 \rangle \neq 0.
\end{equation}

\noindent Therefore, the new vacuum state, $\vert 0(\theta)
\rangle$, is the vacuum state of another Fock space,
$\mathcal{H'}_F$, which is unitary related to the original one.

\smallskip

\noindent For a single mode,  the {\it Bogoliubov transformation}
is defined as: %%
\begin{equation}
b = u a  + v a^{\dag}, \quad b^{\dag} = u^{\ast} a^{\dag}  +
v^{\ast} a,
\end{equation}

\noindent where $[a, a^{\dag}] = 1$ allows for the new bosonic
operators $[b, b^{\dag}] = 1$ if $\vert u\vert^2 - \vert v\vert^2
= 1$. The later condition allows for the following
parametrization: $u = e^{\imath \theta_1} \cosh r$ and $v =
e^{\imath \theta_2} \sinh r$.

\smallskip

\noindent For the fermion system there are analogous
transformations with the condition $\vert u\vert^2 + \vert
v\vert^2 = 1$, i.e. with the parametrization: $u = e^{\imath
\theta_1} \cos r$ and $v = e^{\imath \theta_2} \sin r$.

\smallskip

\noindent The {\it multimode generalization} is straightforward.
e.g. For the boson system, for which $a_i \vert 0 \rangle = 0,
\forall{i}$, the operators: %%
\begin{equation}
a'_i = \sum_j (u_{ij} a_j + v_{ij} a^{\dag}_j)
\end{equation}

\noindent satisfy the bosonic commutator relations, $[a'_i,
a^{'\dag}_j] = \delta_{ij}$, if the condition, $\sum_p (u_{ip}
u^{\ast}_{jp} - v_{ip} v^{\ast}_{jp}) = 1$, is fulfilled. For the
fermion system, the analogous condition reads as: $\sum_p (u_{ip}
u^{\ast}_{jp} + v_{ip} v^{\ast}_{jp}) = 1$.

\pagebreak

\noindent {\bf S.3 Some spin-system related transformations}

\bigskip

\noindent Below, we consider  nontrivial\footnote{\small By
"trivial" transformations, we assume regrouping, fine- or
coarse-graining of the constituent particles.} transformations of
variables.

\smallskip

\noindent The {\it Holstein-Primakoff transformation} targets the
spin observable $\vec S$ (of the spin quantum number $s$) with the
standard basis $\vert s, m_s\rangle$; $m_s = -s, -s+1, -s+2, ,...
s-2, s-1, s$. The leading idea of the transformation is the
following correspondence: %%
\begin{equation}
\vert s, -s + n \rangle \to (n!)^{-1/2} a^{\dag n} \vert 0\rangle
\end{equation}

\noindent where appear the bosonic creation operator $a^{\dag}$
and the vaccum state $\vert 0 \rangle$. Then the transformation is
defined by the following prescription: %%
\begin{equation}
S_z = \hbar (s - a^{\dag}a), S_+ = \hbar \sqrt{2s - a^{\dag}a} a,
S_+ = \hbar a^{\dag} \sqrt{2s - a^{\dag}a}.
\end{equation}

\noindent Given the standard spin commutator relations, $[S_i,
S_j] = \imath\hbar \epsilon_{ijk} S_k$, the bosonic commutator
relation is satisfied, $[a, a^{\dag}] = I$. Generalization to a
system of spins (i.e. to a multimode prescription) is
straightforward.

\smallskip

\noindent The {\it Jordan-Wigner} transformation exhibits
important prescription between the one-dimensional spin-1/2 chain
and a fermion system. If $\vec S_i$ distinguishes a set of $N$
one-dimensional-chain of spins, the spin projections
$S_{i\alpha}$, $i=1,2,...N, \alpha=x,y,z$, allows introduction of
the fermionic annihilation and creation opperators as: %%
\begin{equation}
a_i = (-2)^{i-1} S_{1z} S_{2z}...S_{(i-1)z} S_{i-} \Leftrightarrow
S_{jz} = a^{\dag}_j a_j - 1/2,
\end{equation}

\noindent while $S_- = S_x + \imath S_y$ and $\{a_i, a^{\dag}_j\}
= \delta_{ij}$.

\smallskip

\noindent {\it Fourier transform for fermion system} connects
mutually the two sets of fermion operators. Consider the fermion
operators $a_i$ and $a^{\dag}_i$  that are transformed as: %%
\begin{equation}
\beta_m = \sum_i d_{mi} a_i.
\end{equation}

\noindent For the $N$-fermions system, if chosen %%
\begin{equation}
d_{mi} = \sqrt{2\over N+1} \sin kj, \quad k = {n\pi\over N+1},
\quad n = 1,2,...,N,
\end{equation}

\noindent then the anticommutator relations are satisfied: %%
\begin{equation}
\{\beta_m, \beta_n^{\dag}\} = \delta_{mn}.
\end{equation}

\end{document}